\documentclass[aps,prx,nobibnotes,twocolumn,superscriptaddress,longbibliography]{revtex4-2}

\usepackage{amsfonts}
\usepackage{mathrsfs}
\usepackage{amsmath}
\usepackage{color}
\usepackage{graphicx}
\usepackage{bm}
\usepackage{amssymb}
\usepackage{xspace}
\usepackage{epstopdf}
\usepackage{dcolumn}
\usepackage{longtable}
\usepackage{multirow}
\usepackage{float}
\usepackage{comment}
\usepackage{lineno}

\providecommand{\tabularnewline}{\\}

\usepackage[colorlinks=true, letterpaper=true, pdfstartview=FitV,  linkcolor=blue, citecolor=blue, urlcolor=blue]{hyperref}

\makeatother

\begin{document}

\title{Encyclopedia of emergent particles in three-dimensional crystals }

\author{Zhi-Ming Yu}
\email{Z. M. Yu and Z. Zhang contributed equally to this work.}
\affiliation{Key Lab of Advanced Optoelectronic Quantum Architecture and Measurement
(MOE), Beijing Key Lab of Nanophotonics \& Ultrafine Optoelectronic
Systems, and School of Physics, Beijing Institute of Technology, Beijing
100081, China}
\author{Zeying Zhang}
\email{Z. M. Yu and Z. Zhang contributed equally to this work.}
\affiliation{College of Mathematics and Physics, Beijing University of Chemical Technology, Beijing 100029, China}
\author{Gui-Bin Liu}
\affiliation{Key Lab of Advanced Optoelectronic Quantum Architecture and Measurement
(MOE), Beijing Key Lab of Nanophotonics \& Ultrafine Optoelectronic
Systems, and School of Physics, Beijing Institute of Technology, Beijing
100081, China}
\author{Weikang Wu}
\affiliation{Research Laboratory for Quantum Materials, Singapore University of
Technology and Design, Singapore 487372, Singapore}
\author{Xiao-Ping Li}
\affiliation{Key Lab of Advanced Optoelectronic Quantum Architecture and Measurement
(MOE), Beijing Key Lab of Nanophotonics \& Ultrafine Optoelectronic
Systems, and School of Physics, Beijing Institute of Technology, Beijing
100081, China}
\author{Run-Wu Zhang}
\affiliation{Key Lab of Advanced Optoelectronic Quantum Architecture and Measurement
(MOE), Beijing Key Lab of Nanophotonics \& Ultrafine Optoelectronic
Systems, and School of Physics, Beijing Institute of Technology, Beijing
100081, China}
\author{Shengyuan A. Yang}
\affiliation{Research Laboratory for Quantum Materials, Singapore University of
Technology and Design, Singapore 487372, Singapore}
\author{Yugui Yao}
\email{ygyao@bit.edu.cn}
\affiliation{Key Lab of Advanced Optoelectronic Quantum Architecture and Measurement
(MOE), Beijing Key Lab of Nanophotonics \& Ultrafine Optoelectronic
Systems, and School of Physics, Beijing Institute of Technology, Beijing
100081, China}

\begin{abstract}
The past decade has witnessed a surge of interest in exploring emergent particles in condensed matter systems.
Novel particles, emerged as excitations around exotic band degeneracy points, continue to be reported in real materials and artificially engineered systems, but so far, we do not have a complete picture on all possible types of particles that can be achieved. Here, via systematic symmetry analysis and modeling, we accomplish a complete list of all possible particles in time-reversal-invariant systems. This includes both spinful particles such as electron quasiparticles in solids, and spinless particles such as phonons or even excitations in electric-circuit and mechanical networks. We establish detailed correspondence between the particle, the symmetry condition, the effective model, and the topological character.
This obtained encyclopedia concludes the search for novel emergent particles and provides concrete guidance to achieve them in physical systems.
\end{abstract}
\maketitle

Unlike elementary particles in high energy physics which are strongly constrained by the Poincar\'{e} symmetry, the emergent particles in condensed matters only need to respect much smaller subgroups of the Poincar\'{e} symmetry---the crystal space group symmetries \cite{Bradley2009Mathematical-Oxford}.
It follows that the emergent particles can have much richer types and dramatically different properties \cite{Chiu2016Classification-RoMP,Yan2017Topological-ARoCMP,Armitage2018Weyl-RoMP,Bradlyn2016Dirac-S,Kruthoff_PRX_2017,LuFOP-review,Beenakker-PRL2016,YuMOPRL2016,Zilberberg-PRL2019,LiuY-PRL2020}.
For example, besides the elementary Weyl, Dirac and Majorana type fermions, 
electron quasiparticles in solids, as emerged around certain band degeneracy points, can exhibit a variety of pseudospin structures \cite{Bradlyn2016Dirac-S,DDirac2016-PRL,Weng2016Topological-PRB,Zhu2016Triple-PRX},
dispersion types \cite{Xu2011Chern-Prl,Yang2014Classification-Nc,Chuanwei_PRL_2015,Soluyanov},
topological charges \cite{Fang2012Multi-Prl,BJYang-PRL2018,TB-Science-2019}, and their distribution in momentum space can form manifolds of various topologies \cite{Weng2015Topological-PRB,Fang2016Topological-CPB,Tomas_NC_2016,Wang2017Hourglass-Nc,Wu2018Nodal-PRB}. The study has also been extended to bosons like phonons, photons and magnons \cite{CooperRMP,Yang1013Sci,Acoustic_NS_SciAdv_xiao_2020,Magnons_PRL_2017}, and excitations in artificial periodic systems like electric circuit arrays and mechanical networks \cite{ImhofNP2018,Kane_NP_2014,Po_PRB_MechNL_2016}.

A central task of the field is to classify all possible emergent particles for each space group. So far, most works are case-by-case type studies, focusing on a specific material or a specific type of emergent particle \cite{Wan2011Topological-PRB,Wang2012Dirac-PRB,WengPRX_WP_2015,
Xing-QiuPRL-2016,Zhong2017Three-Nc,Zhang_PRBHNL_2018,ZhangRW_NL_2020,nie-six-PRR2021}.
There do exist a few systematic attempts \cite{Bradlyn2016Dirac-S,ChangNM2018,Yu2019Quadratic-PRB,WuWK-prb2020,Liu_WPhons_2020}. For example, Bradlyn \emph{et al.} classified emergent fermions beyond Weyl and Dirac types at high-symmetry points in the Brillouin zone (BZ) in the presence of spin-orbit coupling (SOC) and time reversal symmetry \cite{Bradlyn2016Dirac-S}. And based on the theory of symmetry indicators,  
thousands of topological semimetal materials have been identified \cite{Feng_Nat_2019,Vergniory_Nat_2019,Fang_Nat_2019}, which possess nontrivial emergent fermions. However, the grand task is still far from being accomplished: the emergent particles not at high-symmetry points (like those on high-symmetry lines) as well as the classification of spinless particles (as appropriate for the large class of bosonic and artificial systems) have not been systematically explored; and while the symmetry indicators are most useful for characterizing gapped phases, they do not give direct classification of the emergent particles.

In this work, we address this challenging task by offering a complete classification of both spinful and spinless emergent particles for three-dimensional (3D) systems with time reversal symmetry. The classification is done for each of the 230 space groups, namely, for a given group, we identify all possible (spinful and spinless, essential and accidental) emergent particles that can appear. Based on this, we obtain a comprehensive list of emergent particles, including 20 types of spinless particles and 23 types of spinful particles (see Table \ref{tab1}), along with the space groups that can host them (see Supplemental Material (SM) \cite{SM}). Furthermore, for each type of emergent particles, we provide its $k\cdot p$ effective model, which characterizes the emergent particle (including its topological charge) and serves as the starting point for any subsequent studies of its properties.

Essentially, the results of this work constitute an encyclopedia of emergent particles in condensed matter, which is not only of fundamental significance but also highly useful for various purposes. For example, for any given system, such as electronic band structure or phonon spectrum of a specific material, the characterization of any spotted emergent particles can be done by directly looking up our table.  The provided effective models can be directly utilized to fit the experimental or first-principles spectra and to study the system properties. Meanwhile, one can also search for a particular type of particle in concrete systems with the guidance of our table. Besides realistic materials, this could be especially useful for realizing spinless emergent particles in artificial systems, which can be precisely engineered and probed with current technology.

\textit{\textcolor{blue}{Rationale.}} We first describe the working procedure that leads to our results. As mentioned, the novel particles emerge around band degeneracy points. The classification of emergent particles is therefore equivalent to the classification of all possible band degeneracies. Any stable degeneracy must require certain protection. For spatially periodic systems, the protection comes from the space group symmetries with the action on the BZ. In this work, we consider the systems with time reversal symmetry $\mathcal{T}$, so the relevant symmetries are the 230 so-called type-II magnetic space groups $\mathcal{M}=\mathcal{G}+\mathcal{TG}$ \cite{Bradley2009Mathematical-Oxford}, where $\mathcal{G}$ is the crystallographic space group. With the exception of Weyl points which only requires the translational subgroup and can be topologically protected at generic ${\bm k}$ points, all other types of degeneracies will require a nontrivial little group at the corresponding ${\bm k}$ points \cite{Armitage2018Weyl-RoMP}. The little group $\text{M}^{\bm k_1}$ at point $\bm k_1$ takes the form of either
\begin{eqnarray}
\boldsymbol{\text{M}}^{\boldsymbol{k}_{1}} & = & \boldsymbol{\text{G}}^{\boldsymbol{k}_{1}},\label{eq1}
\end{eqnarray}
or
\begin{eqnarray}
\boldsymbol{\text{M}}^{\boldsymbol{k}_{1}} & = & \boldsymbol{\text{G}}^{\boldsymbol{k}_{1}}+{\cal{A}}\boldsymbol{\text{G}}^{\boldsymbol{k}_{1}},\label{eq2}
\end{eqnarray}
where $\boldsymbol{\text{G}}^{\boldsymbol{k}_{1}}$ is the corresponding crystallographic little group and ${\cal{A}}$ is certain anti-unitary operator. The band degeneracies are associated with the irreducible representations (IRs) of the group $\text{M}^{\bm k_1}$ \cite{Bradley2009Mathematical-Oxford}. At high-symmetry points of the BZ, band degeneracies correspond to the IRs with dimensions $n>1$. At high-symmetry lines, band degeneracies may involve nodal lines which correspond to IRs with $n>1$, or nodal points  which correspond  to the crossing of bands with different IRs. Similarly, at high-symmetry planes, the generic degeneracies are the nodal surfaces corresponding to IRs with $n>1$, and nodal lines corresponding to two different IRs.

Hence, to obtain the possible band degeneracies for each  space group $\mathcal{M}$, we first find its little groups for all high-symmetry points, lines, and planes. Then for each little group, we find its matrix IRs. For $\boldsymbol{\text{M}}^{\boldsymbol{k}_{1}}$ having the form of Eq.~(\ref{eq1}), the matrix IRs are available in Ref. \cite{Bradley2009Mathematical-Oxford}. For the cases in (\ref{eq2}), the representations (known as corepresentations) can be derived from those of $\boldsymbol{\text{G}}^{\boldsymbol{k}_{1}}$ whenever $\mathcal{A}$ is known. This is discussed in detail in SM \cite{SM}.
Our obtained complete list of the single-valued (for spinless particles) and double-valued (for spinful particles) matrix representations for the little groups of the 230  space groups are presented in Sec.~S7A and S8A of SM \cite{SM}, respectively. Based on this information, we find all possible band degeneracies, and we characterize the emergent particles around these degeneracies by constructing the $k\cdot p$ effective models. The model Hamiltonian ${\cal H}_{\boldsymbol{k}_{1}}$ expanded at the degeneracy point $\bm k_1$ is solved from the constraint equations
\begin{eqnarray}\label{ham-cons}
D({\cal O}_{i}){\cal H}_{\boldsymbol{k}_{1}}(\boldsymbol{q})\left[D({\cal O}_{i})\right]^{-1} & = & {\cal H}_{\boldsymbol{k}_{1}}(\hat{{\cal O}}_{i}\boldsymbol{q}),
\end{eqnarray}
where ${\cal O}_{i}$ runs through the generators of the little group, $D({\cal O}_{i})$ denotes its matrix representation
at the considered band degeneracy, and $\bm q$ is the momentum measured from $\bm k_1$. With the help of the effective models, we can classify the emergent particles by the dimension of degeneracy manifold, the degree of degeneracy, the type of dispersion, and the topological charge, as in Table \ref{tab1}. The list of all possible emergent particles along with their effective models for each space group is presented in Sec. S7 and S8 of SM \cite{SM}.

We have two remarks before proceeding. First, certain degeneracies at high-symmetry points (lines) may not be isolated nodal points, instead, they could be residing on a higher-dimensional degeneracy manifold (nodal lines or surfaces). The effective models are helpful for distinguishing such cases.

Second, nodal lines may form various connection patterns in the BZ, such as nodal chains, crossed nodal loops, nodal boxes, and etc \cite{Weng2015Topological-PRB,Tomas_NC_2016,Sheng_JPCL_2017}. Such connectivity cannot be directly inferred from the effective modeling which is valid locally in BZ. Hence, in this work, we do not make a differentiation based on
the connectivity for multiple nodal lines, but refer to them collectively as nodal-line nets.

{
\global\long\def\arraystretch{1.4}%

\begin{table*}
\caption{\label{Tab-na} Notation of the emergent particles listed in this work.  Abbr. indicates the abbreviation of notation, $d_{c}$ and $d$  indicates the dimensionality and degeneracy of the particles,  Ld indicates the leading order of the band splitting of  particles in BZ, and ${\cal C}$ indicates the maximum topological charge of the particles.  The possible Hamiltonian and typical band structure of the emergent particles are given in Sec. S3 of SM \cite{SM}.\label{tab1}}
\begin{ruledtabular} %
\begin{tabular}{llllllll}
Notation & Abbr. & $d_{c}$ & $d$ & Ld & $|{\cal{C}}|$ & Realization & \tabularnewline
 &  &  &  &  &  & w/o SOC & with SOC\tabularnewline
\hline
Charge-1 Weyl point & C-1 WP & 0 & 2 & (111) & $1$ & $\surd$ & $\surd$\tabularnewline
Charge-2 Weyl point & C-2 WP & 0 & 2 & (122) & $2$ & $\surd$ & $\surd$\tabularnewline
Charge-3 Weyl point & C-3 WP & 0 & 2 & (133) & $3$ & $\surd$ & $\surd$\tabularnewline
Charge-4 Weyl point & C-4 WP & 0 & 2 & (223) & $4$ & $\surd$ & $\times$\tabularnewline
 &  &  &  &  &  &  & \tabularnewline
Triple point & TP & 0 & 3 & (111) &  & $\surd$ & $\surd$\tabularnewline
Charge-2 triple point & C-2 TP & 0 & 3 & (111) & $2$ & $\surd$ & $\surd$\tabularnewline
Quadratic triple point & QTP & 0 & 3 & (122) &  & $\surd$ & $\times$\tabularnewline
Quadratic contact triple point & QCTP & 0 & 3 & (222) & 0 & $\surd$ & $\times$\tabularnewline
 &  &  &  &  &  &  & \tabularnewline
Dirac point & DP & 0 & 4 & (111) & 0 & $\surd$ & $\surd$\tabularnewline
Charge-2 Dirac point & C-2 DP & 0 & 4 & (111) & $2$ & $\surd$ & $\surd$\tabularnewline
Charge-4 Dirac point & C-4 DP & 0 & 4 & (111) & $4$ & $\times$ & $\surd$\tabularnewline
Quadratic Dirac point & QDP & 0 & 4 & (122) & 0 & $\surd$ & $\surd$\tabularnewline
Charge-4 quadratic Dirac point & C-4 QDP & 0 & 4 & (122) & $4$ & $\times$ & $\surd$\tabularnewline
Quadratic contact Dirac point & QCDP & 0 & 4 & (222) & 0 & $\times$ & $\surd$\tabularnewline
Cubic Dirac point & CDP & 0 & 4 & (133) & 0 & $\times$ & $\surd$\tabularnewline
Cubic crossing Dirac point & CCDP & 0 & 4 & (223) & 0 & $\surd$ & $\times$\tabularnewline
 &  &  &  &  &  &  & \tabularnewline
Sextuple point & SP & 0 & 6 & (111) & 0 & $\surd$ & $\surd$\tabularnewline
Charge-4 sextuple point & C-4 SP & 0 & 6 & (111) & $4$ & $\times$ & $\surd$\tabularnewline
Quadratic contact sextuple point & QCSP & 0 & 6 & (222) & 0 & $\times$ & $\surd$\tabularnewline
 &  &  &  &  &  &  & \tabularnewline
Octuple point & OP & 0 & 8 & (111) & 0 & $\times$ & $\surd$\tabularnewline
 &  &  &  &  &  &  & \tabularnewline
Weyl nodal line & WNL & 1 & 2 & (11) & $\pi$ & $\surd$ & $\surd$\tabularnewline
Weyl nodal-line net & WNL net & 1 & 2 & (11) & $\pi$ & $\surd$ & $\surd$\tabularnewline
Quadratic nodal line & QNL & 1 & 2 & (22) & 0 & $\surd$ & $\surd$\tabularnewline
Cubic nodal line & CNL & 1 & 2 & (33) & $\pi$ & $\times$ & $\surd$\tabularnewline
 &  &  &  &  &  &  & \tabularnewline
Dirac nodal line & DNL & 1 & 4 & (11) & 0 & $\surd$ & $\surd$\tabularnewline
Dirac nodal-line net & DNL net & 1 & 4 & (11) & 0 & $\surd$ & $\surd$\tabularnewline
 &  &  &  &  &  &  & \tabularnewline
Nodal surface & NS & 2 & 2 & (1) & 1 & $\surd$ & $\surd$\tabularnewline
\end{tabular}\end{ruledtabular}
\end{table*}

}

{
\global\long\def\arraystretch{1.4}%

\begin{table*}
\caption{Part of the single-valued corepresentation of SG 211, including
both the essential and accidental degeneracies. At the top,
the columns from left to right list the high-symmetry momentum $\boldsymbol{k}$,
the position of $\boldsymbol{k}$, the generating elements of the
little group of $\boldsymbol{k}$, the irreducible representation
of little group of $\boldsymbol{k}$, the dimension of the representation,
the matrix representations of the generating elements, and the effective
Hamiltonian, the species and the topological charge of the degeneracies.
The bottom part lists the accidental degeneracy.  \label{tab2}}
\begin{raggedright}
\begin{tabular}{lllllllll}
\multicolumn{9}{l}{SG 211}\tabularnewline
\hline
\multicolumn{9}{l}{$\Gamma_{c}^{v}$; $\{C_{31}^{-}|000\},\{C_{2z}|000\},\{C_{2x}|000\},\{C_{2a}|000\},\mathcal{T}$;
Non-Centrosymmetric; without SOC}\tabularnewline
\multicolumn{9}{l}{}\tabularnewline
$\Gamma$; & $(000)$; & $C_{31}^{-}$,$C_{2z}$,$C_{2x}$,$C_{2a}$,$\mathcal{T}$; & $R_{1};$ & $1;$ & $1,1,1,1,1;$ &  &  & \tabularnewline
 &  &  & $R_{2};$ & $1;$ & $1,1,1,-1,1;$ &  &  & \tabularnewline
 &  &  & $R_{3};$ & $2;$ & $\frac{-\sigma_{0}+i\sqrt{3}\sigma_{2}}{2},\sigma_{0},\sigma_{0},\sigma_{3},\sigma_{0};$ & $(c_{1}+c_{2}k^{2})\sigma_{0}+c_{3}k_{x}k_{y}k_{z}\sigma_{2}$ & C-4 WP; & 4\tabularnewline
 &  &  &  &  &  & $+c_{4}\left[\sqrt{3}(k_{x}^{2}-k_{y}^{2})\sigma_{1}+(k^{2}-3k_{z}^{2})\sigma_{3}\right]$; &  & \tabularnewline
 &  &  & $R_{4};$ & $3;$ & $A_{11},-\frac{A_{0}+2\sqrt{3}A_{8}}{3},A_{13},A_{12},A_{0};$ & $A_{0}c_{1}+c_{2}\left(A_{3}k_{x}-A_{2}k_{y}+A_{1}k_{z}\right);$ & C-2 TP; & 2\tabularnewline
 &  &  & $R_{5};$ & $3;$ & $A_{11},-\frac{A_{0}+2\sqrt{3}A_{8}}{3},A_{13},-A_{12},A_{0};$ & $A_{0}c_{1}+c_{2}\left(A_{3}k_{x}-A_{2}k_{y}+A_{1}k_{z}\right);$ & C-2 TP; & 2\tabularnewline
$P;$ & $(\frac{1}{4}\frac{1}{4}\frac{1}{4})$; & $C_{31}^{+}$,$C_{2z}$,$C_{2y}$,$C_{2c}\mathcal{T}$; & $R_{1};$ & $1;$ & $1,1,1,1;$ &  &  & \tabularnewline
 &  &  & $R_{2};$ & $1;$ & $(-1)^{2/3},1,1,1;$ &  &  & \tabularnewline
 &  &  & $R_{3};$ & $1;$ & $(-1)^{4/3},1,1,1;$ &  &  & \tabularnewline
 &  &  & $R_{4}$ & $3;$ & $A_{9},-\frac{A_{0}+2\sqrt{3}A_{8}}{3},A_{10},A_{15};$ & $c_{1}A_{0}+c_{2}\left(A_{7}k_{x}+A_{6}k_{y}+A_{4}k_{z}\right)$ & C-2 TP; & 2\tabularnewline
 &  &  &  &  &  & $+c_{3}\left(A_{3}k_{x}-A_{2}k_{y}+A_{1}k_{z}\right);$ &  & \tabularnewline
$\Sigma$; & $\text{\ensuremath{\Gamma}N}$; & $C_{2a}$,$C_{2b}\mathcal{T}$; & $R_{1};$ & $1;$ & $1,1;$ &  &  & \tabularnewline
 &  &  & $R_{2};$ & $1;$ & $-1,1;$ &  &  & \tabularnewline
\end{tabular}
\par\end{raggedright}
\raggedright{}%
\begin{tabular}{lllllllll}
\multicolumn{9}{l}{}\tabularnewline
\hline
\multicolumn{9}{l}{Accidental degeneracies on high-symmetry line}\tabularnewline
\multicolumn{9}{l}{}\tabularnewline
\multirow{1}{*}{$\Sigma$;} & \multirow{1}{*}{$\text{\ensuremath{\Gamma}N}$;} & \multirow{1}{*}{$C_{2a}$,$C_{2b}\mathcal{T}$;} & $\left\{ R_{1}\right\} ,\left\{ R_{2}\right\} ;$ & $2;$ & $\sigma_{3},\sigma_{0};$ & $\sigma_{0}\left[c_{1}+c_{2}\left(k_{x}+k_{y}\right)\right]+c_{3}\sigma_{3}\left(k_{x}+k_{y}\right)+c_{4}\sigma_{1}k_{z}+c_{5}\sigma_{2}\left(k_{x}-k_{y}\right);$ & C-1 WP; & 1\tabularnewline
\hline
\end{tabular}
\end{table*}

}

\textit{\textcolor{blue}{An example: SG 211.}} To illustrate the usage of our results, we discuss
one concrete example---a spinless system with space group No.~211 (SG 211).

Table~\ref{tab2} shows the excerpt from our result in Sec. S7 of SM  \cite{SM} for this group.  SG 211 ($O^{5}$)
is a cubic space group with a body-centred cubic Bravais lattice. The first line in Table~\ref{tab2} presents several basic information, including the BZ type, the symmetry generators (translations are not included here), whether centrosymmetry $I$ is contained in the group, and whether SOC is considered (i.e., spinful or spinless). Here, the $I$ symmetry is highlighted, because the presence of $I$ would lead to a combined $I\mathcal{T}$ symmetry for every ${\bm{k}}$ point. For spinless systems, this combined symmetry prevents the existence of Weyl points but can protect Weyl nodal lines \cite{Weng2015Topological-PRB}.

\begin{figure}
\includegraphics[width=8.8cm]{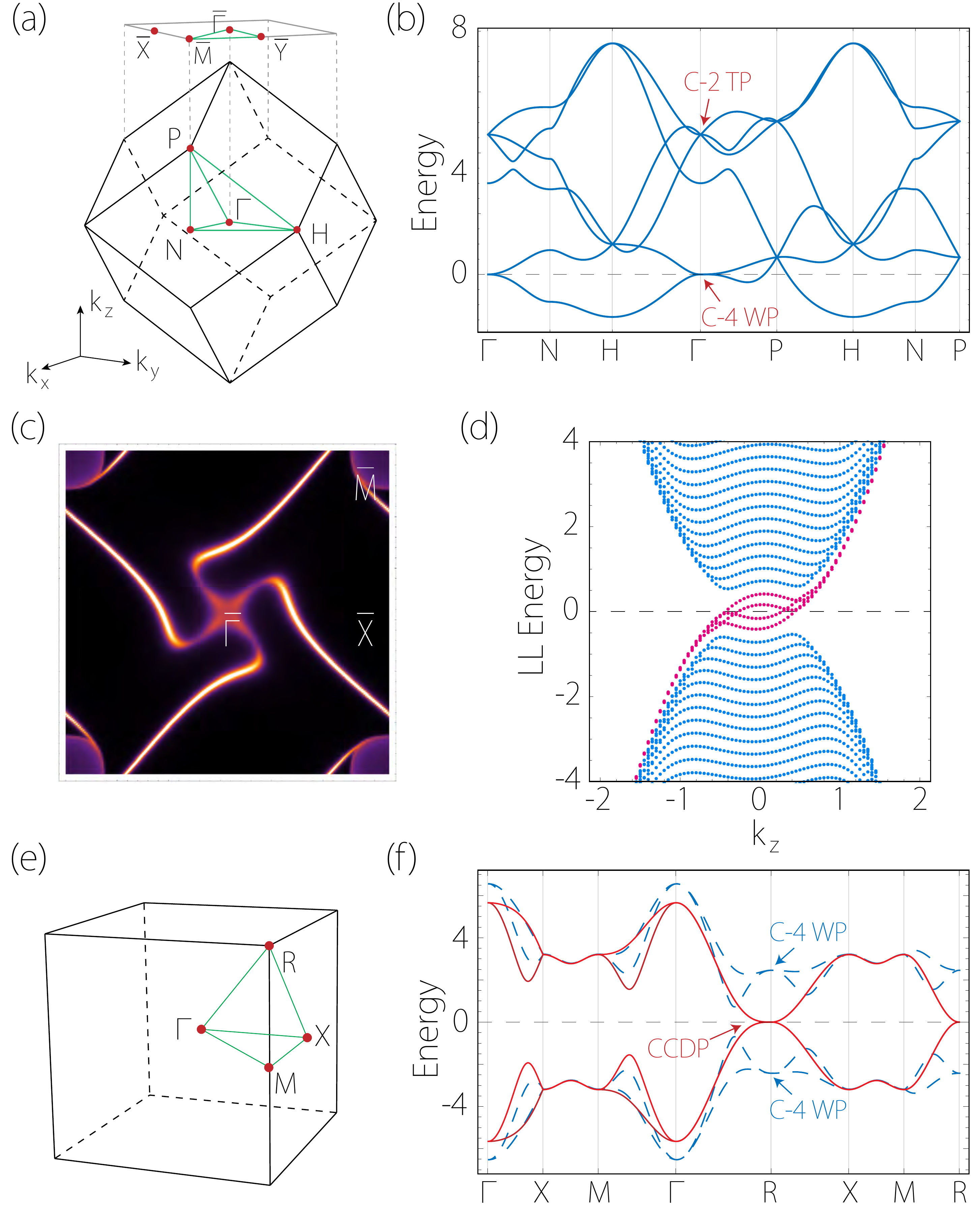}\caption{(a) Bulk and (001) surface BZ of SG 211. (b) The bulk band
structure and (c) the Fermi surface contours at $E_{F}=0$ on the
$(001)$ surface for a TB model with SG 211. The C-4 WP is located
at the $\Gamma$ point formed by the lowest two bands and it gives
four topological Fermi arcs at the surface. For the parameters of
TB model see SM \cite{SM}. (d) Landau spectrum of C-4 WP (obtained by using the $k\cdot p$ effective model) with $B$ field along the $z$ direction.
(e) Bulk BZ  and (f) band structure  of the  TB models with SG 208 (dashed curves) and SG 223 (solid curves).
\label{fig:1}}
\end{figure}

The BZ for SG 211 is illustrated in Fig. \ref{fig:1}(a). In Table~\ref{tab2}, we show the results for high-symmetry points $\Gamma$ and $P$, and high-symmetry line $\Sigma$ ($\Gamma$$N$). The results for other ${\bm{k}}$ points can be found in  Sec. S7 of SM \cite{SM}.
First, let's look at the $\Gamma$
point. Here, the different pieces of information are separated by the semicolons. After the $\Gamma$ symbol, we give the coordinate for the $\Gamma$ point. Then there are the generating elements of the little group at $\Gamma$, including five elements $C_{31}^-$, $C_{2z}$, $C_{2x}$, $C_{2a}$, and $\mathcal{T}$.
Here, $C_{31}^-$ denotes $C_{3,111}^-$, and $C_{2a}$ denotes $C_{2,110}$~\cite{Bradley2009Mathematical-Oxford}.
This little group has five distinct irreducible corepresentations, labelled as $R_{i}$ with $i=1,2,\cdots,5$.
(A correspondence between $R_{i}$ and the band-representation notations can be found in Refs.~\citep{Bradley2009Mathematical-Oxford,GBLiu_2020}.) The number following
$R_{i}$  gives the dimension of $R_{i}$. Here, one observes that $R_1$ and $R_2$ are $1d$ representations,
$R_{3}$ gives a twofold degeneracy, and
$R_{4}$ and $R_{5}$ represent threefold degeneracies. After that, the
matrix representations of the five generating elements are given. Here, the $\sigma$'s (for $R_3$) are the Pauli matrices, and the $A$'s (for $R_4$ and $R_5$) are $3\times 3$ matrices defined in Sec. S6B in SM \cite{SM}. Note that the complex conjugation operator $\mathcal{K}$ in the anti-unitary element $\mathcal{T}$ is not explicitly written out in the table. The next column shows the effective model constructed according to Eq. (\ref{ham-cons}) based on each matrix representation. For example, the emergent particles around the double degeneracy of $R_3$ are described by
\begin{equation}\begin{split}\label{C-4}
  \mathcal{H}_\text{C-4 WP}=&(c_{1}+c_{2}k^{2})\sigma_{0}+c_{3}k_{x}k_{y}k_{z}\sigma_{2}\\
  &+c_{4}[\sqrt{3}(k_{x}^{2}-k_{y}^{2})\sigma_{1}+(k^{2}-3k_{z}^{2})\sigma_{3}],
  \end{split}
\end{equation}
where the $c$'s are real model parameters. Interestingly, from the effective model (\ref{C-4}), one finds that the dispersion of the emergent particle is cubic along the (111) direction and quadratic in the plane perpendicular to (111). Furthermore, the nodal point carries a large topological charge (Chern number) ${\cal{C}}=\pm 4$, hence it is named as a charge-4 Weyl point (C-4 WP). Here, we follow the convention to use Weyl/Dirac to denote twofold/fourfold degeneracy. Notably, the dispersion and the topological charge of this C-4 WP are distinct from the previously known multi-Weyl points \cite{Fang2012Multi-Prl}, which invariably have a direction with linear dispersion and have topological charges less than 4. In addition, we find that C-4 WP can only exist in spinless systems. As shown Sec. S2 in SM \cite{SM}, it can appear in space groups with $T$ or $O$ point groups, including space group No.~195-199 and 207-214. The type of the band degeneracy and the topological charge are presented in the last two columns in Table~\ref{tab2}.

From Table~\ref{tab2}, one can see that $R_4$ and $R_5$ at $\Gamma$ both give a charge-2 triple point (C-2 TP), i.e., a threefold nodal point with topological charge $\pm 2$. This type of nodal point also can appear for the $R_4$ representation at $P$.
As summarized in Sec. S2 in SM \cite{SM}, C-2 TP can appear in both spinless and spinful systems. However, it only occurs at non-${\cal T}$-invariant
momenta for the spinful systems, while it can occur at both ${\cal T}$-invariant
and non-${\cal T}$-invariant momenta for spinless systems.

The similar analysis is performed for the high-symmetry line $\Sigma$. The little
group at $\Sigma$ has two distinct irreducible representations $R_{1}$
and $R_{2}$. These are $1d$ representations, indicating that there
is no essential degeneracy along the $\Sigma$ line. However, there could exist an
accidental degeneracy formed by crossing between $R_{1}$ and $R_{2}$ bands, as described in the last row in Table~\ref{tab2}. The effective model for this accidental crossing is constrained by
\begin{eqnarray}
\Delta({\cal O}){\cal H}(\boldsymbol{k})\left[\Delta({\cal O})\right]^{-1} & = & {\cal H}(\hat{{\cal O}}\boldsymbol{k}),
\end{eqnarray}
where ${\cal O}\in \{C_{2a},C_{2b}{\cal T}\}$ ($C_{2b}$ denotes $C_{2,\bar{1}10}$) is one of the two generators, and $\Delta({\cal O})=D_{1}({\cal O})\oplus D_{2}({\cal O})$,
which is $\sigma_{3}$ for $C_{2a}$ and $\sigma_{0}{\cal K}$ for
$C_{2b}{\cal T}$. The obtained model in Table~\ref{tab2} shows that this crossing is a conventional Weyl point with $|{\cal{C}}|=1$.

\textit{\textcolor{blue}{Signatures of emergent particles.}}
Methods for detecting the emergent particles have been well established. For example, electronic excitations in solid materials can be directly
probed by the angle-resolved photoemission spectroscopy (ARPES) \cite{Armitage2018Weyl-RoMP}. Similar spectroscopy techniques for detecting excitations in photonic, phononic, cold-atom optical lattices, and mechanical systems have also been developed \cite{CooperRMP,Yang1013Sci,Acoustic_NS_SciAdv_xiao_2020,Magnons_PRL_2017,ImhofNP2018}. These methods can directly detect the band degeneracy and map out the dispersion for the emergent particles.

Of most interest are the chiral particles in Table \ref{tab1} with nontrivial topological charges. For example, let's consider the C-4 WP discussed above for spinless systems with SG 211. To confirm the existence of such a nodal point, we explicitly construct a tight-binding (TB) model based on a body-centred cubic lattice with SG 211 (see SM \cite{SM}).
From the calculated band structure in Fig.~\ref{fig:1}(b), one finds that at $\Gamma$, there exist a twofold nodal point and a threefold nodal point, as indicated by the arrows. By checking Table~\ref{tab2} (or Sec.~S7 in SM), one immediately knows that they
correspond to C-4 WP and C-2 TP, respectively.  Similarly,
the two threefold degeneracies at $P$ are two C-2 TPs, and the two nodal points at $H$ are C-4 WP and C-2 TP. The effective models for the emergent particles around these degeneracies can be directly read off from our tables and used to fit the spectrum in Fig. \ref{fig:1}(b). This demonstrates the powerfulness of our established encyclopedia.

Focusing on the C-4 WP, its nontrivial topological charge dictates that at the surface of the system, there must be four Fermi arcs emanating from the projection of the C-4 WP. This is confirmed by the calculation using the tight-binding model, as shown in Fig. \ref{fig:1}(c). The topological charge also can manifest in the Landau spectrum \cite{ZhaoLL_2020}. In Fig. \ref{fig:1}(d), we consider a magnetic field along the $z$ direction and plot the Landau spectrum along $k_z$ around the energy of the C-4 WP (set to be zero here). One clearly observes that there are four chiral Landau bands crossing the zero energy, showing that the original nodal point has a topological charge of 4. Experimentally, the surface Fermi arcs can be probed by scanning tunneling spectroscopy (STS)  or ARPES,  and the chiral Landau bands can be probed by  STS or  magneto-infrared spectroscopy.

Moreover, we find that two C-4 Weyl points may merge together and form a fourfold Dirac points with cubic band crossing, which has not been reported before. For example, in Fig.~\ref{fig:1}(f), the blue dashed curves show a band structure for SG~208, where two C-4 Weyl points of opposite charges appear at the $R$ point. When the inversion symmetry ($I$) is further imposed to promote the space group to SG~223, one finds that the two C-4 Weyl points will merge into the mentioned Dirac point [see Fig.~\ref{fig:1}(f)]. The leading order band splitting is cubic along the (111) direction and is quadratic in the plane normal to this direction. This novel Dirac point is termed as the cubic crossing Dirac point (CCDP) in our Table \ref{tab1}.  For the case in Fig.~\ref{fig:1}(f), an effective model for the CCDP is derived as
\begin{eqnarray}\label{CCDP}
  \mathcal{H}_\text{CCDP}& = & c_0 \Gamma _{0,0}+k_x k_y k_z \left(c_1 \Gamma _{2,2} -c_2 \Gamma _{1,0}\right) \nonumber \\
  &&+\frac{c_5}{2}  \left[\sqrt{3}\left(k_y^2-k_x^2\right) \Gamma _{3,3}   +\left(k^2-3 k_z^2\right)\Gamma _{3,1} \right],
\end{eqnarray}
where $\Gamma_{i,j}=\sigma_i \otimes \sigma_j$ and  the $c$'s are real model parameters.


\textit{\textcolor{blue}{Discussion.}} In this work,
we have systematically investigated all possible spinless and spinful emergent particles protected by space group and time reversal symmetries.
With a unified classification scheme and unified notations, a complete characterization is performed for each space group and for each type of emergent particles. This offers an extremely useful toolbox for subsequent studies. For example, the possible emergent particles in a concrete system can be directly checked by looking up our tables in Sec. S7 and S8 of SM \cite{SM}. The effective models there provide a starting point for understanding the system properties. Using Tables in Sec. S2 of SM \cite{SM}, one can search for a specific type of particles in material database (with the detailed guidance regarding the location in BZ and the band symmetry).  Recently, several high-throughput computation works have searched over all stable non-magnetic solid materials \cite{Feng_Nat_2019,Vergniory_Nat_2019,Fang_Nat_2019}, resulting in thousands of possible topological semimetals. However, the detailed characterization of the emergent particles is lacking. It will be important to make a cross-reference between our encyclopedia and that list of materials, which helps to fully characterize all existing realistic materials.
Particularly, some of us have established a SpaceGroupIrep package~\cite{GBLiu_2020} to analyze the band representation based on the notation of Ref.~\cite{Bradley2009Mathematical-Oxford}, which greatly facilitates the study.
Probably the more intriguing application of our result is in the design of artificial systems with novel emergent particles. For example, the photonic and acoustic crystals can be well controlled with current technology. With the guidance of the encyclopedia, any spinless particles can be readily realized in these systems.

Our approach here can be extended to systems with broken time reversal symmetry, namely, the so-called type-III and type-IV magnetic space groups. And it can extended to systems with additional particle-hole symmetry or chiral symmetry, which appear, e.g., in superfluids and superconductors. Adding such symmetries will change the classification in a fundamental way. For example, a type of $\mathbb{Z}_2$-charged nodal surfaces with variable shapes can exist for spinless systems with chiral symmetry \cite{Wu2018Nodal-PRB}; but without this symmetry, we find that nodal surfaces can only exist at certain boundary planes of the BZ (see Sec.~S2 of SM \cite{SM}).

\acknowledgments
The authors thank Yang Wang and D. L. Deng for helpful discussions. This work is supported by the National Key R\&D Program of China (Grant Nos. 2020YFA0308800, 2016YFA0300600, 2017YFB0701600), the NSF of China (Grants Nos. 11734003, 12061131002, 12004028, 12004035),
the Strategic Priority Research Program of Chinese Academy of Sciences (Grant No. XDB30000000), the China Postdoctoral Science Foundation (Grant No. 2020M670106),  the Singapore Ministry of Education AcRF Tier 2 (Grants No. MOE2019-T2-1-001) and Beijing Institute of Technology Research Fund Program for Young Scholars.

\bibliography{emergent_fermion_ref}

\end{document}


\title{Supplemental Material for ``Encyclopedia of emergent particles in three-dimensional crystals''}

\author{Zhi-Ming Yu}
\email{Z. M. Yu and Z. Zhang contributed equally to this work.}
\affiliation{Key Lab of Advanced Optoelectronic Quantum Architecture and Measurement
(MOE), Beijing Key Lab of Nanophotonics \& Ultrafine Optoelectronic
Systems, and School of Physics, Beijing Institute of Technology, Beijing
100081, China}
\author{Zeying Zhang}
\email{Z. M. Yu and Z. Zhang contributed equally to this work.}
\affiliation{College of Mathematics and Physics, Beijing University of Chemical Technology, Beijing 100029, China}
\author{Gui-Bin Liu}
\affiliation{Key Lab of Advanced Optoelectronic Quantum Architecture and Measurement
(MOE), Beijing Key Lab of Nanophotonics \& Ultrafine Optoelectronic
Systems, and School of Physics, Beijing Institute of Technology, Beijing
100081, China}
\author{Weikang Wu}
\affiliation{Research Laboratory for Quantum Materials, Singapore University of
Technology and Design, Singapore 487372, Singapore}
\author{Xiao-Ping Li}
\affiliation{Key Lab of Advanced Optoelectronic Quantum Architecture and Measurement
(MOE), Beijing Key Lab of Nanophotonics \& Ultrafine Optoelectronic
Systems, and School of Physics, Beijing Institute of Technology, Beijing
100081, China}
\author{Run-Wu Zhang}
\affiliation{Key Lab of Advanced Optoelectronic Quantum Architecture and Measurement
(MOE), Beijing Key Lab of Nanophotonics \& Ultrafine Optoelectronic
Systems, and School of Physics, Beijing Institute of Technology, Beijing
100081, China}
\author{Shengyuan A. Yang}
\affiliation{Research Laboratory for Quantum Materials, Singapore University of
Technology and Design, Singapore 487372, Singapore}
\author{Yugui Yao}
\email{ygyao@bit.edu.cn}
\affiliation{Key Lab of Advanced Optoelectronic Quantum Architecture and Measurement
(MOE), Beijing Key Lab of Nanophotonics \& Ultrafine Optoelectronic
Systems, and School of Physics, Beijing Institute of Technology, Beijing
100081, China}


\begin{abstract}
\end{abstract}


\maketitle

\tableofcontents

\clearpage

\section{Derivation of lattice models in main text}
In the main text, we construct two spinless TB models based on a body-centred cubic lattice with SG 211 and a primitive cubic lattice with SG 223 to demonstrate the existence of C-4 WP and CCDP, respectively.
For the lattice model with SG 211, we assume a unit cell contains six sites locating at $12d$ Wyckoff positions and put one $s$-like orbital state on each site \cite{BradlynNature-2017,Cano-Ber-prb2018}. Following  the method by Wieder and Kane \cite{WiederPRB_2016} as implemented in the MagneticTB package by Zhang \emph{et. al} \cite{Zeying2021}, the lattice Hamiltonian then can be established as
\begin{eqnarray*}
H_{\text{C-4 WP}} & = & c_{0}+\left[\begin{array}{ccc}
f(0,k_{y},k_{z}) & h_{12} & h_{13}\\
h_{12}^{\dagger} & f(k_{x},0,k_{z}) & h_{23}\\
h_{13}^{\dagger} & h_{23}^{\dagger} & f(k_{x},k_{y},0)
\end{array}\right],
\end{eqnarray*}
with $f(\boldsymbol{k})=c_{1}\cos\frac{k_{x}}{2}\cos\frac{k_{y}}{2}\cos\frac{k_{z}}{2}\sigma_{1}$ and
\begin{eqnarray*}
h_{12} & = & c_{2}\left[\begin{array}{cc}
e^{i\frac{k_{x}+k_{y}}{4}} & e^{i\frac{k_{x}-k_{y}}{4}}\\
e^{i\frac{-k_{x}+k_{y}}{4}} & e^{i\frac{-k_{x}-k_{y}}{4}}
\end{array}\right]
    +c_{3}\left[\begin{array}{cc}
e^{-i\frac{k_{x}+k_{y}+2k_{z}}{4}} & e^{i\frac{-k_{x}+k_{y}+2k_{z}}{4}}\\
e^{i\frac{k_{x}-k_{y}+2k_{z}}{4}} & e^{i\frac{k_{x}+k_{y}-2k_{z}}{4}}
\end{array}\right],\\
h_{13} & = & c_{2}\left[\begin{array}{cc}
e^{-i\frac{k_{x}+k_{z}}{4}} & e^{i\frac{-k_{x}+k_{z}}{4}}\\
e^{i\frac{k_{x}-k_{z}}{4}} & e^{i\frac{k_{x}+k_{z}}{4}}
\end{array}\right]
    +c_{3}\left[\begin{array}{cc}
e^{i\frac{k_{x}+2k_{y}+k_{z}}{4}} & e^{i\frac{k_{x}-2k_{y}-k_{z}}{4}}\\
e^{i\frac{-k_{x}-2k_{y}+k_{z}}{4}} & e^{i\frac{-k_{x}+2k_{y}-k_{z}}{4}}
\end{array}\right],\\
h_{23} & = & c_{2}\left[\begin{array}{cc}
e^{i\frac{k_{y}+k_{z}}{4}} & e^{i\frac{k_{y}-k_{z}}{4}}\\
e^{i\frac{-k_{y}+k_{z}}{4}} & e^{i\frac{-k_{y}-k_{z}}{4}}
\end{array}\right]
    +c_{3}\left[\begin{array}{cc}
e^{i\frac{-2k_{x}-k_{y}-k_{z}}{4}} & e^{i\frac{2k_{x}-k_{y}+k_{z}}{4}}\\
e^{i\frac{2k_{x}+k_{y}-k_{z}}{4}} & e^{i\frac{-2k_{x}+k_{y}+k_{z}}{4}}
\end{array}\right].
\end{eqnarray*}
The coefficient $c_{i}$ with $i=0,1,2,3$ is real model parameter. For the results shown in the main text, we have taken the following parameter values: $c_0=2.8$, $c_1=-1.8$, $c_2=1.$ and $c_3=-0.5$.

For the lattice model with SG 223, we assume a unit cell contains two sites locating at $2a$ Wyckoff positions and put two basis states $d_{z^2}$  and $d_{x^2-y^2}$ orbital on each site \cite{BradlynNature-2017,Cano-Ber-prb2018}. Following  the method by Wieder and Kane \cite{WiederPRB_2016} as implemented in the MagneticTB package by Zhang \emph{et. al} \cite{Zeying2021}, the lattice Hamiltonian then can be established as
\begin{eqnarray*}
H_{\text{CCDP}} & = & c_{0}+\left[\begin{array}{cc}
h_{11} & e^{-i\frac{k_{x}+k_{y}+k_{z}}{2}}h_{12}\\
e^{i\frac{k_{x}+k_{y}+k_{z}}{2}}h_{12}^{\dagger} & h_{22}
\end{array}\right],
\end{eqnarray*}
 with
\begin{eqnarray*}
h_{11} & = & c_{3}\left[\begin{array}{cc}
\sqrt{3}\left(\cos k_{y}-\cos k_{x}\right) & \cos k_{x}+\cos k_{y}-2\cos k_{z}\\
\cos k_{x}+\cos k_{y}-2\cos k_{z} & \sqrt{3}\left(\cos k_{x}-\cos k_{y}\right)
\end{array}\right],\\
h_{12} & = & \left(e^{i\left(k_{x}+k_{y}+k_{z}\right)}+e^{ik_{x}}+e^{ik_{y}}+e^{ik_{z}}\right)\left[\begin{array}{cc}
c_{2} & c_{1}\\
-c_{1} & c_{2}
\end{array}\right]\\
 &  & +\left(1+e^{i\left(k_{x}+k_{y}\right)}+e^{i\left(k_{x}+k_{z}\right)}+e^{i\left(k_{y}+k_{z}\right)}\right)\left[\begin{array}{cc}
c_{2} & c_{1}\\
-c_{1} & c_{2}
\end{array}\right],\\
h_{22} & = & c_{3}\left[\begin{array}{cc}
\sqrt{3}\left(\cos k_{x}-\cos k_{y}\right) & -\left(\cos k_{x}+\cos k_{y}-2\cos k_{z}\right)\\
-\left(\cos k_{x}+\cos k_{y}-2\cos k_{z}\right) & \sqrt{3}\left(\cos k_{y}-\cos k_{x}\right)
\end{array}\right].
\end{eqnarray*}
Also the coefficient $c_{i}$ is real model parameter. For the band structure of CCDP [solid curves in Fig. 1(f)] in the main text, we have taken the following parameter values: $c_0=0.$, $c_1=0.5$, $c_2=-0.5$ and $c_3=-0.8$.
A possible term breaking $I$ symmetry may be written as
\begin{eqnarray*}
H^{\prime} & = & \left[\begin{array}{cc}
\boldmath{0} & e^{-i\frac{k_{x}+k_{y}+k_{z}}{2}}h_{12}^{\prime}\\
e^{i\frac{k_{x}+k_{y}+k_{z}}{2}}h_{12}^{\prime \dagger} & \boldmath{0}
\end{array}\right],
\end{eqnarray*}
 with
\begin{eqnarray*}
h_{12}^{\prime} & = & \left(1+e^{i\left(k_{x}+k_{y}\right)}+e^{i\left(k_{x}+k_{z}\right)}+e^{i\left(k_{y}+k_{z}\right)}\right)\left[\begin{array}{cc}
c_{4} & c_{3}\\
-c_{3} & c_{4}
\end{array}\right].
\end{eqnarray*}
For the results [dashed curves in Fig. 1(f)] shown in the main text, we have set $c_3=-0.5c_1$ and $c_4=c_2$.

\clearpage

\section{Quantitative mapping between emergent particles and type-II MSGs}
For each emergent particle listed in main text, we explicitly  present the number of the type-II magnetic space group (MSG) that can host it.
Table \ref{Tab1} is for the 3D crystals without SOC and Table \ref{Tab3} is for the crystals with SOC.

{\renewcommand{\arraystretch}{1.3}
\begin{table*}[h]
\caption{\label{Tab1} List of the type-II MSGs  hosting  the symmetry-protected degeneracies in the crystals without SOC. }
\begin{ruledtabular}
\begin{tabular}{ll}

Species  & SG No. \tabularnewline
\hline
\multicolumn{2}{l}{The essential degeneracies at high-symmetry point and high-symmetry line }\tabularnewline
\tabularnewline

C-1 WP  & 24, 80, 98, 150, 152, 154, 168-173, 177-182, 199, 210, 214 \tabularnewline
 C-2 WP & 75-80, 89-98, 143-146, 149-155, 168-173, 177-182, 196, 207-210 \tabularnewline
 C-4 WP & 195-199, 207-214 \tabularnewline
 TP & 204, 217, 229 \tabularnewline
  C-2 TP & 195-199, 207-214 \tabularnewline
  QCTP & 200-206, 215-230 \tabularnewline
  DP & 29, 33, 52, 54, 56, 60, 73, 103, 104, 106, 110, 130, 138, 142, 158, 159, 161, 163, 165, 167, 184,  185, 192, 193, 206,\tabularnewline
 &   219, 220, 226, 228, 230 \tabularnewline
  C-2 DP & 19, 92, 96, 198, 212, 213 \tabularnewline
  QDP & 114, 124, 126, 128, 130, 133, 135, 137, 176, 184-186, 188, 190, 192-194, 222  \tabularnewline
  CCDP & 218, 220, 222, 223, 230 \tabularnewline
  SP & 218, 220, 222, 223, 230 \tabularnewline
  WNL & 7, 9, 13-15, 27-34, 37, 39-41, 43, 45, 46, 48-50, 52-54, 56,  58, 60, 64, 66-68, 70, 72-74, 84-88,  100-110, 112,\tabularnewline
   &  114, 116-118, 120-122, 124-126, 128, 130-142, 147, 148, 156-167, 175, 176, 183-186, 188-194, 200-206, 215-230\tabularnewline
  WNL net & 7, 9, 13-15, 27-34, 37, 39-41, 43, 45, 46, 48-50, 52-54, 56,  58, 60, 64, 66-68, 70, 72-74, 84-86, 88, 100-110, 112,  \tabularnewline
   &  114, 116-118, 120-122, 124-126, 128, 130-142, 147, 148, 156-167, 184, 188, 190, 192, 200-206, 215-230  \tabularnewline
  QNL & 81-88, 99-142, 174-176, 183-194, 215-230 \tabularnewline
  DNL & 57, 60-62, 205 \tabularnewline
  DNL net & 61, 205 \tabularnewline
 (one) NS  & 4, 11, 14, 17, 20, 26, 29, 31, 33, 36, 51-54, 63, 64, 76, 78, 91, 95, 169, 170, 173, 176, 178, 179, 182, 185, 186,\tabularnewline
   &   193, 194  \tabularnewline
 (two) NSs  & 18, 55-60, 90, 94, 113, 114, 127-130, 135-138  \tabularnewline
  (three) NSs & 19, 61, 62, 92, 96, 198, 205, 212, 213  \tabularnewline
\hline
\tabularnewline
\hline
\multicolumn{2}{l}{The accidental degeneracies  on high-symmetry line}\tabularnewline
\tabularnewline
C-1 WP  & 3-5, 16-24, 35-37, 44-46, 75-82, 89-98, 111-122, 139, 143-146, 149-156, 158, 168-174, 177-182, 187-188, 195-199, \tabularnewline
& 207-216, 218, 219 \tabularnewline
C-2 WP& 75-80, 89-98, 168-173, 177-182, 207-214 \tabularnewline
C-3 WP & 168-173, 177-182 \tabularnewline
 TP & 81-88, 99-142, 147, 148, 156-167, 174-176, 183-194, 200-206, 215-230 \tabularnewline
 QTP & 175, 176, 183-186, 191-194 \tabularnewline
DP & 26, 29, 31, 33, 36, 51-64, 113, 114, 127-130, 135-138, 175, 176, 183-186, 191-194, 205 \tabularnewline
C-2 DP & 18, 19, 90, 92, 94, 96, 198, 212, 213 \tabularnewline
QDP & 113, 114, 127-130, 135-138 \tabularnewline
WNL  & 10-15, 25-74, 83-88, 99-142, 156-167, 174-176, 183-194, 200-206, 215-230 \tabularnewline
WNL net & 25-74, 83-88, 99-142, 156-167, 175-176, 183-194, 200-206, 215-230 \tabularnewline
\end{tabular}
\end{ruledtabular}
\end{table*}
}

\clearpage

{\renewcommand{\arraystretch}{1.3}

\begin{table*}[h]
\raggedright{}
\caption{\label{Tab3} List of the type-II MSGs  hosting  the symmetry-protected degeneracies in the crystals with SOC.}
\begin{ruledtabular}
\begin{tabular}{ll}
Species  & SG No. \tabularnewline
\hline
\multicolumn{2}{l}{The essential degeneracies at high-symmetry point and high-symmetry line}\tabularnewline
\tabularnewline
C-1 WP  & 1, 3-5, 8-9, 16-24, 35-37, 42-46, 75-82, 89-98, 111, 112, 119-122, 143-146, 149-155, 168-173, 177-182, 195-199, \tabularnewline
& 207-214   \tabularnewline
C-2 WP & 80, 98, 210 \tabularnewline
C-3 WP & 143-146, 149-155, 168-173, 177-182, 196, 209, 210 \tabularnewline
TP & 220 \tabularnewline
C-2 TP & 199, 214 \tabularnewline
DP & 11, 13-15, 26-27, 29-34, 36-37, 43, 48-50, 52-54, 56, 58,  60, 64, 66-68, 70, 72-74, 84-86, 88,  100-106,  108-110, 112, \tabularnewline
& 114, 116-118, 122, 124-126, 128, 130-138, 140-142, 158, 159, 161, 163, 165, 167, 184-186, 188, 190, 192, 198, 201, \tabularnewline
&  203-204, 206, 215-220, 222-224, 226-230  \tabularnewline
C-2 DP & 18, 19, 90, 92, 94, 96, 198, 212, 213 \tabularnewline
C-4 DP & 195-199, 207-214 \tabularnewline
QDP & 142, 228 \tabularnewline
C-4 QDP & 92, 96 \tabularnewline
QCDP & 200-206, 221-230 \tabularnewline
CDP & 163, 165, 167, 184-186, 192, 226, 228 \tabularnewline
SP & 206, 230 \tabularnewline
C-4 SP & 198, 212, 213 \tabularnewline
QCSP & 205 \tabularnewline
OP & 130, 135, 218, 220, 222, 223, 230 \tabularnewline
WNL & 6-9, 25-46, 81-82, 99-122, 156-161, 174, 183-190, 215-220 \tabularnewline
WNL net & 45, 109, 110, 120, 122, 156-161, 174, 184, 187-190, 215-220 \tabularnewline
QNL & 174, 187-190 \tabularnewline
CNL & 183-186 \tabularnewline
DNL & 51-64, 113-114, 127-130, 135-138, 176, 193-194, 205 \tabularnewline
DNL net & 52, 54-56, 58, 60, 62, 127-128, 130, 135-136, 138, 176, 193-194, 205 \tabularnewline
(one) NS & 4, 17, 20, 26, 29, 31, 33, 36, 76, 78, 91, 95, 169, 170, 173, 178, 179, 182, 185, 186 \tabularnewline
(two) NSs & 18, 90, 94, 113, 114 \tabularnewline
(three) NSs & 19, 92, 96, 198, 212, 213 \tabularnewline
\hline
\tabularnewline
\hline
\multicolumn{2}{l}{The accidental degeneracies  on high-symmetry line}\tabularnewline
\tabularnewline
C-1 WP  & 3-5, 16-24, 35-37, 44-46, 75-82, 89-98, 111-122, 143-146, 149-156, 158, 168-174, 177-182, 187-188, 195-199, \tabularnewline
& 207-216, 218, 219 \tabularnewline
C-2 WP& 75-80, 89-98, 168-173, 177-182, 207-214 \tabularnewline
C-3 WP & 168-173, 177-182 \tabularnewline
TP & 156-161, 174, 183-190, 215-220 \tabularnewline
DP & 13-15, 26, 29, 31, 33, 36, 48-50, 52-54, 56, 58-60, 64, 66-68, 70, 72-74, 83-88, 99-110, 123-142, 147-148, 162-167,\tabularnewline
&    175, 176, 183-186, 191-194, 200-206, 221-230 \tabularnewline
C-2 DP & 18-19, 90, 92, 94, 96, 198, 212, 213 \tabularnewline
QDP & 175-176, 191-194 \tabularnewline
WNL & 25-46, 99-122, 156-161, 174, 183-190, 215-220 \tabularnewline
WNL net & 28-34, 40-41, 43, 100, 102, 104, 106, 109-110, 117, 118, 122, 156-161, 183-190, 215-220 \tabularnewline
DNL & 51-64, 128-130, 136-138, 176, 193-194, 205\tabularnewline
DNL net & 57, 60-62, 205 \tabularnewline
\end{tabular}
\end{ruledtabular}
\end{table*}
}

\clearpage

\section{Emergent particles in 3D crystals}

As discussed in main text,  we have  performed  an exhaustive investigation over all the symmetry-protected band degeneracies in 3D crystals with ${\cal{T}}$ symmetry, as shown in Sec. \ref{S3}. Thus it becomes possible to introduce a uniform notation to label the degeneracies and then the emergent particles.
In this work, we classify the band degeneracy from four perspectives: the dimension of degeneracy manifold, the degree of degeneracy, the type of dispersion, and the topological charge.
For dimension of degeneracy manifold, we term the 0D, 1D and 2D band degeneracy as point, line and surface, respectively, and for degree of degeneracy, we term the two-, three-, four-, six- and eight-fold degeneracy as Weyl, triple, Dirac, sextuple, and octuple point/line/surface, respectively.
Moreover, we find that the leading order band energy splitting of  emergent particles along certain direction can be linear, quadratic and cubic.
At last, a $d$-D degeneracy in 3D systems can be  topologically characterized  by a topological charge defined on a $(3-d-1)$-D sphere enclosing the  degeneracy \cite{Chiu2016Classification-RoMP}.  Specifically, the nodal point is characterized  by Chern number, which is a $\mathbb{Z}$-valued  topological charge, the nodal line is characterized  by Berry phase, which is a ${\mathbb{Z}}_2$-valued  topological charge, and the nodal surface also is characterized  a ${\mathbb{Z}}_2$-valued  topological charge \cite{Wu2018Nodal-PRB,TB-PRB2017}.

A complete list of the  emergent particles according to this classification has been presented  in Table I in main text. In the following, we discuss  the possible Hamiltonian and the typical band structure of the emergent particles one by one.

\subsection{Twofold degeneracy point}
\subsubsection{Charge-1 Weyl point}

The charge-1 Weyl point (C-1 WP) is a 0D two-fold  band degeneracy. It features a linear energy splitting along any direction in momentum space, and can occur at a generic $\boldsymbol{k}$ point in BZ. Moreover, the topological charge (Chern number) of  C-1 WP is ${\cal{C}}=\pm 1$.
A typical band structure of C-1 WP is schematically shown in Fig. \ref{fig:WP}(a).

A general  Hamiltonian of C-1 WP is
\begin{eqnarray}
H_{\text{C-1\ WP}} & = & \sum_{i=0}^{3}\left(c_{i,1}k_{x}+c_{i,2}k_{y}+c_{i,3}k_{z}\right)\sigma_{i}. \label{eq:C-1 ham}
\end{eqnarray}
With additional symmetry, the Hamiltonian (\ref{eq:C-1 ham}) would be simplified.
A possible simple  Hamiltonian for C-1 WP may be written as
\begin{eqnarray}
H_{\text{C-1\ WP}} & = & c_{1}k_{z}+c_{2}k_{z}\sigma_{3}+\left(\alpha k_{-}\sigma_{+}+h.c.\right),\label{eq:C-1 ham1}
\end{eqnarray}
with $k_{\pm}=k_{x}\pm ik_{y}$ and $\sigma_{\pm}=(\sigma_{x}\pm i\sigma_{y})/2$.
The C-1 WP can be further classified as type I and type II  \cite{Soluyanov}. For example, the C-1 WP described by Eq. (\ref{eq:C-1 ham1}) is type I when $|c_{2}|>|c_{1}|$ and is type II when $|c_{2}|<|c_{1}|$. However, when  C-1 WP locates at ${\cal T}$-symmetric point, it can not be type II, as  ${\cal T}$-symmetry requires $c_1=0$.

\subsubsection{Charge-2 Weyl point}

The charge-2 Weyl point (C-2 WP) is a 0D two-fold  band degeneracy with a topological charge ${\cal{C}}=\pm 2$. It features a linear dispersion along one  direction and a quadratic energy splitting in the plane normal to the direction.
The C-2 WP can occur on high-symmetry line or at high-symmetry point in BZ.
A typical band structure of C-2 WP is schematically shown in Fig. \ref{fig:WP}(b).

A possible Hamiltonian for C-2 WP may be written as
\begin{eqnarray}
H_{\text{C-2\ WP}} & = & c_{1}k_{z}+c_{2}k_{\parallel}^{2}+c_{3}k_{z}\sigma_{3}+\left(\alpha k_{+}^{2}\sigma_{+}+h.c.\right),\label{eq:C-2 ham1}
\end{eqnarray}
with $k_{\parallel}=\sqrt{k_{x}^{2}+k_{y}^{2}}$.
The C-2 WP can be further classified as type I, type II and type III \cite{type-iii-li}. For example, the C-2 WP described by Eq. (\ref{eq:C-2 ham1}) is type I when $|c_{3}|>|c_{1}|$ and $|\alpha|>|c_{2}|$, is type II when $|c_{3}|<|c_{1}|$, and is type III when $|c_{3}|>|c_{1}|$ and $|\alpha|<|c_{2}|$. Similarly, the C-2 WP at a ${\cal T}$-symmetric point cannot be type II.

\subsubsection{Charge-3 Weyl point}

The charge-3 Weyl point (C-3 WP)  is a 0D two-fold  band degeneracy with a topological charge ${\cal{C}}=\pm 3$.
It features a linear dispersion along one  direction and a cubic energy splitting in the plane normal to the direction.
The C-3 WP can occur on high-symmetry line or at high-symmetry point in BZ.
A typical band structure of C-3 WP is schematically shown in Fig.  \ref{fig:WP}(c).

A possible Hamiltonian for C-3 WP may be written as
\begin{eqnarray}
H_{\text{C-3\ WP}} & = & c_{1}k_{z}+c_{2}k_{\parallel}^{2}+c_{3}k_{z}\sigma_{3}+\left(\alpha k_{+}^{3}\sigma_{+}+h.c.\right).\label{eq:C-3 ham1}
\end{eqnarray}
 The C-3 WP can be further classified as type I, type II and type III, depending on model parameters.

\subsubsection{Charge-4 Weyl point}

\begin{figure}
\includegraphics[width=16cm]{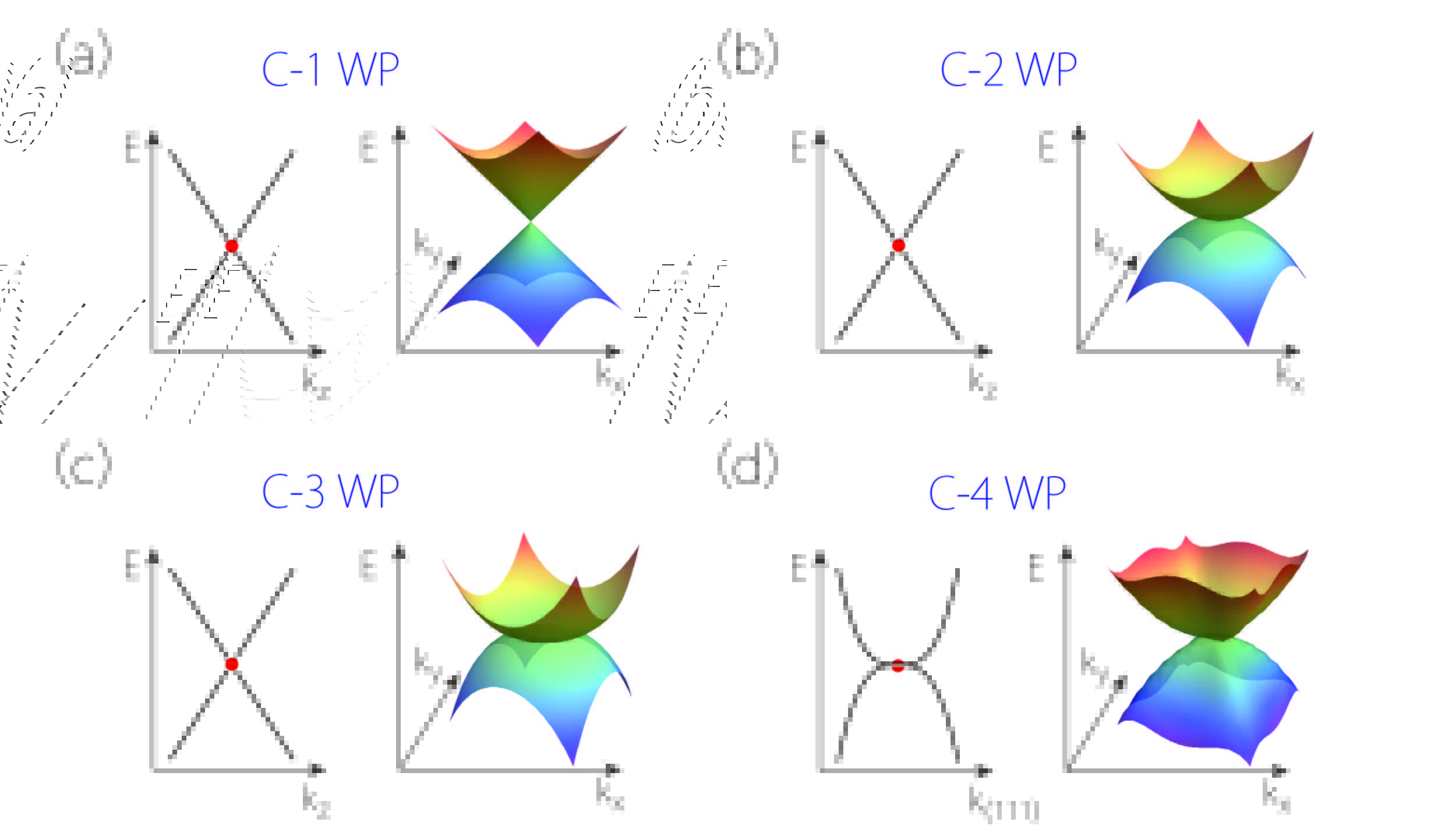}\caption{Typical band structure of  possible Weyl points in 3D crystals.\label{fig:WP}}
\end{figure}

The charge-4 Weyl point (C-4 WP) is a 0D two-fold  band degeneracy with a topological charge ${\cal{C}}=\pm 4$.
It features a cubic  energy splitting along one  direction and a quadratic energy splitting in the plane normal to the direction.
The C-4 WP only occurs  at certain ${\cal T}$-symmetric points in spinless systems.
Besides this work, C-4 WP was also recently unveiled in Ref. \cite{Tiantian_PRB_2020} and Ref. \cite{Liu_WPhons_2020}. 
A typical band structure of C-4 WP is schematically shown in Fig. \ref{fig:WP}(d).

A possible Hamiltonian for C-4 WP may be written as
\begin{eqnarray}
H_{\text{C-4\ WP}} & = & c_{1}k^{2}+c_{2}\left[\sqrt{3}(k_{x}^{2}-k_{y}^{2})\sigma_{1}+(k_{x}^{2}+k_{y}^{2}-2k_{z}^{2})\sigma_{3}\right]+c_{3}k_{x}k_{y}k_{z}\sigma_{2},\label{eq:C-4 ham1}
\end{eqnarray}
with $k=\sqrt{k_x^2+k_y^2+k_z^2}$, which shows a cubic energy splitting along $k_{(111)}$ direction and quadratic energy splitting along $k_{x,y,z}$ direction.

\subsection{Threefold degeneracy point}

\subsubsection{Triple point}

\begin{figure}
\includegraphics[width=16cm]{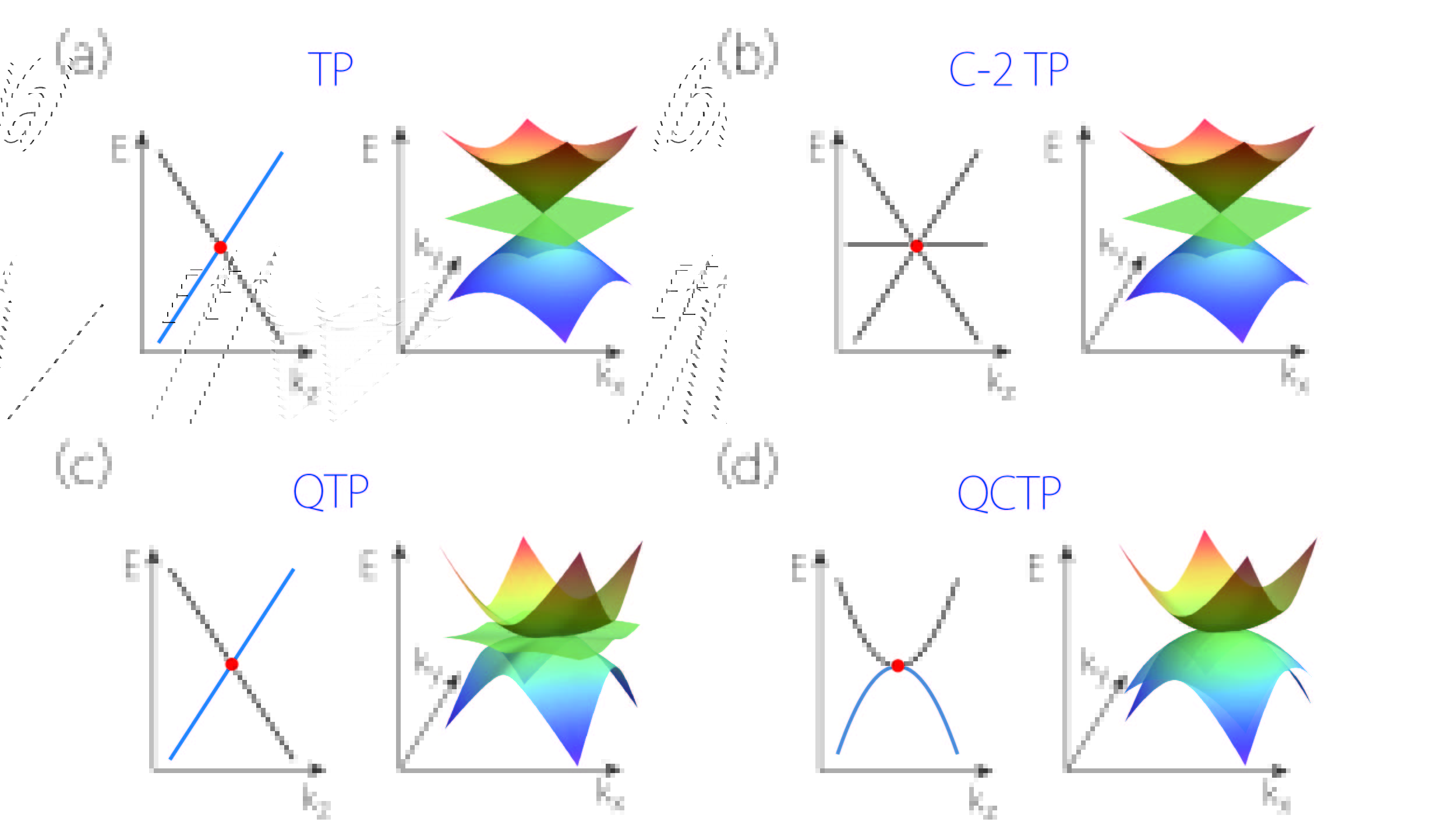}\caption{Typical band structure of possible three-fold nodal points in 3D crystals. The black and blue curves denote non-degenerate and doubly degenerate bands, respectively. \label{fig:TP}}
\end{figure}

The triple point (TP) is a 0D three-fold  band degeneracy, formed by a linear crossing between  a doubly degenerate band and a non-degenerate band.
The TP does not have a well-defined topological charge of Chern number, as there does not exist a fully gapped sphere surrounding TP in BZ.
It features a linear energy splitting along any  direction in momentum space, and can occur on high-symmetry line or at high-symmetry point in BZ.
A typical band structure of TP is schematically shown in Fig. \ref{fig:TP}(a).

A possible Hamiltonian for  TP may be written as
\begin{eqnarray}
H_{\text{TP}} & = & c_{1}k_{z}+\left[\begin{array}{ccc}
c_{2}k_{z} & \alpha k_{y} & \alpha k_{x}\\
\alpha^{*}k_{y} & -c_{2}k_{z} & 0\\
\alpha^{*}k_{x} & 0 & -c_{2}k_{z}
\end{array}\right].\label{eq:TP ham1}
\end{eqnarray}
The TP can be further classified as type I and type II depending on model parameters. For example, the TP described by Eq. (\ref{eq:TP ham1}) is type I when $|c_{2}|>|c_{1}|$ and is type II when $|c_{2}|<|c_{1}|$.

\subsubsection{Charge-2 Triple point}

The charge-2 triple point (C-2 TP)  is a 0D three-fold  band degeneracy with a topological charge ${\cal{C}}=\pm 2$. It features a linear energy splitting along any direction in momentum space.
The C-2 TP only occurs at high-symmetry point in BZ.
A typical band structure of C-2 TP is schematically shown in Fig. \ref{fig:TP}(b).

A  possible   Hamiltonian for C-2 TP may be written as
\begin{eqnarray}
H_{\text{C-2\ TP}} & = & \left[\begin{array}{ccc}
0 & ik_{z} & ik_{y}\\
-ik_{z} & 0 & ik_{x}\\
-ik_{y} & -ik_{x} & 0
\end{array}\right],\label{eq:C-2 TP ham1}
\end{eqnarray}
which actually describes the conventional spin-1 Weyl fermion.

\subsubsection{Quadratic triple point}

The quadratic triple point (QTP)  is a 0D three-fold  band degeneracy, formed by a linear crossing between  a doubly degenerate band and a non-degenerate band along certain high-symmetry line. The QTP also does  not have a well-defined topological charge of Chern number.
However, in contrast to TP,  QTP features a quadratic energy splitting in the plane normal to the high-symmetry line.
The QTP only occurs on high-symmetry line in spinless systems.
A typical band structure of  QTP is schematically shown in Fig. \ref{fig:TP}(c).

A possible Hamiltonian for  QTP may be written as
\begin{eqnarray}
H_{\text{QTP}} & = &c_{1}k_{z}+c_{2}k_{\parallel}^{2}+\left[\begin{array}{ccc}
c_{3}k_{z}+c_{4}k_{\parallel}^{2} & 0 & 0\\
0 & -c_{3}k_{z}-c_{4}k_{\parallel}^{2} & 0\\
0 & 0 & -c_{3}k_{z}-c_{4}k_{\parallel}^{2}
\end{array}\right]\nonumber \\
 &  & +c_{5}\left[\begin{array}{ccc}
0 & 2ik_{x}k_{y} & i(k_{x}^{2}-k_{y}^{2})\\
-2ik_{x}k_{y} & 0 & 0\\
-i(k_{x}^{2}-k_{y}^{2}) & 0 & 0
\end{array}\right]+c_{6}\left[\begin{array}{ccc}
0 & 0 & 0\\
0 & k_{y}^{2} & -k_{x}k_{y}\\
0 & -k_{x}k_{y} & k_{x}^{2}
\end{array}\right].
\end{eqnarray}
The QTP  can be further classified as type I, type II and type III, depending on model parameters.

\subsubsection{Quadratic contact triple point}

The quadratic contact triple point (QCTP) is a 0D three-fold  band degeneracy with  a topological charge ${\cal{C}}=0$. It features a quadratic energy splitting along any direction in momentum space, and splits into a doubly degenerate band and a non-degenerate band along certain high-symmetry line(s), and three non-degenerate bands at generic momentum points.
The QCTP only occurs at high-symmetry point in spinless systems.
A typical band structure of  QCTP is schematically shown in Fig. \ref{fig:TP}(d).

A possible Hamiltonian that captures the essential physics of QCTP may be written as
\begin{eqnarray}
H_{\text{QCTP}} & = & \left[\begin{array}{ccc}
c_{1}k_{x}^{2}+c_{2}\left(k_{y}^{2}+k_{z}^{2}\right) & c_{3}k_{x}k_{y} & c_{3}k_{x}k_{z}\\
c_{3}k_{x}k_{y} & c_{1}k_{y}^{2}+c_{2}\left(k_{x}^{2}+k_{z}^{2}\right) & c_{3}k_{y}k_{z}\\
c_{3}k_{x}k_{z} & c_{3}k_{y}k_{z} & c_{1}k_{z}^{2}+c_{2}\left(k_{x}^{2}+k_{y}^{2}\right)
\end{array}\right].
\end{eqnarray}

\subsection{Fourfold degeneracy point}

\subsubsection{Dirac point}

The (charge-0) Dirac point (DP) is a 0D four-fold band degeneracy with a topological charge ${\cal{C}}=0$.  It features a linear dispersion along any direction in momentum space.
The DP can occur on high-symmetry line or at high-symmetry point in BZ.
The DP in a spinful system with spatial inversion symmetry $I$ and ${\cal{T}}$ splits into two doubly degenerate bands at each $\boldsymbol{k}$ in BZ.
For the other cases, the DP  splits into four non-degenerate  bands at a generic $\boldsymbol{k}$, but into two doubly degenerate bands along certain high-symmetry line(s).
A typical band structure of  DP is schematically shown in Fig. \ref{fig:DP}(a).

A possible  Hamiltonian for DP may be written as
\begin{eqnarray}
H_{\text{DP}} & = & c_{1}k_{z}+\left[\begin{array}{cccc}
c_{2}k_{z} & -ic_{3}k_{+} & 0 & 0\\
ic_{3}k_{-} & -c_{2}k_{z} & 0 & 0\\
0 & 0 & c_{2}k_{z} & ic_{3}k_{-}\\
0 & 0 & -ic_{3}k_{+} & -c_{2}k_{z}
\end{array}\right].\label{eq:DP ham1}
\end{eqnarray}
The DP also can be further classified as type I and type II, depending on model parameters. However, the DP at ${\cal T}$-symmetric points cannot be type II.

\subsubsection{Charge-2 Dirac point}

\begin{figure}
\includegraphics[width=16cm]{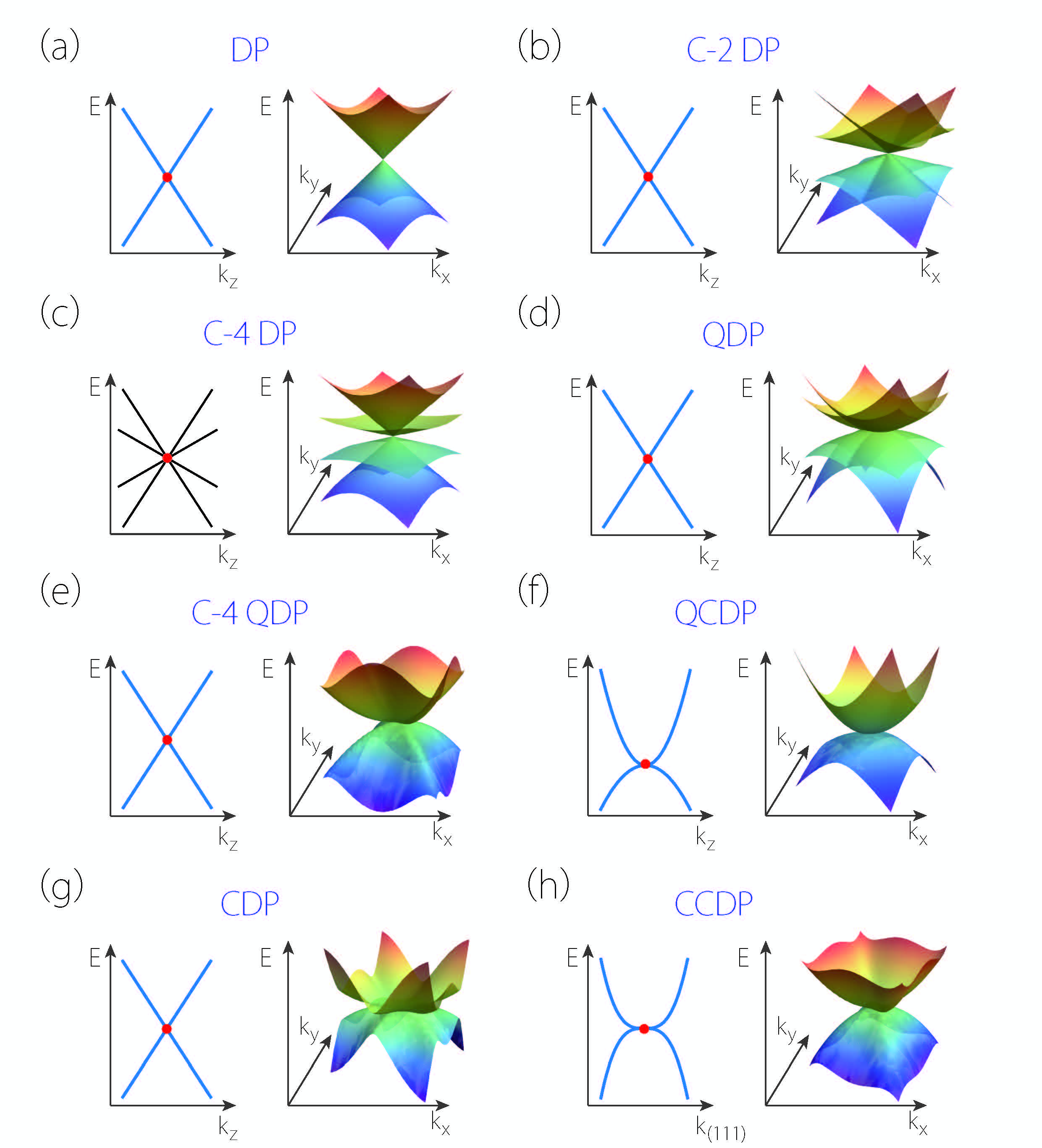}\caption{Typical band structure of  possible Dirac points in 3D crystals. The black and blue curves denote non-degenerate and doubly degenerate bands, respectively. \label{fig:DP}}
\end{figure}

The charge-2 Dirac point (C-2 DP) is a 0D four-fold band degeneracy with a topological charge ${\cal{C}}=\pm 2$.
It also features a linear dispersion along any direction in momentum space, and can occur on high-symmetry line or at high-symmetry point in BZ.
Contrast to DP which can be considered as a combination of two  C-1 WPs with opposite topological charge, the C-2 DP contains two  C-1 WPs with same topological charge.
The C-2 DP splits into four bands at a generic  $\boldsymbol{k}$  but into two doubly degenerate bands along certain high-symmetry line.
A typical band structure of   C-2 DP is schematically shown in Fig. \ref{fig:DP}(b).

A possible  Hamiltonian for C-2 DP may be written as
\begin{eqnarray}
H_{\text{C-2 DP}} & = & c_{1}k_{z}+\left[\begin{array}{cccc}
c_{2}k_{z} & 0 & \alpha k_{-} & \beta k_{+}\\
0 & c_{2}k_{z} & -\beta^{*}k_{+} & \alpha^{*}k_{-}\\
\alpha^{*}k_{+} & -\beta k_{-} & -c_{2}k_{z} & 0\\
\beta^{*}k_{-} & \alpha k_{+} & 0 & -c_{2}k_{z}
\end{array}\right].\label{eq:DP ham1-1}
\end{eqnarray}
The C-2 DP  can be further classified as type I and type II, depending on model parameters, while it at ${\cal T}$-symmetric point cannot be type II.

\subsubsection{Charge-4 Dirac point}

The charge-4 Dirac point (C-4 DP) is a 0D four-fold band degeneracy with a topological charge ${\cal{C}}=\pm 4$.
It features a linear dispersion along any direction in momentum space with all the four bands being fully splitted at each  generic  $\boldsymbol{k}$.
The C-4 DP only occurs at high-symmetry point in spinful systems.
A typical band structure of   C-4 DP is schematically shown in Fig. \ref{fig:DP}(c).

A possible Hamiltonian for C-4 DP may be written as
\begin{eqnarray}
H_{\text{C-4 DP}} & = & \left[\begin{array}{cccc}
-c_{1}k_{z} & c_{1}k_{+} & \alpha\frac{k_{-}+k_{x}-\sqrt{3}k_{y}}{2} & e^{-i2\pi/3}\alpha k_{z}\\
c_{1}k_{-} & c_{1}k_{z} & e^{-i2\pi/3}\alpha k_{z} & -\alpha\frac{k_{+}+k_{x}+\sqrt{3}k_{y}}{2}\\
\alpha^{*}\frac{k_{+}+k_{x}-\sqrt{3}k_{y}}{2} & e^{i2\pi/3}\alpha^{*}k_{z} & c_{1}k_{z} & -c_{1}k_{-}\\
e^{i2\pi/3}\alpha^{*}k_{z} & -\alpha^{*}\frac{k_{-}+k_{x}+\sqrt{3}k_{y}}{2} & -c_{1}k_{+} & -c_{1}k_{z}
\end{array}\right],
\end{eqnarray}
which essentially describes the conventional  spin-$3/2$ Weyl  fermion.

\subsubsection{Quadratic Dirac point}

The (charge-0) quadratic Dirac point (QDP) is a 0D four-fold band degeneracy with a topological charge ${\cal{C}}=0$.
It is formed by a linear crossing between two doubly degenerate bands along certain high-symmetry line, and has a quadratic energy splitting in the plane normal to the high-symmetry line.
The QDP in a spinful system with $I$ and ${\cal{T}}$ symmetries splits into two doubly degenerate bands at each $\boldsymbol{k}$ in BZ. For the other cases, it  splits into four non-degenerate bands at a generic $\boldsymbol{k}$.
The QDP can occur  on high-symmetry line or at high-symmetry point in BZ.
A typical band structure of QDP is schematically shown in Fig. \ref{fig:DP}(d).

A possible  Hamiltonian  for QDP may be written as
\begin{eqnarray}
H_{\text{QDP}} & = & c_{1}k_{z}+c_{2}k_{\parallel}^{2}+\left[\begin{array}{cccc}
1 & 0 & 0 & 0\\
0 & 1 & 0 & 0\\
0 & 0 & -1 & 0\\
0 & 0 & 0 & -1
\end{array}\right]\left(c_{3}k_{z}+c_{4}k_{\parallel}^{2}\right)\nonumber \\
 &  & +\left[\begin{array}{cccc}
c_{5} & \alpha_1 & 0 & 0\\
\alpha_1^{*} & -c_{5} & 0 & 0\\
0 & 0 & c_{6} & \alpha_2\\
0 & 0 & \alpha_2^{*} & -c_{6}
\end{array}\right]k_{x}k_{y}+\left[\begin{array}{cccc}
0 & 0 & \alpha_3 & \alpha_4\\
0 & 0 & \alpha_4^{*} & -\alpha_3^{*}\\
\alpha_3^{*} & \alpha_4 & 0 & 0\\
\alpha_4^{*} & -\alpha_3 & 0 & 0
\end{array}\right]\left(k_{x}^{2}-k_{y}^{2}\right),\label{eq:QDP ham1}
\end{eqnarray}
The QDP can be further classified as type I, type II and type III depending on model parameters, while it at  ${\cal T}$-symmetric point cannot be type II.

\subsubsection{Charge-4 quadratic Dirac point}

The charge-4 quadratic Dirac point (C-4 QDP)  is a 0D four-fold band degeneracy with a topological charge ${\cal{C}}=\pm4$.
It has a linear dispersion along certain  high-symmetry line and a quadratic energy splitting in the plane normal to the line.
The C-4 QDP can be considered as a combination of two C-2 WPs with same topological charge, and  only occurs at certain  high-symmetry points in spinful systems.
The  C-4 QDP splits into two doubly degenerate bands at three high-symmetry planes and along certain high-symmetry line. However, it  would split into four non-degenerate bands at a generic $\boldsymbol{k}$.
A typical band structure of C-4 QDP is schematically shown in Fig. \ref{fig:DP}(e).

A possible  Hamiltonian for C-4 QDP may be written as
\begin{eqnarray}
H_{\text{C-4 QDP}} & = & c_{1}k_{\parallel}^{2}+\left[\begin{array}{cccc}
c_{3}k_{x}k_{y} & ic_{2}k_{z} & \beta k_{x}k_{y} & \alpha k_{z}\\
-ic_{2}k_{z} & -c_{3}k_{x}k_{y} & -\alpha k_{z} & -\beta k_{x}k_{y}\\
\beta^{*}k_{x}k_{y} & -\alpha^{*}k_{z} & -c_{3}k_{x}k_{y} & -ic_{2}k_{z}\\
\alpha^{*}k_{z} & -\beta^{*}k_{x}k_{y} & ic_{2}k_{z} & c_{3}k_{x}k_{y}
\end{array}\right]+c_{4}\left(k_{x}^{2}-k_{y}^{2}\right)\left[\begin{array}{cccc}
0 & 1 & 0 & 0\\
1 & 0 & 0 & 0\\
0 & 0 & 0 & 1\\
0 & 0 & 1 & 0
\end{array}\right].
\end{eqnarray}

\subsubsection{Quadratic contact Dirac point}

The quadratic contact Dirac point (QCDP) is a  0D four-fold band degeneracy with a topological charge ${\cal{C}}=0$. It feature a quadratic energy splitting  along any direction in momentum space.
The QCDP in a  spinful system with  $I$ and ${\cal{T}}$ symmetries splits into two doubly degenerate bands at each $\boldsymbol{k}$ in BZ. For the other cases, it splits into four bands at a generic $\boldsymbol{k}$ but into two doubly degenerate bands along certain high-symmetry line.
The QCDP only occurs at high-symmetry point in BZ.
A typical band structure of QCDP is schematically shown in Fig. \ref{fig:DP}(f).

A possible  Hamiltonian  for QCDP may be written as
\begin{eqnarray}
H_{\text{QCDP}} & = & c_{1}k^{2}+c_{2}(3k_{x}^{2}-k^{2})\left[\begin{array}{cccc}
1 & 0 & 0 & 0\\
0 & 1 & 0 & 0\\
0 & 0 & -1 & 0\\
0 & 0 & 0 & -1
\end{array}\right]+c_{3}\left(k_{y}^{2}-k_{z}^{2}\right)\left[\begin{array}{cccc}
0 & 0 & 0 & 1\\
0 & 0 & -1 & 0\\
0 & -1 & 0 & 0\\
1 & 0 & 0 & 0
\end{array}\right]\nonumber \\
 &  & +c_{4}k_{x}k_{y}\left[\begin{array}{cccc}
0 & 0 & 1 & i\\
0 & 0 & i & 1\\
1 & -i & 0 & 0\\
-i & 1 & 0 & 0
\end{array}\right]+c_{4}k_{x}k_{z}\left[\begin{array}{cccc}
0 & 0 & 1 & -i\\
0 & 0 & -i & 1\\
1 & i & 0 & 0\\
i & 1 & 0 & 0
\end{array}\right]+\sqrt{2}c_{4}k_{y}k_{z}\left[\begin{array}{cccc}
0 & 0 & i & 0\\
0 & 0 & 0 & -i\\
-i & 0 & 0 & 0\\
0 & i & 0 & 0
\end{array}\right].
\end{eqnarray}

\subsubsection{Cubic Dirac point}

The (charge-0) cubic Dirac point (CDP) is a 0D four-fold band degeneracy with a topological charge ${\cal{C}}=0$.
It is formed by a linear crossing between two doubly degenerate bands along certain high-symmetry line, and has a cubic energy splitting in the plane normal to the high-symmetry line.
The CDP only occurs at certain high-symmetry points in spinful systems.
When the systems have  $I$ and ${\cal{T}}$ symmetries, the CDP would split into two doubly degenerate bands at each $\boldsymbol{k}$ in BZ. For the other cases, the CDP would  split into four non-degenerate bands at a generic $\boldsymbol{k}$, but into two doubly degenerate bands along certain high-symmetry line.
 A typical band structure of CDP is schematically shown in Fig. \ref{fig:DP}(g).

A possible  Hamiltonian  for CDP may be written as
\begin{eqnarray}
H_{\text{CDP}} & = & \left[\begin{array}{cccc}
c_{1}k_{\parallel}^{2} & i\left(c_{3}k_{+}^{3}+c_{4}k_{-}^{3}\right) & 0 & k_{z}c_{2}\\
-i\left(c_{3}k_{-}^{3}+c_{4}k_{+}^{3}\right) & c_{1}k_{\parallel}^{2} & k_{z}c_{2} & 0\\
0 & c_{2}k_{z} & c_{1}k_{\parallel}^{2} & i\left(c_{4}k_{+}^{3}+c_{3}k_{-}^{3}\right)\\
c_{2}k_{z} & 0 & -i\left(c_{4}k_{-}^{3}+c_{3}k_{+}^{3}\right) & c_{1}k_{\parallel}^{2}
\end{array}\right].
\end{eqnarray}

\subsubsection{Cubic crossing Dirac point}

The (charge-0) cubic crossing Dirac point (CCDP) is a 0D four-fold band degeneracy with a topological charge ${\cal{C}}=0$.
It is formed by a cubic (not linear) crossing between two doubly degenerate bands along certain high-symmetry line, and  has a quadratic energy splitting in the plane normal to the high-symmetry line. It would  split into four non-degenerate bands at a generic $\boldsymbol{k}$.
The CCDP can be considered as a combination of two C-4 WP with opposite topological charge, and hence only appears in spinless systems.
The CCDP is a new emergent particle that has never been reported before. A typical band structure of CCDP is schematically shown in Fig. \ref{fig:DP}(h).

A possible  Hamiltonian  for CCDP may be written as
\begin{eqnarray}
H_{\text{CCDP}} & = & \left[
\begin{array}{cc}
 0 & h_{12} \\
 h_{12}^{\dagger} & 0 \\
\end{array}
\right],
\end{eqnarray}
with
\begin{eqnarray}
h_{12} & = & \left[
\begin{array}{cc}
 \sqrt{3} \alpha _1 \left(k_x^2-k_y^2\right)+i \alpha _1
   \left(k_x^2+k_y^2-2 k_z^2\right) & \alpha _2 k_x k_y k_z \\
 -\alpha_2 k_x k_y k_z & -\sqrt{3} \alpha_1   \left(k_x^2-k_y^2\right)+i \alpha_1   \left(k_x^2+k_y^2-2 k_z^2\right) \\
\end{array}
\right],
\end{eqnarray}
which shows a cubic energy splitting along $k_{(111)}$ direction and quadratic energy splitting along $k_{x,y,z}$ direction, similar to the band splitting of the C-4 WP.

\subsection{Sixfold degeneracy point}

\subsubsection{Sextuple point}
The (charge-0) sextuple point (SP) is a 0D six-fold band degeneracy with a topological charge ${\cal{C}}=0$.
It features a linear energy splitting  along any direction in momentum space. The SP only occurs at high-symmetry point in BZ.
A typical band structure of SP is schematically shown in Fig. \ref{fig:SP}(a).

A possible Hamiltonian for SP may be written as
\begin{eqnarray}
H_{\text{SP}} & = & \left[\begin{array}{cccccc}
0 & c_{1}k_{x} & -c_{1}k_{y} & 0 & \alpha k_{x} & \alpha k_{y}\\
c_{1}k_{x} & 0 & -c_{1}k_{z} & -\alpha k_{x} & 0 & -\alpha k_{z}\\
-c_{1}k_{y} & -c_{1}k_{z} & 0 & -\alpha k_{y} & \alpha k_{z} & 0\\
0 & \alpha^{*}k_{x} & \alpha^{*}k_{y} & 0 & -c_{1}k_{x} & c_{1}k_{y}\\
-\alpha^{*}k_{x} & 0 & -\alpha^{*}k_{z} & -c_{1}k_{x} & 0 & c_{1}k_{z}\\
-\alpha^{*}k_{y} & \alpha^{*}k_{z} & 0 & c_{1}k_{y} & c_{1}k_{z} & 0
\end{array}\right].
\end{eqnarray}

\subsubsection{Charge-4 sextuple point}

\begin{figure}
\includegraphics[width=16cm]{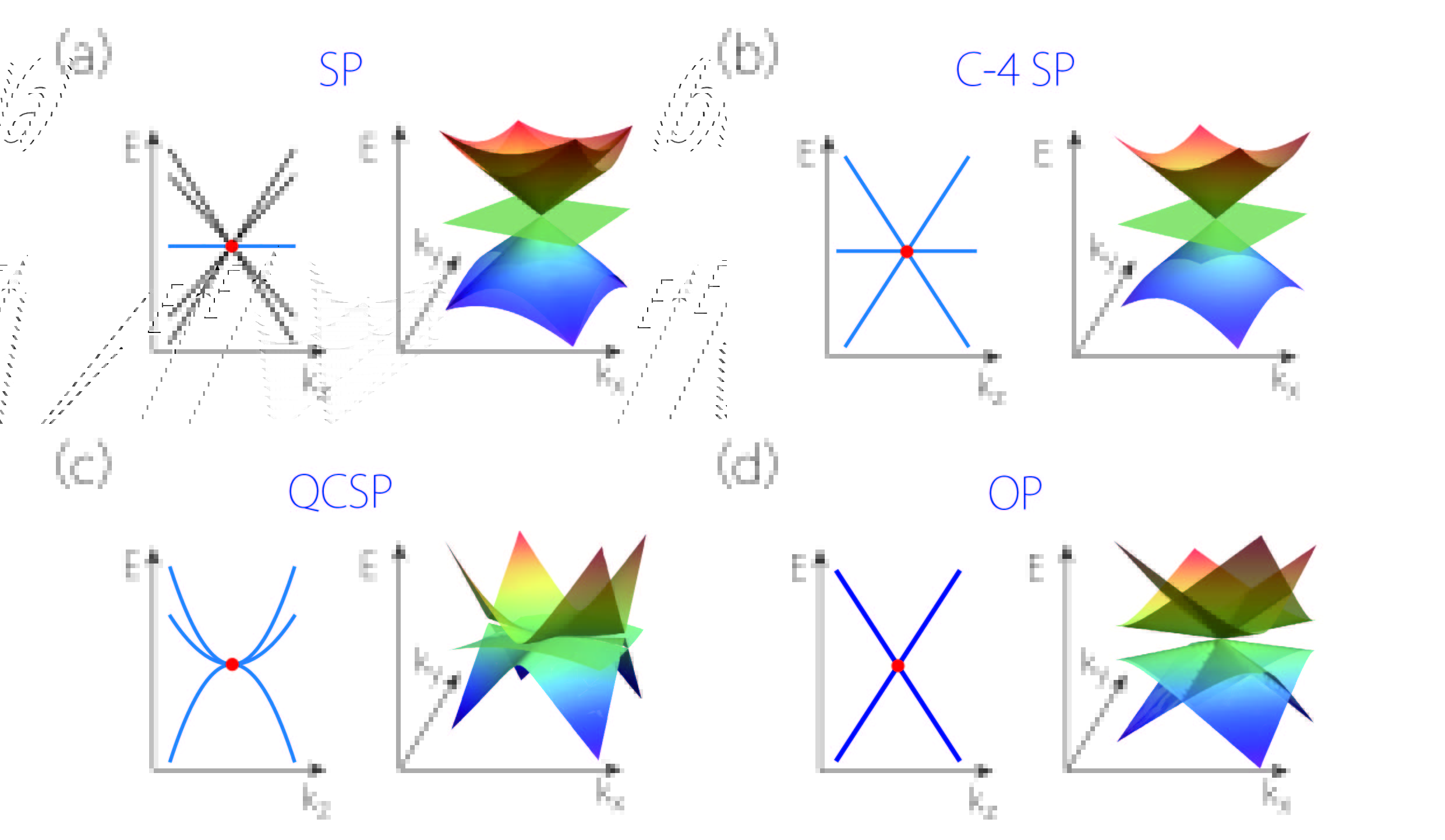}\caption{Typical band structure of  possible six-fold and eight-fold nodal points in 3D crystals. The black, blue and dark blue curves denote non-degenerate, doubly degenerate and  four-fold degenerate bands, respectively. \label{fig:SP}}
\end{figure}

The charge-4 sextuple point (C-4 SP) is a 0D six-fold band degeneracy with a topological charge ${\cal{C}}=\pm 4$.
It linearly splits into six bands at a generic $\boldsymbol{k}$, but into three doubly degenerate bands along certain high-symmetry line and in three high-symmetry planes.
The C-4 SP can be considered as a combination of two C-2 TPs with same topological charge, and  only occurs at certain high-symmetry points in spinful systems.
A typical band structure of C-4 SP is schematically shown in Fig. \ref{fig:SP}(b).

A possible  Hamiltonian for C-4 SP may be written as
\begin{eqnarray}
H_{\text{C-4 SP}} & = & \left[\begin{array}{cccccc}
0 & \alpha_1 k_{x} & \alpha_1^{*}k_{y} & 0 & \alpha_2 k_{x} & \alpha_2 k_{y}\\
\alpha_1^{*}k_{x} & 0 & \alpha_1 k_{z} & \alpha_2 k_{x} & 0 & \alpha_2 k_{z}\\
\alpha_1 k_{y} & \alpha_1^{*}k_{z} & 0 & \alpha_2 k_{y} & \alpha_2 k_{z} & 0\\
0 & \alpha_2^{*}k_{x} & \alpha_2^{*}k_{y} & 0 & -\alpha_1^{*}k_{x} & -\alpha_1 k_{y}\\
\alpha_2^{*}k_{x} & 0 & \alpha_2^{*}k_{z} & -\alpha_1 k_{x} & 0 & -\alpha_1^{*}k_{z}\\
\alpha_2^{*}k_{y} & \alpha_2^{*}k_{z} & 0 & -\alpha_1^{*}k_{y} & -\alpha_1 k_{z} & 0
\end{array}\right].
\end{eqnarray}

\subsubsection{Quadratic contact sextuple point}

The quadratic contact sextuple point (QCSP) is a 0D six-fold band degeneracy with a topological charge ${\cal{C}}=0$.
It features a quadratic dispersion along any direction in momentum space.
Interestingly, the QCSP only occurs in spinful systems with SG 205 locating at  high-symmetry point $R$.
The QCSP splits into three doubly degenerate bands at each $\boldsymbol{k}$ in BZ as SG 205 has $I$ symmetry.
A typical band structure of QCSP is schematically shown in Fig. \ref{fig:SP}(c).

A possible  Hamiltonian for QCSP may be written as
\begin{eqnarray}
H_{\text{QCSP}} & = & c_{1}\left[\begin{array}{cccccc}
k_{x}^{2} & 0 & 0 & 0 & 0 & 0\\
0 & k_{z}^{2} & 0 & 0 & 0 & 0\\
0 & 0 & k_{y}^{2} & 0 & 0 & 0\\
0 & 0 & 0 & k_{x}^{2} & 0 & 0\\
0 & 0 & 0 & 0 & k_{z}^{2} & 0\\
0 & 0 & 0 & 0 & 0 & k_{y}^{2}
\end{array}\right]+c_{2}\left[\begin{array}{cccccc}
k_{y}^{2} & 0 & 0 & 0 & 0 & 0\\
0 & k_{x}^{2} & 0 & 0 & 0 & 0\\
0 & 0 & k_{z}^{2} & 0 & 0 & 0\\
0 & 0 & 0 & k_{y}^{2} & 0 & 0\\
0 & 0 & 0 & 0 & k_{x}^{2} & 0\\
0 & 0 & 0 & 0 & 0 & k_{z}^{2}
\end{array}\right]+c_{3}\left[\begin{array}{cccccc}
k_{z}^{2} & 0 & 0 & 0 & 0 & 0\\
0 & k_{y}^{2} & 0 & 0 & 0 & 0\\
0 & 0 & k_{x}^{2} & 0 & 0 & 0\\
0 & 0 & 0 & k_{z}^{2} & 0 & 0\\
0 & 0 & 0 & 0 & k_{y}^{2} & 0\\
0 & 0 & 0 & 0 & 0 & k_{x}^{2}
\end{array}\right]\nonumber \\
 &  & +\left[\begin{array}{cccccc}
0 & \alpha_1 k_{y}k_{z} & \alpha_1^{*}k_{x}k_{z} & 0 & \alpha_2 k_{y}k_{z} & -\alpha_2 k_{x}k_{z}\\
\alpha_1^{*}k_{y}k_{z} & 0 & \alpha_1 k_{x}k_{y} & -\alpha_2 k_{y}k_{z} & 0 & \alpha_2 k_{x}k_{y}\\
\alpha_1 k_{x}k_{z} & \alpha_1^{*}k_{x}k_{y} & 0 & \alpha_2 k_{x}k_{z} & -\alpha_2 k_{x}k_{y} & 0\\
0 & -\alpha_2^{*}k_{y}k_{z} & \alpha_2^{*}k_{x}k_{z} & 0 & \alpha_1^{*}k_{y}k_{z} & \alpha_1 k_{x}k_{z}\\
\alpha_2^{*}k_{y}k_{z} & 0 & -\alpha_2^{*}k_{x}k_{y} & \alpha_1 k_{y}k_{z} & 0 & \alpha_1^{*}k_{x}k_{y}\\
-\alpha_2^{*}k_{x}k_{z} & \alpha_2^{*}k_{x}k_{y} & 0 & \alpha_1^{*}k_{x}k_{z} & \alpha_1 k_{x}k_{y} & 0
\end{array}\right].
\end{eqnarray}

\subsection{Eightfold degeneracy point}
\subsubsection{Octuple point}

The (charge-0) octuple point (OP) is a 0D eight-fold band degeneracy with a topological charge ${\cal{C}}=0$.
It features a linear energy dispersion  along any direction in momentum space, and is formed by a linear crossing between two fourfold degenerate bands along certain high-symmetry line.
When the systems have $I$ and ${\cal{T}}$ symmetries, the OP splits into four doubly degenerate bands at a generic  $\boldsymbol{k}$ in BZ. For the other cases, it would split into eight non-degenerate bands at a generic  $\boldsymbol{k}$.
The OP only occurs at  certain high-symmetry points in spinful systems.
A typical band structure of SP is schematically shown in Fig. \ref{fig:SP}(d).

A possible  Hamiltonian for OP may be written as
\begin{eqnarray}
H_{\text{OP}} & = & \left[\begin{array}{cc}
h_{11} & h_{12}\\
h_{12}^{\dagger} & h_{22}
\end{array}\right],
\end{eqnarray}
with
\begin{eqnarray}
h_{11} & = & \left[\begin{array}{cccc}
c_{1}k_{y} & c_{1}k_{x}+ic_{2}k_{z} & -ic_{3}k_{y} & ic_{3}k_{x}\\
c_{1}k_{x}-ic_{2}k_{z} & -c_{1}k_{y} & ic_{3}k_{x} & ic_{3}k_{y}\\
ic_{3}k_{y} & -ic_{3}k_{x} & -c_{1}k_{y} & -c_{1}k_{x}-ic_{2}k_{z}\\
-ic_{3}k_{x} & -ic_{3}k_{y} & -c_{1}k_{x}+ic_{2}k_{z} & c_{1}k_{y}
\end{array}\right], \\
h_{12} & = & \left[\begin{array}{cccc}
0 & \alpha_1 k_{z} & -\alpha_2 k_{y} & \alpha_2 k_{x}\\
-\alpha_1 k_{z} & 0 & \alpha_2 k_{x} & \alpha_2 k_{y}\\
\alpha_2 k_{y} & -\alpha_2 k_{x} & 0 & -\alpha_1 k_{z}\\
-\alpha_2 k_{x} & -\alpha_2 k_{y} & \alpha_1 k_{z} & 0
\end{array}\right],\\
h_{22} & = & \left[\begin{array}{cccc}
c_{1}k_{y} & c_{1}k_{x}-ic_{2}k_{z} & ic_{3}k_{y} & -ic_{3}k_{x}\\
c_{1}k_{x}+ic_{2}k_{z} & -c_{1}k_{y} & -ic_{3}k_{x} & -ic_{3}k_{y}\\
-ic_{3}k_{y} & ic_{3}k_{x} & -c_{1}k_{y} & -c_{1}k_{x}+ic_{2}k_{z}\\
ic_{3}k_{x} & ic_{3}k_{y} & -c_{1}k_{x}-ic_{2}k_{z} & c_{1}k_{y}
\end{array}\right].
\end{eqnarray}

\subsection{Twofold degeneracy line}

\subsubsection{Weyl nodal line}

The Weyl nodal line (WNL) is a 1D two-fold band degeneracy. The WNL features a linear energy dispersion in the plane normal to the line.
Moreover, the topological charge (Berry phase) for WNL is ${\cal{C}}=\pi$.
The WNL  generally   appears along high-symmetry line or in  high-symmetry plane in BZ.
A typical band structure of WNL is schematically shown in Fig. \ref{fig:NL-NS}(a).

A possible  Hamiltonian expanded at a general  point $\boldsymbol{K}$ on  WNL may be written as
\begin{eqnarray}
H_{\text{WNL}} & = & c_{1}q_{x}+c_{2}q_{y}+c_{3}q_{x}\sigma_{1}+c_{4}q_{y}\sigma_{2},
\end{eqnarray}
with the wave vector $\boldsymbol{q}$ measured from $\boldsymbol{K}$. Here, we assume the $q_{x}$-$q_{y}$ plane passes through $\boldsymbol{K}$ point and is normal to the WNL. The WNL can be further classified as type I, type II \cite{Li2017Type-PRB}  and hybrid type \cite{Zhang_PRBHNL_2018}, depending on model parameters.

\subsubsection{Weyl nodal-line net}

The Weyl nodal-line net (WNL net) contains multiple two-fold NLs, which  share (at least) one nodal point in momentum space, as schematically shown in Fig. \ref{fig:NL-NS}(b).
The  joint nodal point of the WNLs  must locates at high-symmetry line or high-symmetry point in BZ, and are termed as P-WNLs in Sec. \ref{S3B} and Sec. \ref{S4B}.
WNL net can take various form in BZ, such as crossed nodal line \cite{Weng2015Topological-PRB}, nodal chain \cite{Tomas_NC_2016}  nodal box \cite{Sheng_JPCL_2017} and so on.

A possible Hamiltonian expanded around a P-WNLs may be written as
\begin{eqnarray}
H_{\text{WNL net}} & = & \left[c_{1}k_{z}+c_{2}k_{\parallel}^{2}+c_{3}k_{x}(k_{x}^{2}-3k_{y}^{2})\right]\sigma_{3}+\alpha k_{y}(3k_{x}^{2}-k_{y}^{2})\sigma_{1}, \label{eq:WNL net}
\end{eqnarray}
which indicates that the  point described by Eq. (\ref{eq:WNL net}) is not isolated but shall be an joint point of three WNLs  lying in  three vertical mirror planes dictated by equation $3k_{x}^{2}-k_{y}^{2}=0$. The band structure
obtained from   Eq. (\ref{eq:WNL net}) is shown in Fig. \ref{fig:NL-NS}(b), where we set $c_2=c_3=0$ for the convenience of a clear presentation.

\subsubsection{Quadratic nodal line}

The quadratic nodal line (QNL) is a 1D two-fold band degeneracy with a topological charge ${\cal{C}}=0$ mod $2\pi$.
The QNL features a quadratic  energy splitting  in the plane normal to the line.
It only appears along high-symmetry line in BZ.
A typical band structure of QNL is schematically shown in Fig. \ref{fig:NL-NS}(c).

A possible   Hamiltonian expanded at a general  point $\boldsymbol{K}$ on  QNL may be written as
\begin{eqnarray}
H_{\text{QNL}} & = & c_{1}k_{\parallel}^{2}+\left(\alpha k_{+}^{2}\sigma_{+}+h.c.\right).
\end{eqnarray}
The QNL can be further classified as type I, type III  and hybrid type  depending on model parameters.

\subsubsection{Cubic nodal line}

The cubic nodal line (CNL) is a 1D two-fold band degeneracy with a topological charge ${\cal{C}}=\pi$ mod $2\pi$.
The CNL features a cubic   energy splitting  in the plane normal to the line.
It only appears along high-symmetry line in BZ.
A typical band structure of CNL is schematically shown in Fig. \ref{fig:NL-NS}(d).

A possible   Hamiltonian expanded at a general  point $\boldsymbol{K}$ on  CNL may be written as
\begin{eqnarray}
H_{\text{CNL}} & = & c_{1}k_{\parallel}^{2}+\left(\alpha k_{+}^{3}\sigma_{+}+h.c.\right).
\end{eqnarray}

\subsection{Fourfold degeneracy line}
\subsubsection{Dirac nodal line}

The Dirac nodal line (DNL) is a 1D four-fold band degeneracy with a topological charge ${\cal{C}}=0$ mod $2\pi$.
It features a linear dispersion  in the plane normal to the line.
The DNL in spinful systems with  $I$ and ${\cal{T}}$ symmetries splits into two doubly degenerate bands at a generic $\boldsymbol{k}$ in BZ. For the other cases, it  splits into four bands at a generic $\boldsymbol{k}$, but would split into two doubly degenerate bands along certain high-symmetry line(s).
It can appear along high-symmetry line or in high-symmetry plane in BZ.
A typical band structure of DNL is schematically shown in Fig. \ref{fig:NL-NS}(e).

A possible  Hamiltonian expanded at a general  point $\boldsymbol{K}$ on  DNL may be written as
\begin{eqnarray}
H_{\text{DNL}} & = & c_{1}k_{z}+\left[\begin{array}{cccc}
0 & c_{2}k_{y}+ic_{3}k_{x} & 0 & \alpha_1 k_{x}+\alpha_2 k_{y}\\
c_{2}k_{y}-ic_{3}k_{x} & 0 & -\alpha_1 k_{x}+\alpha_2 k_{y} & 0\\
0 & -\alpha_1^{*}k_{x}+\alpha_2^{*}k_{y} & 0 & -c_{2}k_{y}-ic_{3}k_{x}\\
\alpha_1^{*}k_{x}+\alpha_2^{*}k_{y} & 0 & -c_{2}k_{y}+ic_{3}k_{x} & 0
\end{array}\right].
\end{eqnarray}
The DNL can be further classified as type I, type II and hybrid type depending on model parameters.

\subsubsection{Dirac nodal-line net}

The Dirac nodal-line net (DNL net) contains multiple four-fold NLs, which share (at least) one nodal point in momentum space, as schematically shown in Fig. \ref{fig:NL-NS}(f).
The joint nodal point of the DNLs  must locates at high-symmetry line or high-symmetry point in BZ, and are termed as P-DNLs in Sec. \ref{S3B} and Sec. \ref{S4B}.

A possible Hamiltonian expanded around a P-DNLs may be written as
\begin{eqnarray}
H_{\text{DNL net}} & = &\left[\begin{array}{cccc}
 c_1 k_x^2+c_2 k_y^2+c_3 k_z & 0 & c_7 k_x k_y & c_8 k_x k_y \\
 0 & c_1 k_x^2+c_2 k_y^2+c_3 k_z & -c_8 k_x k_y & c_7 k_x k_y \\
 c_7 k_x k_y & -c_8 k_x k_y & -c_4 k_x^2-c_5 k_y^2-c_6 k_z & 0 \\
 c_8 k_x k_y & c_7 k_x k_y & 0 & -c_4 k_x^2-c_5 k_y^2-c_6 k_z \\
\end{array}\right], \label{eq:DNL}
\end{eqnarray}
which indicates that the point described by Eq. (\ref{eq:DNL}) is not isolated but shall be an joint point of two DNLs lying in $k_x=0$ and $k_y=0$ mirror planes.  The band structure obtained from Eq. (\ref{eq:DNL}) is shown in Fig. \ref{fig:NL-NS}(f).

\subsection{Twofold degeneracy surface}
\subsubsection{Nodal surface}
\begin{figure}
\includegraphics[width=16cm]{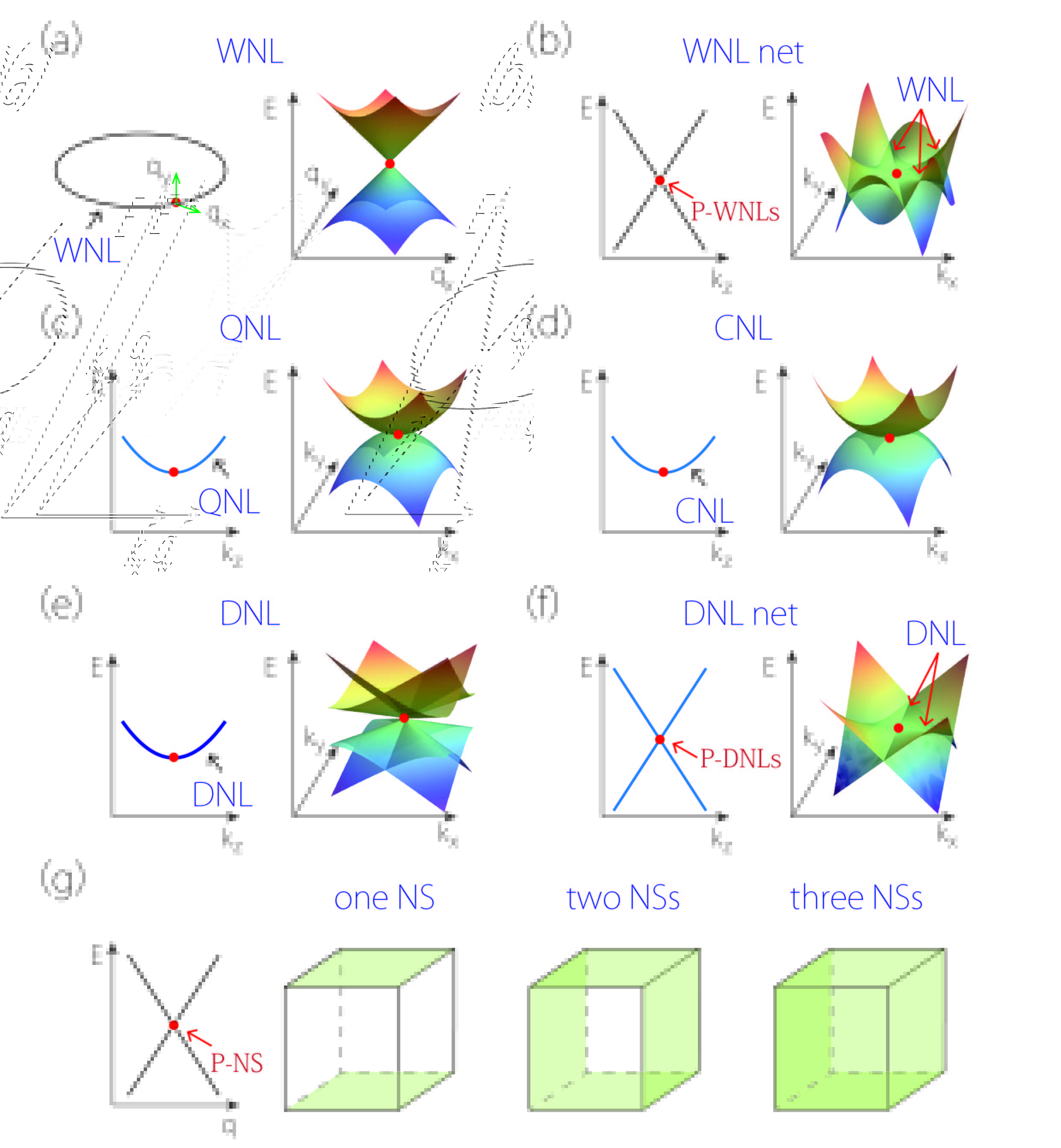}\caption{Typical band structure of  possible nodal line and nodal surface in 3D crystals. The black, blue and dark blue curves denote non-degenerate, doubly degenerate and four-fold degenerate bands, respectively.
\label{fig:NL-NS}}
\end{figure}

The nodal surface (NS) is a 2D two-fold band degeneracy.
The NS only appears at the boundary plane of BZ and has linear dispersion along the direction normal to the surface.
Moreover, the 2D nodal surface in 3D systems can be   topologically characterized  by a ${\mathbb{Z}}_2$-valued  topological charge, which is defined on (a 0D sphere) two points surrounding the surface in BZ \cite{Chiu2016Classification-RoMP,Wu2018Nodal-PRB}.
A typical band structure of NS is schematically shown in the left picture  of Fig. \ref{fig:NL-NS}(g).

A possible  Hamiltonian expanded around  a general  point $\boldsymbol{K}$ in NS may be written as
\begin{eqnarray}
H_{\text{NS}} & = & c_{1}q+c_{2}q\sigma_{1},
\end{eqnarray}
with the wave vector $q$ measured from $\boldsymbol{K}$. Here, we assume the $q$ axis passes through $\boldsymbol{K}$ point and is normal to the surface.

There only exist three possibilities for the NS semimetals, namely, the systems exhibits one NS, two NSs or three NSs, as illustrated in Fig. \ref{fig:NL-NS}(g).

\section{Derivation of the coreps of type-II MSGs}
\subsection{Abstract group}
It is known that for a generic momentum $\boldsymbol{k}$ in BZ of a certain SG, its symmetry is described by a little group $\text{\textbf{G}}^{\boldsymbol{k}}$.  For high-symmetry point $\boldsymbol{k}_1$, the reps of  $\text{\textbf{G}}^{\boldsymbol{k}_1}$  generally is  related to those of Herring little group (see Sec. 3.8 in Ref. \cite{Bradley2009Mathematical-Oxford}) and for high-symmetry line $\boldsymbol{k}_1$, the reps of  $\text{\textbf{G}}^{\boldsymbol{k}_1}$ is related to those of the  central extension of the corresponding little co-group (see Sec. 3.7 in Ref. \cite{Bradley2009Mathematical-Oxford}).
Notice that, different from the notation used in Ref. \cite{Bradley2009Mathematical-Oxford}, we will abuse the symbol $\text{\textbf{G}}^{\boldsymbol{k}_1}$ to denote  Herring little group (at high-symmetry points)  and central extension (on high-symmetry lines) in the following discussion.
$\text{\textbf{G}}^{\boldsymbol{k}_1}$ (the Herring little group or the central extension) for different high-symmetry momenta in one SG or for same high-symmetry momentum in different SGs may be isomorphic to one certain abstract group, leading  to a great simplification for investigating the reps of 230 SGs.
In  Ref. \cite{Bradley2009Mathematical-Oxford}, the abstract group for each $\text{\textbf{G}}^{\boldsymbol{k}_1}$ in 230 SGs are explicitly given in Table 5.7 and Table 6.13, respectively, and the rep information of the relevant abstract groups are presented in Table 5.1. Moreover, the correspondence between the reps of abstract group (labelled as $R_{i}$ with $i=1,2,3 ...$ in Ref. \cite{Bradley2009Mathematical-Oxford}) and the conventional notations of the reps of electronic bands can be found in Table 5.8 and Table 6.14 of Ref. \cite{Bradley2009Mathematical-Oxford}, and  a recent work by Liu et. al. \cite{GBLiu_2020}.

\subsection{Magnetic space groups and corepresentations} \label{MSG-corep}
The MSGs also is called as Shubnikov SGs, and  can be subdivided into four types \cite{Bradley2009Mathematical-Oxford}. There exist in total  1651 MSGs.
The type I MSGs are the ordinary 230 SGs, containing only unitary operators. A type I MSG, $\text{\textbf{M}}$, is given by
\begin{equation}
\text{\textbf{M}}=\text{\textbf{G}},
\end{equation}
where $\text{\textbf{G}}$ is any ordinary SG. In the following, we use $\text{\textbf{G}}$ and $\text{\textbf{M}}$ to denote ordinary SG and MSG, respectively. The single- and double-valued reps of type I MSGs (e.g. the ordinary SGs) have been given  in Ref. \cite{Bradley2009Mathematical-Oxford}.

All the other three types of MSGs contain  anti-unitary operators, e.g. the operators involve time-reversal symmetry ${\cal T}$. A type II MSG, $\text{\textbf{M}}$, is given by
\begin{equation}
\text{\textbf{M}}=\text{\textbf{G}}+{\cal T}\text{\textbf{G}}.
\end{equation}
Clearly, there are 230 type II MSGs. Since ${\cal T}$ commutes with all the SG operators, one then can rewrite the  type II MSG $\text{\textbf{M}}$, as $\text{\textbf{M}}=\text{\textbf{G}}\otimes\left\{ E+{\cal T}\right\} $ with $E$ the identity operator. A type III MSG, $\text{\textbf{M}}$, is given by
\begin{equation}
\text{\textbf{M}}=\text{\textbf{H}}+{\cal T}(\text{\textbf{G}}-\text{\textbf{H}}),
\end{equation}
where $\text{\textbf{H}}$ is a halving subgroup of  $\text{\textbf{G}}$. The total number of type III MSG is 674, more than the number of the ordinary SG. At last, the type IV MSGs are defined on black and white Bravais lattices, which include two interpenetrating sublattices occupied by up-spin (black) and down-spin (white), respectively. A type IV MSG, $\text{\textbf{M}}$, is given by
\begin{equation}
\text{\textbf{M}}=\text{\textbf{G}}+{\cal T}\left\{ E|\boldsymbol{t}_{0}\right\} \text{\textbf{G}},
\end{equation}
where $\boldsymbol{t}_{0}$ is a translation connecting black and white Bravais lattices. There are 517 type IV MSGs. Generally, one can use a general form to rewrite the type II, III, and IV MSG, expressed as
\begin{equation}\label{eq:magG}
\text{\textbf{M}}=\text{\textbf{G}}^{\prime}+{\cal A}\text{\textbf{G}}^{\prime},
\end{equation}
with $\text{\textbf{G}}^{\prime}$ a unitary subgroup of $\textbf{M}$ and ${\cal A}$ an anti-unitary element of $\textbf{M}$. For type II MSG, ${\cal A}$ is ${\cal T}$ symmetry, and for type  III MSG, ${\cal A}$ is a combined operator containing ${\cal T}$  and a spatial operator $O$. Notice that $O$ is not an element of $\text{\textbf{G}}^{\prime}$. For type IV MSG, the anti-unitary operator ${\cal A}$  is  ${\cal T}\left\{ E|\boldsymbol{t}_{0}\right\}$, namely, ${\cal T}$ symmetry followed by a pure translation.

While one uses the representation theory to study the type I MSGs, e.g. the ordinary SGs, it is necessary to extend the representation theory to corepresentation theory \cite{Wigner-coreps} for studying type II, III, and IV MSGs, as they contain anti-unitary operators.

Without loss of generality, we consider a magnetic group
\begin{equation}\label{eq:magG}
\text{\textbf{M}}_0=\text{\textbf{G}}_0+{\cal A}\text{\textbf{G}}_0,
\end{equation}
and assuming  the rep information  of $\text{\textbf{G}}_{0}$ is known. For a irreducible rep $\Gamma$ of $\text{\textbf{G}}_{0}$ and an operator $R\in\text{\textbf{G}}_{0}$, one has
\begin{eqnarray}
R\langle\psi| & = & \langle\psi|\boldsymbol{\Delta}(R),
\end{eqnarray}
with $\langle\psi|$ the basis state of $\Gamma$ and $\boldsymbol{\Delta}(R)$ the matrix representation of $R$ in $\Gamma$. Then the basis state of $\textbf{M}_0$ can be written as $\langle\psi,\phi|$ with $\langle\phi|={\cal A}\langle\psi|$.
A straightforward calculation gives \cite{Bradley2009Mathematical-Oxford}
\begin{equation}
R\langle\psi,\phi|=\langle\psi,\phi|\left[\begin{array}{cc}
\boldsymbol{\Delta}(R) & \boldsymbol{0}\\
\boldsymbol{0} & \boldsymbol{\Delta}^{*}({\cal A}^{-1}R{\cal A})
\end{array}\right],
\end{equation}
for $R\in\text{\textbf{G}}_{0}$, and
\begin{equation}
B\langle\psi,\phi|=\langle\psi,\phi|\left[\begin{array}{cc}
\boldsymbol{0} & \boldsymbol{\Delta}(B{\cal A})\\
\boldsymbol{\Delta}^{*}({\cal A}^{-1}B) & \boldsymbol{0}
\end{array}\right],
\end{equation}
for $B\in {\cal A}\text{\textbf{G}}_{0}$. The matrices
\begin{eqnarray}
\boldsymbol{D}(R)=\left[\begin{array}{cc}
\boldsymbol{\Delta}(R) & \boldsymbol{0}\\
\boldsymbol{0} & \boldsymbol{\Delta}^{*}({\cal A}^{-1}R{\cal A})
\end{array}\right], & \ \  & \boldsymbol{D}(B)=\left[\begin{array}{cc}
\boldsymbol{0} & \boldsymbol{\Delta}(B{\cal A})\\
\boldsymbol{\Delta}^{*}({\cal A}^{-1}B) & \boldsymbol{0}
\end{array}\right], \label{eq: DRB}
\end{eqnarray}
are the coreps of $\textbf{M}_0$  derived from  $\Gamma$ of $\text{\textbf{G}}_{0}$, which are denoted as $\boldsymbol{D}\Gamma$.
The relations of coreps are different from those of reps.
The relations of reps are
\begin{equation}
\boldsymbol{\Delta}(R)\boldsymbol{\Delta}(S)=\boldsymbol{\Delta}(RS),
\end{equation}
with $R,\ S\in\text{\textbf{G}}_{0}$. In contrast, the relations of coreps have a dependence on the first rep, given as
\begin{eqnarray}
\boldsymbol{D}(R)\boldsymbol{D}(S)=\boldsymbol{D}(RS), & \ \  & \boldsymbol{D}(B)\boldsymbol{D}^{*}(S)=\boldsymbol{D}(BS),
\end{eqnarray}
with $R\in\text{\textbf{G}}_{0}$, $B\in {\cal A}\text{\textbf{G}}_{0}$ and $S\in\textbf{M}_0$.

Notice that although  $\Gamma$ is a irreducible rep of $\text{\textbf{G}}_{0}$, it does not mean the deduced corep $\boldsymbol{D}\Gamma$ also is irreducible. If there exist a unitary transformation changing $\boldsymbol{D}\Gamma$ to $\boldsymbol{D}'\Gamma$ and all the matrices of $\boldsymbol{D}'\Gamma$ are in the same block diagonal form, then the corep $\boldsymbol{D}\Gamma$ is reducible. In contrast, if there does not exist such transformation, the corep $\boldsymbol{D}\Gamma$ is irreducible.
A irreducible (reducible) corep $\boldsymbol{D}\Gamma$ means that for an electronic band belonging to the rep $\Gamma$, its degeneracy is doubled (not affected) by the addition of ${\cal A}$ symmetry to the group $\text{\textbf{G}}_{0}$. There are  three different cases for the deduced corep $\boldsymbol{D}\Gamma$ \cite{Bradley2009Mathematical-Oxford}.

Case (a): The deduced corep $\boldsymbol{D}\Gamma$ is reduciable, corresponding to the case that the addition of ${\cal A}$ symmetry does not change the degeneracy of the electronic band with $\Gamma$ rep. The matrix representation  of the operators in $\textbf{M}_0$ for the  corep $\boldsymbol{D}\Gamma$ then is
\begin{eqnarray}
\boldsymbol{D}(R)=\boldsymbol{\Delta}(R), & \ \  & \boldsymbol{D}(B)=\boldsymbol{\Delta}(B{\cal A}^{-1})\boldsymbol{N},\label{eq:case-a}
\end{eqnarray}
where  $\boldsymbol{N}$ is determined by the following equations $\boldsymbol{N}\boldsymbol{N}^{*}=\boldsymbol{\Delta}({\cal A}^{2})$ and $\boldsymbol{\Delta}(R)=\boldsymbol{N}\boldsymbol{\Delta}^{*}({\cal A}^{-1}R{\cal A})\boldsymbol{N}^{-1}$. From Eq. (\ref{eq:case-a}), one directly knows
\begin{eqnarray}
\boldsymbol{D}({\cal A}) &=& \boldsymbol{N}.
\end{eqnarray}

Case (b): The deduced corep $\boldsymbol{D}\Gamma$ is irreduciable, and $\boldsymbol{\Delta}(R)$ is equivalent to $\boldsymbol{\Delta}^{*}({\cal A}^{-1}R{\cal A})$.
Then the matrix set for $\boldsymbol{D}\Gamma$ (\ref{eq: DRB}) under unitary transformation can be written as
\begin{eqnarray}
\boldsymbol{D}(R)=\left[\begin{array}{cc}
\boldsymbol{\Delta}(R) & \boldsymbol{0}\\
\boldsymbol{0} & \boldsymbol{\Delta}(R)
\end{array}\right], & \ \  & \boldsymbol{D}(B)=\left[\begin{array}{cc}
\boldsymbol{0} & -\boldsymbol{\Delta}(B{\cal A}^{-1})\boldsymbol{N}\\
\boldsymbol{\Delta}(B{\cal A}^{-1})\boldsymbol{N} & \boldsymbol{0}
\end{array}\right],\label{eq:case-b}
\end{eqnarray}
where  $\boldsymbol{N}$ is determined by the following equations $\boldsymbol{N}\boldsymbol{N}^{*}=-\boldsymbol{\Delta}({\cal A}^{2})$
and $\boldsymbol{\Delta}(R)=\boldsymbol{N}\boldsymbol{\Delta}^{*}({\cal A}^{-1}R{\cal A})\boldsymbol{N}^{-1}$. The matrix representation of  ${\cal A}$ in this case  can be written as
\begin{eqnarray}
\boldsymbol{D}({\cal A})=\left[\begin{array}{cc}
\boldsymbol{0} & -\boldsymbol{N}\\
\boldsymbol{N} & \boldsymbol{0}
\end{array}\right].
\end{eqnarray}

Case (c): The deduced corep $\boldsymbol{D}\Gamma$ is irreduciable, and $\boldsymbol{\Delta}(R)$ is not equivalent to $\boldsymbol{\Delta}^{*}({\cal A}^{-1}R{\cal A})$. The matrix set for $\boldsymbol{D}\Gamma$  then is
\begin{eqnarray}
\boldsymbol{D}(R)=\left[\begin{array}{cc}
\boldsymbol{\Delta}(R) & \boldsymbol{0}\\
\boldsymbol{0} & \boldsymbol{\Delta}^{*}({\cal A}^{-1}R{\cal A})
\end{array}\right], & \ \  & \boldsymbol{D}(B)=\left[\begin{array}{cc}
\boldsymbol{0} & \boldsymbol{\Delta}(B{\cal A})\\
\boldsymbol{\Delta}^{*}({\cal A}^{-1}B) & \boldsymbol{0}
\end{array}\right],\label{eq:case-c}
\end{eqnarray}
and the matrix representation  of  ${\cal A}$   in this case is
\begin{eqnarray}
\boldsymbol{D}({\cal A})=\left[\begin{array}{cc}
\boldsymbol{0} & \boldsymbol{\Delta}({\cal A}^2)\\
\boldsymbol{1} & \boldsymbol{0}
\end{array}\right],
\end{eqnarray}
with $\boldsymbol{1}$ denoting identity matrix. Hence, whenever the matrix $\boldsymbol{N}$ is solved, the  coreps then are obtained.
Notice that whether or not the degeneracy of the electronic bands is doubled  by the addition of ${\cal A}$, the corresponding effective Hamiltonian of the bands would be inevitably  affected, as ${\cal A}$ imposes additional symmetry constraint on the effective Hamiltonian.

The single- and double-valued reps of 230 SGs has been fully investigated in previous works and can be conveniently accessed, such as from the classic book ``The mathematical theory of symmetry in solids: Representation theory for point groups and space groups" \cite{Bradley2009Mathematical-Oxford} and the Bilbao Crystallographic Server (BCS) \cite{web}. However  the notations  of the reps used in Ref. \cite{Bradley2009Mathematical-Oxford} and Ref. \cite{web}  are different, and a correspondence between these two notations are given in a recent work by Liu et. al. \cite{GBLiu_2020}.
The single-valued and double-valued reps of  type II, III and IV MSGs based on BCS notation have been  established \cite{BradlynNature-2017,Elcoro:ks5574,Elcoro-MTQC,XuYuanfeng-Nature-2020} and uploaded in BCS website \cite{web2}.
In contrast,  both single-valued and double-valued reps of  type II, III and IV MSGs based on the notation of Ref. \cite{Bradley2009Mathematical-Oxford} are still missing.

\subsection{Concrete  steps of the derivation of corep}
In this work we adopt the notations of reps used in Ref. \cite{Bradley2009Mathematical-Oxford} and establish the complete tables for both single-valued and double-valued coreps of  type II MSGs from the reps of type I MSGs (the ordinary SGs) presented in Ref. \cite{Bradley2009Mathematical-Oxford}.
As aforementioned, the corresponding abstract group, generating elements   and  the generating relations of the little groups of the high-symmetry momenta ${\boldsymbol{k}_1}$ in BZ of each space group are given in Ref. \cite{Bradley2009Mathematical-Oxford}, which is very  helpful for obtaining  the corep information of each type II MSG.  While the type II MSG has one additional ${\cal T}$ symmetry comparing with the corresponding SG, the high-symmetry momenta ${\boldsymbol{k}_1}$ in BZ of one certain type II MSG may or may not have additional symmetries comparing with that in corresponding SG. Nevertheless, the little groups  $\text{\textbf{M}}^{\boldsymbol{k}_1}$  must take the form of either
\begin{eqnarray}
\text{\textbf{M}}^{\boldsymbol{k}_1}=\text{\textbf{G}}^{\boldsymbol{k}_1}, \label{eq:CS-X}
\end{eqnarray}
or
\begin{eqnarray}
\text{\textbf{M}}^{\boldsymbol{k}_1}=\text{\textbf{G}}^{\boldsymbol{k}_1}+{\cal{A}} \text{\textbf{G}}^{\boldsymbol{k}_1}, \label{eq:CS-abc}
\end{eqnarray}
with $\text{\textbf{G}}^{\boldsymbol{k}_1}$  the corresponding crystallographic little group, for which the rep information is known and available,  and ${\cal{A}}$ a certain anti-unitary operator. Starting from here, the  subsequent  steps for obtaining the (co)rep information of $\text{\textbf{M}}^{\boldsymbol{k}_1}$ are as follows.

First, for each $\text{\textbf{M}}^{\boldsymbol{k}_1}$, we  find out all the  symmetry operators to determine its suitable form [Eq.~\eqref{eq:CS-X} or Eq.~(\ref{eq:CS-abc})].

Second, if $\text{\textbf{M}}^{\boldsymbol{k}_1}$  has the form of Eq.~\eqref{eq:CS-X}, its reps $\Gamma_{i}$ and the corresponding matrix representations $\boldsymbol{\Delta}_{\Gamma_{i}}$ are available in Ref. \cite{Bradley2009Mathematical-Oxford}.

Third, if $\text{\textbf{M}}^{\boldsymbol{k}_1}$ has the form of Eq.~(\ref{eq:CS-abc}), we need to obtain the specific form of ${\cal{A}}$ and derive the coreps $\boldsymbol{D}\Gamma$ of  $\text{\textbf{M}}^{\boldsymbol{k}_1}$ from those ($\Gamma$) of $\text{\textbf{G}}^{\boldsymbol{k}_1}$ (which is available in Ref. \cite{Bradley2009Mathematical-Oxford}) using the method discussed in Sec. \ref{MSG-corep}. A crucial point here is to determine which of the three cases, namely, Case (a)-(c) in Sec. \ref{MSG-corep},  is appropriate for a given  rep $\Gamma_{i}^{\boldsymbol{k}_1}$ of a SG $\text{\textbf{G}}$.  Generally, this can be inferred from the reality of the induced SG reps ($\Gamma_{i}^{\boldsymbol{k}_1}\uparrow \text{\textbf{G}}$) (see Sec. 4.6, 5.2 and 7.6 in Ref. \cite{Bradley2009Mathematical-Oxford}). Fortunately, this has already been done by  Ref. \cite{Bradley2009Mathematical-Oxford}, where Table 5.7 (Table 6.13)  explicitly indicate the case type for all the single-valued (double-valued) reps $\Gamma_{i}^{\boldsymbol{k}_1}$ of each $\text{\textbf{G}}$ when adding  ${\cal{T}}$ symmetry to $\text{\textbf{G}}$.

Finally, by tedious  but  straightforward calculations, we  establish  complete tables of single-valued and double-valued coreps  of  type II MSGs, along with the matrix representations of the generating elements, and the effective Hamiltonian of the symmetry-protected band degeneracies.

\subsection{Examples of deriving coreps}

Here, we use two examples to show the details of deriving coreps of type II MSG from the reps of the corresponding SG, and discuss the  underneath  physics of  extra degeneracy caused by  the addition of ${\cal{T}}$ symmetry.
For single-valued or double-valued rep $\Gamma_{i}^{\bf{k}_1}$, the relation between the reality of its induced SG reps $\Gamma_{i}^{\bf{k}_1}\uparrow\text{\textbf{G}}$ and the three cases of extra degeneracies described in Sec. \ref{MSG-corep} is as follow \cite{Bradley2009Mathematical-Oxford}
{
\global\long\def\arraystretch{1.4}%
\begin{table}[H]
\begin{ruledtabular}
\begin{tabular}{lll|ll}
\multicolumn{2}{l}{single-valued reps} &  & double-valued reps & \tabularnewline
\hline
Reality  & Degeneracy case &  & Reality  & Degeneracy case\tabularnewline
1 & a &  & 1 & b\tabularnewline
2 & b &  & 2 & a\tabularnewline
3 & c &  & 3 & c\tabularnewline
\end{tabular}
\end{ruledtabular}
\end{table}}

\begin{figure*}
\includegraphics[width=5 cm]{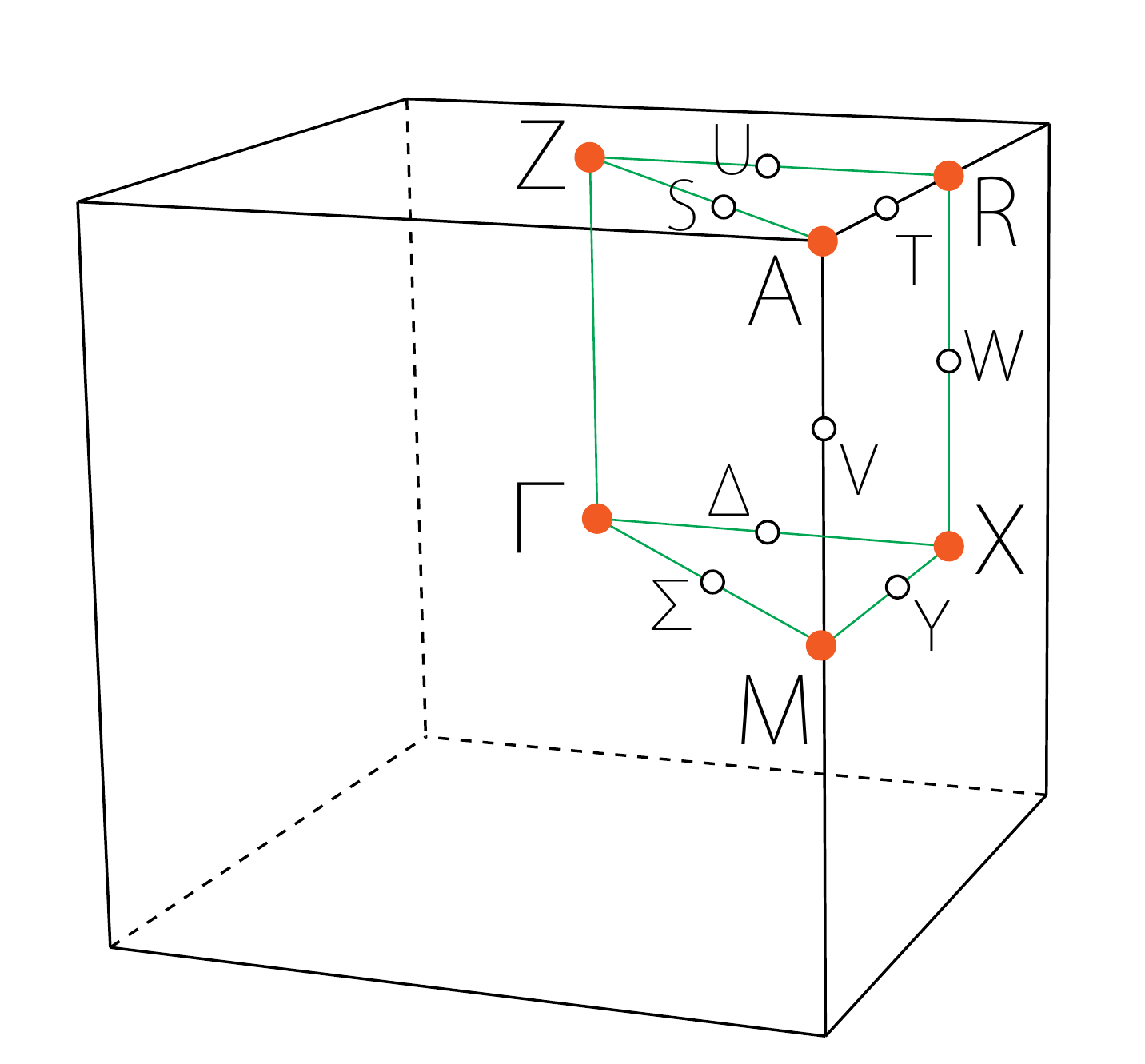}

\caption{BZ of SG 76 and SG 92. We adopt the notation of Ref. \cite{Bradley2009Mathematical-Oxford}  to  label the  high-symmetry points and high-symmetry lines.  \label{fig:exam}}

\end{figure*}

\subsubsection{Single-valued coreps of type II MSG 76}
The single-valued reps of SG 76 can be found in  Ref. \cite{Bradley2009Mathematical-Oxford}, which is partially reproduced in Table \ref{tab:exam1}.
Table \ref{tab:exam1} also  lists the corresponding part of  the established single-valued coreps of type II MSG 76.
We present the defining relations and the character tables of two related  abstract groups $G_1^1$ and $G_4^1$ in Table \ref{tab:exam1-AG}. The BZ of SG 76 is shown in Fig. \ref{fig:exam}, along with the labels of high-symmetry points and high-symmetry lines.
The single-valued reps describes  the  spinless systems or the systems with negligible SOC effect, where a $2\pi$ rotation equals to identity operator and ${\cal T}^{2}=1$. We discuss the high-symmetry points $\Gamma$ and $R$, and  high-symmetry line $U$ in turn.

{
\global\long\def\arraystretch{1.4}%
\begin{table}[H]
\caption{Part of single-valued reps of SG 76 (reproduced from Ref. \cite{Bradley2009Mathematical-Oxford}) and the corresponding  single-valued coreps of type II MSG 76 (see Sec. \ref{S3}).
For the left part, the columns from left to right list the high-symmetry momentum $\mathbf{k}_1$, the little group at $\mathbf{k}_1$: $\text{\textbf{G}}^{\mathbf{k}_1}$ (presented as an abstract group  $\text{\textbf{G}}^{n}_{|\textbf{G}|}$), the generating elements of  $\text{\textbf{G}}^{\mathbf{k}_1}$ and  a series of code $(R_i,j)$ separated by semicolons. $R_i$ is the rep of $\text{\textbf{G}}^{n}_{|\textbf{G}|}$ corresponding to an allowed rep $\Gamma_{p}^{\mathbf{k}_1}$ of $\text{\textbf{G}}^{\mathbf{k}_1}$, and the integer $j$ denotes  the reality  of the induced SG reps  $\Gamma_{p}^{\mathbf{k}_1}\uparrow\text{\textbf{G}}$.
For the right part, the columns from left to right list $\mathbf{k}_1$, the position of $\mathbf{k}_1$, the generating elements of the little
group  at $\mathbf{k}_1$ (only point-group operator of the elements are presented), the  corep derived from the corresponding  $\Gamma_{p}^{\mathbf{k}_1}$ ($R_i$), the dimension of the corep, the matrix representations of the generating elements,  the species and the topological charge of the degeneracies.
\label{tab:exam1}}
\begin{ruledtabular}
\begin{tabular}{lllll|llllllll}
\multicolumn{4}{l}{SG 76} & & \multicolumn{4}{l}{type II MSG 76} \tabularnewline
\hline
$\Gamma$ & $G_{4}^{1}$: & $\{C_{4z}^{+}|00\frac{1}{4}\}$: & $R_{1}$, 1; $R_{2}$, 3; $R_{3}$, 1; $R_{4}$, 3. &  & $\Gamma;$ & $(000)$ & $C_{4z}^{+}$, ${\cal T}$; & $R_{1}$; & 1; & 1, 1; &  & \tabularnewline
 &  &  &  &  &  &  &  & \{$R_{2}$, $R_{4}$\}; & 2; & $i\sigma_{3}$, $\sigma_{1}$; & C-2 WP; & 2\tabularnewline
 &  &  &  &  &  &  &  & $R_{3}$; & 1; & $-1$, 1; &  & \tabularnewline
$R$ & $G_{4}^{1}$: & $\{C_{2z}^{+}|00\frac{1}{2}\}$: & $R_{2}$, 3; $R_{4}$, 3. &  & $R;$ & $(0\frac{1}{2}\frac{1}{2})$ & $C_{2z}$, ${\cal T}$; & \{$R_{2}$, $R_{4}$\}; & 2; & $i\sigma_{3}$, $\sigma_{1}$; & P-NS$_{ZAR}$; & \tabularnewline
$U$ & $G_{1}^{1}$: & $(E,0)$: & $R_{1}$, 2. &  & $U$; & ZR & $E$, $C_{2z}{\cal T}$; & \{$R_{1}$, $R_{1}$\}; & 2; & $\sigma_{0}$, $-i\sigma_{2}$; & L-NS$_{ZAR}$; & \tabularnewline
\end{tabular}
\end{ruledtabular}
\end{table}
}

{
\global\long\def\arraystretch{1.4}%
\begin{table}[H]
\caption{The defining relation and character tables of abstract groups $G_{1}^{1}$ and $G_{4}^{1}$ (reproduced from Ref. \cite{Bradley2009Mathematical-Oxford}). \label{tab:exam1-AG} }
\begin{ruledtabular}
\begin{tabular}{lll|lllll}
\multicolumn{2}{l}{$G_{1}^{1}$} &  &  $G_{4}^{1}$  \tabularnewline

\multicolumn{2}{l}{$C_{1}=E$} &  &  $P^{4}=E$ &  &  &  & \tabularnewline
\hline
& &  & \multicolumn{5}{l}{$C_{1}=E$; $C_{2}=P$; $C_{3}=P^{2}$; $C_{4}=P^{3}$.}\tabularnewline
 & $C_{1}$ &  &  & $C_{1}$ & $C_{2}$ & $C_{3}$ & $C_{4}$\tabularnewline
$R_{1}$ & 1 &  & $R_{1}$ & 1 & 1 & 1 & 1\tabularnewline
 &  &  & $R_{2}$ & 1 & $i$ & -1 & $-i$\tabularnewline
 &  &  & $R_{3}$ & 1 & -1 & 1 & -1\tabularnewline
 &  &  & $R_{4}$ & 1 & $-i$ & -1 & $i$\tabularnewline
\end{tabular}
\end{ruledtabular}
\end{table}
}

$\Gamma$: $\boldsymbol{k}=(000)$.

As shown in the left part of Table \ref{tab:exam1}, $\text{\textbf{G}}^{\Gamma}$, the (Herring) little group at $\Gamma$ point in SG 76, is  abstract group $G_{4}^{1}$, which is isomorphic to point group $C_{4}$, and its generating element is $P=\{C_{4z}^{+}|00\frac{1}{4}\}$. After generating element is  a series of  code $(R_i,j)$ separated by semicolons. $R_i$ is the rep of abstract group $G_{4}^{1}$ corresponding to an allowed rep $\Gamma_{p}^{\Gamma}$ of $\text{\textbf{G}}^{\Gamma}$, and the integer $j$ denotes  the reality  of the induced SG reps  $\Gamma_{p}^{\Gamma}\uparrow\text{\textbf{G}}$.
Since $\Gamma$  is invariant under ${\cal T}$ symmetry,  the little group $\text{\textbf{M}}^{\Gamma}$ of type II MSG 76 is
\begin{eqnarray}
\text{\textbf{M}}^{\Gamma} & = & \text{\textbf{G}}^{\Gamma}+{\cal T}\text{\textbf{G}}^{\Gamma},
\end{eqnarray}
which indicates the generating elements of $\text{\textbf{M}}^{\Gamma}$ can be chosen as $P$ ($\{C_{4z}^{+}|00\frac{1}{4}\}$) and $\cal{T}$.

All the four 1D reps of $G_{4}^{1}$ are the allowed reps of $\text{\textbf{G}}^{\Gamma}$. For $R_{1}$ rep, it belongs to Case (a), as the reality of its induced SG rep is $1$. According to the discussions in Sec. \ref{MSG-corep}, the solution of $\boldsymbol{N}$ can be calculated as
\begin{eqnarray}
\boldsymbol{N} & = & 1.
\end{eqnarray}
Then the matrix representations of the generating elements in $\text{\textbf{M}}^{\Gamma}$ for the  corep derived from $R_1$ are
\begin{eqnarray}
\boldsymbol{D}_{1}(P)=\boldsymbol{\Delta}_{1}(P)=1, & \ \ \ \  & \boldsymbol{D}_{1}({\cal T})=\boldsymbol{N}=1,
\end{eqnarray}
which are listed in the right part  of Table \ref{tab:exam1}.

For $R_{2}$ rep, it belongs to Case (c), as the reality of its induced SG rep is $3$. Hence, the ${\cal T}$ symmetry makes $R_{2}$ and one another rep degenerate in energy. A direct calculation gives the matrix representations of  the  corep derived from $R_2$,
\begin{eqnarray}
\boldsymbol{D}_{2}(P)=\left[\begin{array}{cc}
\boldsymbol{\Delta}_{2}(P) & 0\\
0 & \boldsymbol{\Delta}_{2}^{*}({\cal T}^{-1}P{\cal T})
\end{array}\right]=\left[\begin{array}{cc}
i & 0\\
0 & -i
\end{array}\right]=i\sigma_3, & \ \ \ \ \ \  & \boldsymbol{D}_{2}({\cal T})=\left[\begin{array}{cc}
0 & \boldsymbol{\Delta}({\cal T}^{2})\\
1 & 0
\end{array}\right]=\left[\begin{array}{cc}
0 & 1\\
1 & 0
\end{array}\right]=\sigma_1.
\end{eqnarray}
One can find that the matrix trace of $\boldsymbol{\Delta}_{2}^{*}({\cal T}^{-1}R{\cal T})$ is identical to that of  $\Delta_{4}(R)$ with $R$ the element of $\text{\textbf{G}}^{\Gamma}$,  indicating that the rep degenerate with $R_{2}$ is $R_{4}$. Actually, in the present simple case, we can easily figure it  out, as only $R_{2}$ and $R_{4}$ belong to Case (c).

We then discuss the physics underneath the degeneracy between $R_{2}$ and $R_{4}$. Since $\Gamma$ point has $P=\{C_{4z}^{+}|00\frac{1}{4}\}$
symmetry, the basis state of electronic bands at $\Gamma$ point can be chosen as the eigenstates of $P$, denoted as $|c_{4z}\rangle$ with $c_{4z}=\pm i,\ \pm1$ as $P^{4}=1$. One observes that
\begin{eqnarray}
P{\cal T}|\pm i\rangle & = & {\cal T}P|\pm i\rangle=\mp i\left({\cal T}|\pm i\rangle\right),
\end{eqnarray}
which means that ${\cal T}|\pm i\rangle=|\mp i\rangle$ and hence the two states $|i\rangle$ ($R_2$) and $|-i\rangle$ ($R_4$) would be degenerate due to the addition of ${\cal{T}}$ symmetry. The low-energy $k\cdot p$ Hamiltonian  expanded  around this double degeneracy with $\{R_2, R_4\}$ rep is
\begin{eqnarray}
H & = & c_{1}+c_{2}k^{2}+c_{3}k_{z}^{2}+\sum_{i=1}^{2}\left[c_{i,1}k_{x}k_{y}+c_{i,2}(k_{x}^{2}-k_{y}^{2})\right]\sigma_{i}+c_{4}\sigma_{3}k_{z},
\end{eqnarray}
from which we know the double degeneracy is a  C-2 WP with a topological charge $|{\cal{C}}|=2$, as listed in  Table \ref{tab:exam1}.


For $R_{3}$ rep, it belongs to Case (a), as the reality of its induced SG rep is $1$. The solution of $\boldsymbol{N}$ can be calculated as
\begin{eqnarray}
\boldsymbol{N} & = & 1,
\end{eqnarray}
and  the matrix representations of   the  corep derived from $R_3$ are obtained as
\begin{eqnarray}
\boldsymbol{D}_{3}(P)=\boldsymbol{\Delta}_{3}(P)=1, & \ \ \ \   & \boldsymbol{D}_{3}({\cal T})=\boldsymbol{N}=1.
\end{eqnarray}

$R$: $\boldsymbol{k}=(0\frac{1}{2}\frac{1}{2})$.

$R$  is invariant under ${\cal T}$ symmetry and the little group $\text{\textbf{M}}^{R}$ is
\begin{eqnarray}
\text{\textbf{M}}^{R} & = & \text{\textbf{G}}^{R}+{\cal T}\text{\textbf{G}}^{R}.
\end{eqnarray}
$\text{\textbf{G}}^{R}$ also is abstract group $G_{4}^{1}$, and its generating element is $P=\{C_{2z}|00\frac{1}{2}\}$. Since $R$ locates at the BZ boundary,  one has  $P^{2}=-1$ at $R$ point, due to the presence of fraction translation in $P$. Consequently, only two of the four 1D reps of $G_{4}^{1}$, $R_{2}$ and $R_{4}$, are the allowed reps of $\text{\textbf{G}}^{R}$.
Both $R_{2}$ and $R_{4}$ belong to Case (c), indicating they would degenerate in energy with the addition of ${\cal T}$ symmetry.
We use $R_{2}$ to calculate the coreps, obtained as
\begin{eqnarray}
\boldsymbol{D}_{2}(P)=\left[\begin{array}{cc}
\boldsymbol{\Delta}_{2}(P) & 0\\
0 & \boldsymbol{\Delta}_{2}^{*}({\cal T}^{-1}P{\cal T})
\end{array}\right]=\left[\begin{array}{cc}
i & 0\\
0 & -i
\end{array}\right]=i\sigma_3, & \ \ \  & \boldsymbol{D}_{2}({\cal T})=\left[\begin{array}{cc}
0 & \boldsymbol{\Delta}({\cal T}^{2})\\
1 & 0
\end{array}\right]=\left[\begin{array}{cc}
0 & 1\\
1 & 0
\end{array}\right]=\sigma_1.
\end{eqnarray}

The physics underneath the degeneracy is that the electronic bands with reps $R_{2}$ and $R_{4}$ respectively correspond to the electronic states of $|p=i\rangle$ and $|p=-i\rangle$, where $|p\rangle$ is the eigenstate of $P$, and one has
\begin{eqnarray}
P{\cal T}|\pm i\rangle & = & {\cal T}P|\pm i\rangle=\mp i\left({\cal T}|\pm i\rangle\right).
\end{eqnarray}
This means that ${\cal T}|\pm i\rangle=|\mp i\rangle$ and the two states $|i\rangle$ ($R_2$) and $|-i\rangle$ ($R_4$) would be degenerate. The low-energy $k\cdot p$ Hamiltonian (up to first order) expanded  around this double degeneracy is
\begin{eqnarray}
H & = & c_{1}+c_{2}\sigma_{3}k_{z}.
\end{eqnarray}
A careful analysis shows this degeneracy is not  isolate but a point  in a nodal surface locating at $ZAR$ plane (P-NS$_{ZAR}$), which will be  further  discussed below in detail.
The corep information and the analysis of the degeneracy at $R$ point are listed in the right part  of Table \ref{tab:exam1}.

$U$: $\boldsymbol{k}=(0\alpha\frac{1}{2})$ with $\alpha\in(0,\frac{1}{2})$, $ZR$ path.

$U$ is invariant under ${\cal A}=\{C_{2z}|00\frac{1}{2}\}{\cal T}$ symmetry and the little group $\text{\textbf{M}}^{U}$  is
\begin{eqnarray}
\text{\textbf{M}}^{U} & = & \text{\textbf{G}}^{U}+{\cal A}\text{\textbf{G}}^{U}.
\end{eqnarray}
$\text{\textbf{G}}^{U}$ is  $G_{1}^{1}$, and its generating element is $P=(E,0)$. The $R_{1}$ rep of $G_{1}^{1}$ is the allowed reps of $\text{\textbf{G}}^{U}$.  It belongs to Case (b), as the reality of its induced SG reps is $2$, which indicates that there exists a nodal line along $U$ (ZR path) formed by the two electronic bands with same rep ($R_1$). For $R_{1}$ rep, the matrix representation of  $\{E|000\}$ is
\begin{eqnarray}
\Delta_{1}(\{E|000\}) & = & \Delta_{1}(P)e^{-i2\pi\boldsymbol{k}\cdot(000)}e^{-i2\pi\times(0/g)}=\Delta_{1}(P),
\end{eqnarray}
where $g=1$ is determined by the factor system for the projected representation of $\text{\textbf{G}}^{U}$. Since ${\cal A}^{2}=-1$ on $U$, we have
\begin{eqnarray}
\boldsymbol{N} & = & 1,
\end{eqnarray}
and the matrix representations of the corep derived from $R_1$ are
\begin{eqnarray}
\boldsymbol{D}_{1}(P)=\left[\begin{array}{cc}
\boldsymbol{\Delta}_{1}(P) & 0\\
0 & \boldsymbol{\Delta}_{1}(P)
\end{array}\right]=\left[\begin{array}{cc}
1 & 0\\
0 & 1
\end{array}\right]=\sigma_0, & \ \ \  & \boldsymbol{D}_{1}({\cal A})=\left[\begin{array}{cc}
0 & -\boldsymbol{N}\\
\boldsymbol{N} & 0
\end{array}\right]=\left[\begin{array}{cc}
0 & -1\\
1 & 0
\end{array}\right]=-i\sigma_2.
\end{eqnarray}

We then discuss the physics of the degeneracy along high-symmetry line $U$. While ${\cal T}^{2}=1$ for spinless systems, the anti-unitary operator ${\cal A}^{2}=-1$ on $U$, due to the half-integer translation following $C_{2z}$. This inevitably leads to Kramers-like degeneracies. Specifically, we can denote the eigenstate $|e\rangle$ by the eigenvalue of $\{E|000\}$, and then the two states $|e=1\rangle$ ($R_1$) and  ${\cal{A}}|e=1\rangle$ ($R_1$)  are linearly independent  and would  be degenerate in energy, due to  ${\cal A}^{2}=-1$.  In fact, all these points in $ZAR$ plane have ${\cal A}$ symmetry and  ${\cal A}^{2}=-1$, which eventually leads to  a nodal surface at $ZAR$ plane \cite{Wu2018Nodal-PRB}. Thus, the nodal line along $U$ is not isolated but resides in a nodal surface locating at $ZAR$ plane (labelled as L-NS$_{ZAR}$ in Table \ref{tab:exam1}).
 The low-energy $k\cdot p$ Hamiltonian (up to first order) expanded  around a generic point on $U$ is
\begin{eqnarray}
H & = & c_{1}+c_{2}k_{x}+c_{3}k_{y}+\sum_{i=1}^{3}c_{i,1}\sigma_{i}k_{z}.
\end{eqnarray}

\subsubsection{Double-valued coreps of type II MSG 92}
The double-valued reps of SG 92 can be found in  Ref. \cite{Bradley2009Mathematical-Oxford}, which is partially reproduced in Table \ref{tab:exam2}.
Table \ref{tab:exam2} also  lists the corresponding part of  the established double-valued coreps of type II MSG 92.
We  present some  relevant reps information  of the related abstract groups $G_8^2$, $G_{14}^{16}$ and $G_{32}^{11}$ in Table \ref{tab:exam2-AG}. The BZ of SG 92 is same with that of SG 76, and  is shown in Fig. \ref{fig:exam}.
The double-valued reps describe the systems with strong SOC effect, where a $2\pi$ rotation leads to a minus sign and ${\cal T}^{2}=-1$. We discuss the high-symmetry  points $\Gamma$ and $M$, and high-symmetry lines $Y$ and $T$  in turn.

{
\global\long\def\arraystretch{1.4}%
\begin{table}[H]
\caption{Part of double-valued reps of SG 92 (reproduced from Ref. \cite{Bradley2009Mathematical-Oxford}) and the corresponding double-valued coreps of type II MSG 92 (see Sec. \ref{S3}). \label{tab:exam2}}
\begin{ruledtabular}
\begin{tabular}{lllll|llllllll}
\multicolumn{4}{l}{SG 92} & & \multicolumn{4}{l}{type II MSG 92} \tabularnewline
\hline
$\Gamma$ & $G_{16}^{14}$: & $\{C_{4z}^{+}|00\frac{1}{4}\}$, $\{C_{2x}|\frac{1}{2}\frac{1}{2}0\}$: & $R_{6}$, 2; $R_{7}$, 2. &  & $\Gamma;$ & $(000)$ & $C_{4z}^{+}$, $C_{2x}$, ${\cal T}$; & $R_{6}$; & 2; & $\frac{\sigma_0+i\sigma_2}{\sqrt{2}}$, $i\sigma_1$, $-i\sigma_2$; & C-1 WP; & 1\tabularnewline
 &  &  &  &  &  &  &  & $R_{7}$; & 2; & $-\frac{\sigma_0-i\sigma_2}{\sqrt{2}}$, $i\sigma_1$, $-i\sigma_2$; & C-1 WP; & 1\tabularnewline
$M$ & $G_{32}^{11}$: & $\{C_{4z}^{+}|00\frac{1}{4}\}$, $\{C_{2x}|\frac{1}{2}\frac{1}{2}0\}$: & $R_{6}$, 3; $R_{7}$, 3. &  & $M;$ & $(\frac{1}{2}\frac{1}{2}0)$ & $C_{4z}$, $C_{2x}$, ${\cal T}$; & \{$R_{6}$, $R_{7}$\}; & 4; & $\frac{\Gamma_{0,1}+i\Gamma_{3,0}}{\sqrt{2}}$, $\Gamma_{0,3}$, $-i\Gamma_{2,0}$; & C-2 DP; & 2 \tabularnewline
$Y$ & $G_{8}^{2}$: & $(C_{2x},0)$, $(E,1)$: & $R_{5}$, 3; $R_{7}$, 3. &  & $Y$; & XM & $C_{2x}$, $E$, $C_{2y}{\cal T}$; & \{$R_{5}$, $R_{7}$\}; & 2; & $-i\sigma_3$, $\sigma_{0}$, $-i\sigma_{2}$; & L-NS$_{ZAR}$; & \tabularnewline
$T$ & $G_{8}^{2}$: & $(C_{2x},0)$, $(E,1)$: & $R_{5}$, 1; $R_{7}$, 1. &  & $T$; & RA & $C_{2x}$, $E$, $C_{2y}{\cal T}$; & \{$R_{5}$, $R_{5}$\}; & 2; & $-i \sigma_{0}$, $\sigma_{0}$, $-i\sigma_{2}$; & L-NSs; & \tabularnewline
 &  &  & &  & & &  & \{$R_{7}$, $R_{7}$\}; & 2; & $i\sigma_{0}$, $\sigma_{0}$, $-i\sigma_{2}$; & L-NSs; & \tabularnewline
\end{tabular}
\end{ruledtabular}
\end{table}
}

{
\global\long\def\arraystretch{1.4}%
\begin{table}[H]
\caption{The defining relation and the matrix representation of the generating elements of abstract groups $G_{8}^{2}$, $G_{16}^{14}$  and $G_{32}^{11}$ (reproduced from Ref. \cite{Bradley2009Mathematical-Oxford}). \label{tab:exam2-AG} }
\begin{ruledtabular}
\begin{tabular}{l|lll|l|lll|l|lll}
\multicolumn{3}{l}{$G_{8}^{2}$} &  & \multicolumn{3}{l}{$G_{16}^{14}$} &  & \multicolumn{3}{l}{$G_{32}^{11}$} & \tabularnewline
\multicolumn{3}{l}{$P^{4}=E$; $Q^{2}=E;$} &  & \multicolumn{3}{l}{$P^{8}=E$; $Q^{4}=E;$} &  & \multicolumn{3}{l}{$P^{8}=E$; $Q^{4}=E;$} & \tabularnewline
\multicolumn{3}{l}{$QP=PQ$.} &  & \multicolumn{3}{l}{$QP=P^{7}Q$.} &  & \multicolumn{3}{l}{$QP=P^{3}Q^{3}$; $Q^{2}P=PQ^{2}$.} & \tabularnewline
\hline
 & $P$ & $Q$ &  &  & $P$ & $Q$ &  &  & $P$ & $Q$ & \tabularnewline
\hline
$R_{5}$ & 1 & -1 &  & $R_{6}$ & $\frac{1}{\sqrt{2}}\left[\begin{array}{cc}
1 & 1\\
-1 & 1
\end{array}\right]$ & $i\left[\begin{array}{cc}
0 & 1\\
1 & 0
\end{array}\right]$ &  & $R_{6}$ & $\frac{1}{\sqrt{2}}\left[\begin{array}{cc}
i & 1\\
1 & i
\end{array}\right]$ & $\left[\begin{array}{cc}
1 & 0\\
0 & -1
\end{array}\right]$ & \tabularnewline
$R_{7}$ & -1 & -1 &  & $R_{7}$ & $\frac{1}{\sqrt{2}}\left[\begin{array}{cc}
-1 & 1\\
-1 & -1
\end{array}\right]$ & $i\left[\begin{array}{cc}
0 & 1\\
1 & 0
\end{array}\right]$ &  & $R_{7}$ & $\frac{1}{\sqrt{2}}\left[\begin{array}{cc}
-i & 1\\
1 & -i
\end{array}\right]$ & $\left[\begin{array}{cc}
1 & 0\\
0 & -1
\end{array}\right]$ & \tabularnewline
\end{tabular}
\end{ruledtabular}
\end{table}
}
$\Gamma$: $\boldsymbol{k}=(000)$.

$\Gamma$ is invariant under ${\cal T}$ symmetry and the little group $\text{\textbf{M}}^{\Gamma}$ of type II double-valued MSG 92 is
\begin{eqnarray}
\text{\textbf{M}}^{\Gamma} & = & \text{\textbf{G}}^{\Gamma}+{\cal T}\text{\textbf{G}}^{\Gamma}.
\end{eqnarray}
 $\text{\textbf{G}}^{\Gamma}$ is abstract group $G_{16}^{14}$, which is isomorphic to point group $D_{4}$, and its generating elements
are $P=\{C_{4z}^{+}|00\frac{1}{4}\}$ and $Q=\{C_{2x}|\frac{1}{2}\frac{1}{2}0\}$. Only two 2D reps of $G_{16}^{14}$, $R_{6}$ and $R_{7}$, are the
allowed reps of $\text{\textbf{G}}^{\Gamma}$, as $P^4=Q^2=-1$ at $\Gamma$ point. For $R_{6}$ rep, the reality of its induced SG rep is $2$ and then it belongs to Case (a).

The solution of $\boldsymbol{N}$  is
\begin{eqnarray}
\boldsymbol{N} & = & \ensuremath{-i\sigma_{2}},
\end{eqnarray}
as ${\cal{T}}^2=-1$ and $R={\cal{T}}^{-1}R{\cal{T}}$ for any element $R$ in $\text{\textbf{G}}^{\Gamma}$.
Therefore, the matrix representations of the corep derived from $R_6$ are
\begin{eqnarray}
\boldsymbol{D}_{6}(P) & = & \boldsymbol{\Delta}_{6}(P)=\frac{1}{\sqrt{2}}\left[\begin{array}{cc}
1 & 1\\
-1 & 1
\end{array}\right]=\frac{\sigma_{0}+i\sigma_{2}}{\sqrt{2}},\\
\boldsymbol{D}_{6}(Q) & = & \boldsymbol{\Delta}_{6}(Q)=i\left[\begin{array}{cc}
0 & 1\\
1 & 0
\end{array}\right]=i\sigma_{1},\\
\boldsymbol{D}_{6}({\cal T}) & = & \boldsymbol{N}=-i\sigma_{2}.
\end{eqnarray}
The low-energy Hamiltonian expanded around this double degeneracy is obtained  as
\begin{eqnarray}
H & = & c_{1}+c_{2}k_{z}\sigma_{2}+c_{3}(k_{x}\sigma_{1}+k_{y}\sigma_{3}),
\end{eqnarray}
from which we know the degeneracy is a C-1 WP with a topological charge $|{\cal{C}}|=1$, as listed in the right part of Table \ref{tab:exam2}.

The $R_{7}$ rep also belongs to Case (a). Similarly, the  solution for $\boldsymbol{N}$ is
\begin{eqnarray}
\boldsymbol{N} & = & \ensuremath{-i\sigma_{2}},
\end{eqnarray}
 and  we have
\begin{eqnarray}
\boldsymbol{D}_{7}(P) & = & \boldsymbol{\Delta}_{7}(P)=\frac{1}{\sqrt{2}}\left[\begin{array}{cc}
-1 & 1\\
-1 & -1
\end{array}\right]=\frac{-\sigma_{0}+i\sigma_{2}}{\sqrt{2}},\\
\boldsymbol{D}_{7}(Q) & = & \boldsymbol{\Delta}_{7}(Q)=i\left[\begin{array}{cc}
0 & 1\\
1 & 0
\end{array}\right]=i\sigma_{1},\\
D_{7}({\cal T}) & = & \boldsymbol{N}=-i\sigma_{2}.
\end{eqnarray}
The low-energy Hamiltonian expanded around this double degeneracy is
\begin{eqnarray}
H & = & c_{1}+c_{2}k_{z}\sigma_{2}+c_{3}(k_{x}\sigma_{1}-k_{y}\sigma_{3}),
\end{eqnarray}
from which we know this degeneracy also is a C-1 WP with a topological charge $|{\cal{C}}|=1$, as listed in the right part of Table \ref{tab:exam2}.

$M$: $\boldsymbol{k}=(\frac{1}{2}\frac{1}{2}0)$.

$M$ is invariant under ${\cal T}$ symmetry and the little group $\text{\textbf{M}}^{M}$  is
\begin{eqnarray}
\text{\textbf{M}}^{M} & = & \text{\textbf{G}}^{M}+{\cal T}\text{\textbf{G}}^{M}.
\end{eqnarray}
$\text{\textbf{G}}^{M}$ is abstract group $G_{32}^{11}$, and its generating elements are identical with those of $\Gamma$ point, namely,
$P=\{C_{4z}^{+}|00\frac{1}{4}\}$ and $Q=\{C_{2x}|\frac{1}{2}\frac{1}{2}0\}$.
However, due to the presence of fractional translation  in the generating elements, the commutation relation between $P$ and $Q$ at $\Gamma$ and $M$ points are completely different, as shown in Table \ref{tab:exam2-AG}. Only two 2D reps of $G_{32}^{11}$, $R_{6}$ and $R_{7}$, are the allowed reps of $\text{\textbf{G}}^{M}$, due to  the physical constraints of $P^{4}=-1$ and $Q^{2}=1$ at $M$ point. The two reps would degenerate in energy with the addition of ${\cal T}$ symmetry, as they belong to Case (c). We use $R_{6}$ to calculate the coreps, obtained as
\begin{eqnarray}
\boldsymbol{D}_{6}(P) & = & \left[\begin{array}{cc}
\boldsymbol{\Delta}_{6}(P) & \boldsymbol{0}\\
\boldsymbol{0} & \boldsymbol{\Delta}_{6}^{*}({\cal T}^{-1}P{\cal T})
\end{array}\right]=\left[\begin{array}{cc}
\boldsymbol{\Delta}_{6}(P) & \boldsymbol{0}\\
\boldsymbol{0} & \boldsymbol{\Delta}_{6}^{*}(P)
\end{array}\right]=\frac{\Gamma_{0,1}+i\Gamma_{3,0}}{\sqrt{2}},\\
\boldsymbol{D}_{6}(Q) & = & \left[\begin{array}{cc}
\boldsymbol{\Delta}_{6}(Q) & \boldsymbol{0}\\
\boldsymbol{0} & \boldsymbol{\Delta}_{6}^{*}({\cal T}^{-1}Q{\cal T})
\end{array}\right]=\left[\begin{array}{cc}
\boldsymbol{\Delta}_{6}(Q) & \boldsymbol{0}\\
\boldsymbol{0} & \boldsymbol{\Delta}_{6}^{*}(Q)
\end{array}\right]=\Gamma_{0,3},\\
\boldsymbol{D}_{6}({\cal T}) & = & \left[\begin{array}{cc}
\boldsymbol{0} & \boldsymbol{\Delta}({\cal T}^2)\\
\boldsymbol{1} & \boldsymbol{0}
\end{array}\right]=\left[\begin{array}{cc}
\boldsymbol{0} & \boldsymbol{-1}\\
\boldsymbol{1} & \boldsymbol{0}
\end{array}\right]=-i\Gamma_{2,0}.
\end{eqnarray}

We then discuss the physics of the four-fold degeneracy with $\{R_{6},R_{7}\}$ rep. According to the defining relation of $G_{32}^{11}$ and the constraints $P^{4}=-1$ and $Q^{2}=1$, we have the following algebra at $M$ point
\begin{equation}
P^{3}Q^{3}=P^{3}Q=QP.
\end{equation}
The Bloch states at $M$ can be chosen as the eigenstates of $P$, which we denote as $|p\rangle$ with $p=e^{\pm i\pi/4},\ e^{\pm i3\pi/4}$ due to $P^{4}=-1$.
Then, we have
\begin{eqnarray}
P^{3}Q|p\rangle & = & QP|p\rangle=pQ|p\rangle,
\end{eqnarray}
which indicates that $Q|e^{\pm i3\pi/4}\rangle\sim|e^{\pm i\pi/4}\rangle$ and $Q|e^{\pm i\pi/4}\rangle\sim|e^{\pm i3\pi/4}\rangle$. Hence, the two states $|e^{i\pi/4}\rangle$ and $|e^{i3\pi/4}\rangle$ would be degenerate, corresponding to rep $R_{6}$, as $e^{i\pi/4}+e^{3i\pi/4}=\text{Tr}[\boldsymbol{\Delta}_{6}(P)]$. Similarly, the two states $|e^{-i\pi/4}\rangle$ and $|e^{-i3\pi/4}\rangle$ would be degenerate, corresponding to rep $R_{7}$, as $e^{-i\pi/4}+e^{-3i\pi/4}=\text{Tr}[\boldsymbol{\Delta}_{7}(P)]$.
Moreover, the state $|e^{i\pi/4}\rangle$ and its time-reversal partner ${\cal T}|e^{i\pi/4}\rangle=|e^{-i\pi/4}\rangle$, as well as the state $|e^{i3\pi/4}\rangle$ and its time-reversal partner ${\cal T}|e^{i3\pi/4}\rangle=|e^{-i3\pi/4}\rangle$, are linearly independent. Therefore, the four states $\{|e^{i\pi/4}\rangle,\ |e^{i3\pi/4}\rangle,\ {\cal T}|e^{i\pi/4}\rangle,\ {\cal T}|e^{i3\pi/4}\rangle\}$ must be degenerate,  leading to a  four-fold degeneracy (formed by reps $R_6$ and $R_7$).

The low-energy Hamiltonian of this four-fold degeneracy is obtained as
\begin{eqnarray}
H & = & c_{1}+c_{2}k_{z}\Gamma_{3,1}+c_{3}(k_{x}\Gamma_{3,3}-k_{y}\Gamma_{0,2})+\left[\alpha(k_{x}\Gamma_{+,0}+ik_{y}\Gamma_{+,1})+h.c.\right],
\end{eqnarray}
from which we know this degeneracy  is a  C-2 DP with a topological charge $|{\cal{C}}|=2$, as listed in the right part of Table \ref{tab:exam2}.
Here, we define  $\Gamma_{i,j}=\sigma_{i}\otimes \sigma_{j}$ with $i,j=0,1,2,3,\pm $ and  $\sigma_{\pm}=(\sigma_1\pm i\sigma_{2})/2$.

$Y$: $\boldsymbol{k}=(\alpha\frac{1}{2}0)$ with $\alpha\in(0,\frac{1}{2})$, $XM$ path.

$Y$ is invariant under ${\cal A}=\{C_{2y}|\frac{1}{2}\frac{1}{2}\frac{1}{2}\}{\cal T}$ symmetry and the little group $\text{\textbf{M}}^{Y}$ is
\begin{eqnarray}
\text{\textbf{M}}^{Y} & = & \text{\textbf{G}}^{Y}+{\cal A}\text{\textbf{G}}^{Y},
\end{eqnarray}
with ${\cal{A}}^2=-1$ at $Y$.
$\text{\textbf{G}}^{Y}$ is $G_{8}^{2}$, and its generating element is $P=(C_{2x},0)$ and $Q=(E,1)$. Only two 1D reps of $G_{8}^{2}$, $R_{5}$ and $R_{7}$, are the allowed reps of $\text{\textbf{G}}^{Y}$, due to the constraint $\{C_{2x}|\frac{1}{2}\frac{1}{2}0\}^2=\{\bar{E}|100\}=-e^{-i2\pi\alpha}$ at $Y$. The two reps would degenerate in energy with the addition of ${\cal A}$ symmetry, as they belong to Case (c). Then one has a nodal line along $Y$ with  $\{R_{5},R_{7}\}$ rep. For $R_{5}$ and $R_{7}$ reps, the matrix representation of $\{C_{2x}|\frac{1}{2}\frac{1}{2}0\}$ and $\{E|000\}$ are \cite{Bradley2009Mathematical-Oxford}
\begin{eqnarray}
\Delta_{5(7)}(\{C_{2x}|\frac{1}{2}\frac{1}{2}0\}) & = & \Delta_{5(7)}(P)e^{-i2\pi\boldsymbol{k}\cdot(\frac{1}{2}\frac{1}{2}0)}e^{-i2\pi\times(0/g)}=-ie^{-i\pi\alpha}\Delta_{5(7)}(P),\\
\Delta_{5(7)}(\{E|000\}) & = & \Delta_{5(7)}(Q)e^{-i2\pi\boldsymbol{k}\cdot(000)}e^{-i2\pi\times(1/g)}=-\Delta_{5(7)}(Q),
\end{eqnarray}
where $g=2$ is determined by the factor system for the projected representation of $\text{\textbf{G}}^{Y}$. One finds that the matrix rep $\Delta_{5(7)}^2(\{C_{2x}|\frac{1}{2}\frac{1}{2}0\})=-e^{-i2\pi\alpha} \Delta_{5(7)}^2(P)=-e^{-i2\pi\alpha}$ satisfies the constraint
$\{C_{2x}|\frac{1}{2}\frac{1}{2}0\}^2=-e^{-i2\pi\alpha}$ at $Y$.
We use rep $R_{5}$ to calculate the coreps, obtained as
\begin{eqnarray}
\boldsymbol{D}_{5}(C_{2x}) & = & \left[\begin{array}{cc}
\boldsymbol{\Delta}_{5}(\{C_{2x}|\frac{1}{2}\frac{1}{2}0\}) & \boldsymbol{0}\\
\boldsymbol{0} & \boldsymbol{\Delta}_{5}^{*}({\cal A}^{-1}\{C_{2x}|\frac{1}{2}\frac{1}{2}0\}{\cal A})
\end{array}\right]=-ie^{-i\pi\alpha}\left[\begin{array}{cc}
1 & 0\\
0 & -1
\end{array}\right]=-i\sigma_{3} e^{-i\pi\alpha},\\
\boldsymbol{D}_{5}(E) & = & \left[\begin{array}{cc}
\boldsymbol{\Delta}_{5}(\{E|000\}) & \boldsymbol{0}\\
\boldsymbol{0} & \boldsymbol{\Delta}_{5}^{*}({\cal A}^{-1}\{E|000\}{\cal A})
\end{array}\right]=-\left[\begin{array}{cc}
-1 & 0\\
0 & -1
\end{array}\right]=\sigma_{0},\\
\boldsymbol{D}_{5}({\cal A}) & = & \left[\begin{array}{cc}
\boldsymbol{0} & \boldsymbol{\Delta}({\cal A}^{2})\\
\boldsymbol{1} & \boldsymbol{0}
\end{array}\right]=\left[\begin{array}{cc}
0 & -1\\
1 & 0
\end{array}\right]=i\sigma_{2}.
\end{eqnarray}
Notice that, for simplification we omit the phase factor containing $\alpha$, e.g. $e^{-i\pi\alpha}$, in Table \ref{tab:exam2} and all the tables in Sec. \ref{S3} and Sec. \ref{S4}.

We then discuss the physics of the  degeneracy between $R_5$ and $R_7$ reps. First, as we have ${\cal{A}}^2=-1$ at $Y$, the electronic bands along this high-symmetry line are  at least doubly degenerate due to the Kramers-like degeneracy.  Then since ${\cal A}^{-1}\{C_{2x}|\frac{1}{2}\frac{1}{2}0\}{\cal A}=\{\bar{C}_{2x}|\bar{\frac{1}{2}}\bar{\frac{1}{2}}1\}$ and
\begin{eqnarray}
\Delta_{5}({\cal A}^{-1}\{C_{2x}|\frac{1}{2}\frac{1}{2}0\}{\cal A}) & = & \Delta_{5}(\{\bar{C}_{2x}|\bar{\frac{1}{2}}\bar{\frac{1}{2}}1\})=
-\Delta_{5}(P)e^{-i2\pi\boldsymbol{k}\cdot(\bar{\frac{1}{2}}\bar{\frac{1}{2}}1)}=-ie^{i\pi\alpha}\Delta_{5}(P),
\end{eqnarray}
we have
\begin{eqnarray}
\Delta_{7}(\{C_{2x}|\frac{1}{2}\frac{1}{2}0\})=\Delta_{5}^{*}({\cal A}^{-1}\{C_{2x}|\frac{1}{2}\frac{1}{2}0\}{\cal A}),
\end{eqnarray}
which  means that the  doubly degenerate bands is formed by the states with $R_5$  and $R_7$ reps.

Moreover, we find all the points in the boundary plane $MARX$ have ${\cal A}$ symmetry and ${\cal A}^{2}=-1$. Hence, the nodal line along  $Y$ is not isolated
but lie in a nodal surface locating at $MARX$ plane (L-NS$_{MARX}$). The low-energy Hamiltonian expanded around a general point on $Y$
is obtained as
\begin{eqnarray}
H & = & c_{1}+c_{2}k_{x}+(c_{3}\sigma_{1}-c_{4}\sigma_{2})k_{y}.
\end{eqnarray}

$T$: $\boldsymbol{k}=(\alpha\frac{1}{2}\frac{1}{2})$ with $\alpha\in(0,\frac{1}{2})$.

$T$ also is invariant under ${\cal A}=\{C_{2y}|\frac{1}{2}\frac{1}{2}\frac{1}{2}\}{\cal T}$ symmetry and the little group $\text{\textbf{M}}^{T}$ is
\begin{eqnarray}
\text{\textbf{M}}^{T} & = & \text{\textbf{G}}^{T}+{\cal A}\text{\textbf{G}}^{T},
\end{eqnarray}
with ${\cal{A}}^2=-1$ at $T$.
$\text{\textbf{G}}^{T}$ is $G_{8}^{2}$, and its generating element is $P=(C_{2x},0)$ and $Q=(E,1)$. Only two 1D reps of $G_{8}^{2}$,
$R_{5}$ and $R_{7}$, are the allowed reps of $\text{\textbf{G}}^{T}$, due to the constraint $\{C_{2x}|\frac{1}{2}\frac{1}{2}0\}^2=\{\bar{E}|100\}=-e^{-i2\pi\alpha}$ at $T$.
Both two reps belong to Case (b), as the reality of their induced SG reps are $1$, indicating there exist two different degenerate pairs: $\{R_{5},R_{5}\}$ and $\{R_{7},R_{7}\}$ at $T$. For $R_{5(7)}$ rep, the matrix representation of $\{C_{2x}|\frac{1}{2}\frac{1}{2}0\}$ and $\{E|000\}$ are
\begin{eqnarray}
\Delta_{5(7)}(\{C_{2x}|\frac{1}{2}\frac{1}{2}0\}) & = & \Delta_{5(7)}(P)e^{-i2\pi\boldsymbol{k}\cdot(\frac{1}{2}\frac{1}{2}0)}e^{-i2\pi\times(0/g)}=-ie^{-i\pi\alpha}\Delta_{5(7)}(P),\\
\Delta_{5(7)}(\{E|000\}) & = & \Delta_{5(7)}(Q)e^{-i2\pi\boldsymbol{k}\cdot(000)}e^{-i2\pi\times(1/g)}=-\Delta_{5(7)}(Q),
\end{eqnarray}
where $g=2$ is determined by the factor system for the projected representation of $\text{\textbf{G}}^{T}$.

The double degeneracies at  $T$ also are Kramers-like degeneracies  caused  by  ${\cal{A}}^2=-1$. However, different from the case at $Y$,  the electronic band at $T$ with rep $R_{5(7)}$ would be degenerate with the band with same rep.  This is because that we have  ${\cal A}^{-1}\{C_{2x}|\frac{1}{2}\frac{1}{2}0\}{\cal A}=\{\bar{C}_{2x}|\bar{\frac{1}{2}}\bar{\frac{1}{2}}1\}$ and
\begin{eqnarray}
\Delta_{5(7)}({\cal A}^{-1}\{C_{2x}|\frac{1}{2}\frac{1}{2}0\}{\cal A}) & = & \Delta_{5(7)}(\{\bar{C}_{2x}|\bar{\frac{1}{2}}\bar{\frac{1}{2}}1\})=-\Delta_{5(7)}(P)e^{-i2\pi\boldsymbol{k}\cdot(\bar{\frac{1}{2}}\bar{\frac{1}{2}}1)}=
ie^{i\pi\alpha}\Delta_{5(7)}(P),
\end{eqnarray}
which leads to
\begin{eqnarray}
 \Delta_{5(7)}(\{C_{2x}|\frac{1}{2}\frac{1}{2}0\}) = \Delta_{5(7)}^{*}({\cal A}^{-1}\{C_{2x}|\frac{1}{2}\frac{1}{2}0\}{\cal A}).
\end{eqnarray}

For $R_5$, the solution of $\boldsymbol{N}$ matrix is
\begin{eqnarray}
\boldsymbol{N} & = & 1.
\end{eqnarray}
and the deduced coreps are
\begin{eqnarray}
\boldsymbol{D}_{5}(C_{2x}) & = & \left[\begin{array}{cc}
\boldsymbol{\Delta}_{5}(\{C_{2x}|\frac{1}{2}\frac{1}{2}0\}) & \boldsymbol{0}\\
\boldsymbol{0} & \boldsymbol{\Delta}_{5}(\{C_{2x}|\frac{1}{2}\frac{1}{2}0\})
\end{array}\right]=-ie^{-i\pi\alpha}\left[\begin{array}{cc}
1 & 0\\
0 & 1
\end{array}\right]=-i\sigma_{0}e^{-i\pi\alpha},\\
\boldsymbol{D}_{5}(E) & = & \left[\begin{array}{cc}
\boldsymbol{\Delta}_{5}(\{E|000\}) & \boldsymbol{0}\\
\boldsymbol{0} & \boldsymbol{\Delta}_{5}(\{E|000\})
\end{array}\right]=-\left[\begin{array}{cc}
-1 & 0\\
0 & -1
\end{array}\right]=\sigma_{0},\\
\boldsymbol{D}_{5}({\cal A}) & = & \left[\begin{array}{cc}
\boldsymbol{0} & \boldsymbol{-N}\\
\boldsymbol{1} & \boldsymbol{0}
\end{array}\right]=\left[\begin{array}{cc}
0 & -1\\
1 & 0
\end{array}\right]=-i\sigma_{2}.
\end{eqnarray}
Again, for simplification we omit the phase factor containing $\alpha$, e.g. $e^{-i\pi\alpha}$, in Table \ref{tab:exam2} and all the tables in Sec. \ref{S3} and Sec. \ref{S4}. Notice that $T$ locates at the hinge of two boundary planes {[}$MARX$ and $ZAR$ plane{]}, which both exhibit nodal surface. The nodal surface at $ZAR$ plane is generated by the $\{C_{2z}|00\frac{1}{2}\}{\cal{T}}$ symmetry. Therefore, the nodal line along $T$ resides at the intersection of two nodal surface (which then is labelled as L-NSs in Table \ref{tab:exam2}). The low-energy Hamiltonian expanded around a general point on the line with $\{R_{5},R_{5}\}$ rep is obtained as
\begin{eqnarray}
H & = & c_{1}+c_{2}k_{x}+\left[(\alpha\sigma_{+}+c_{3}\sigma_{3})k_{y}k_{z}+h.c.\right].
\end{eqnarray}

Similarly, the coreps deduced from $R_{7}$ rep can be obtained as
\begin{eqnarray}
\boldsymbol{D}_{7}(C_{2x}) & = & \left[\begin{array}{cc}
\boldsymbol{\Delta}_{7}(\{C_{2x}|\frac{1}{2}\frac{1}{2}0\}) & \boldsymbol{0}\\
\boldsymbol{0} & \boldsymbol{\Delta}_{7}(\{C_{2x}|\frac{1}{2}\frac{1}{2}0\})
\end{array}\right]=-ie^{-i\pi\alpha}\left[\begin{array}{cc}
-1 & 0\\
0 & -1
\end{array}\right]=i\sigma_{0}e^{-i\pi\alpha},\\
\boldsymbol{D}_{7}(E) & = & \left[\begin{array}{cc}
\boldsymbol{\Delta}_{7}(\{E|000\}) & \boldsymbol{0}\\
\boldsymbol{0} & \boldsymbol{\Delta}_{7}(\{E|000\})
\end{array}\right]=-\left[\begin{array}{cc}
-1 & 0\\
0 & -1
\end{array}\right]=\sigma_{0},\\
\boldsymbol{D}_{7}({\cal A}) & = & \left[\begin{array}{cc}
\boldsymbol{0} & \boldsymbol{-N}\\
\boldsymbol{1} & \boldsymbol{0}
\end{array}\right]=\left[\begin{array}{cc}
0 & \-1\\
1 & 0
\end{array}\right]=-i\sigma_{2}.
\end{eqnarray}
And the low-energy Hamiltonian expanded around a general point on the line with $\{R_{7},R_{7}\}$ rep is obtained as
\begin{eqnarray}
H & = & c_{1}+c_{2}k_{x}+\left[(\alpha\sigma_{+}+c_{3}\sigma_{3})k_{y}k_{z}+h.c.\right].
\end{eqnarray}

\section{Spatial operators of 230 SGs} \label{SGs-opers}

We use the Seitz symbols $\left\{ \boldsymbol{R}|\boldsymbol{v}\right\} $ to express the elements of the space group, with $\boldsymbol{R}$ a point-group operator and $\boldsymbol{v}$ a translation vector following $\boldsymbol{R}$. The space group operator denoted by Seitz symbol is an active operator, that
\begin{equation}
\left\{ \boldsymbol{R}|\boldsymbol{v}\right\} \boldsymbol{r}=\boldsymbol{R}\ \boldsymbol{r}+\boldsymbol{v}.
\end{equation}
Reference \cite{Bradley2009Mathematical-Oxford} lists all the elements of the 32 point group labeled by the Schonflies notation in  Tables  1.3, as well as the effect of  $\boldsymbol{R}$ acting on a generic vector $(xyz)$ in Tables 1.4.
In the following, we present all the  elements (not including the elements of translation group) of each of 230 SGs based on the convention used in Ref. \cite{Bradley2009Mathematical-Oxford}.
\include{SGopers}
\clearpage

\section{Notations and defined matrices used in Sec. \ref{S3} and Sec. \ref{S4}}
\subsection{Notations}
(i) We adopt the notations used in Ref. \cite{Bradley2009Mathematical-Oxford} to describe the information of SG and SG rep. The tables and figures in Ref. \cite{Bradley2009Mathematical-Oxford} that are relevant to this work   is listed as follows.
\begin{itemize}
\item Figure 1.1-1.3:  identifying the point-group operators;
\item Figure 3.2-3.15:  the BZ of each of the 14 Bravais lattices, along with the label of high symmetry momenta;
\item Table 1.4:   the effect of each  point-group operator on a vector $(xyz)$;
\item Table 3.1 (3.3):  list of all 14 Bravais lattice and the corresponding basic vectors in real (reciprocal) space;
\item Table 3.6:  list of the coordinates of high-symmetry point and line with respect to reciprocal vectors $(\boldsymbol{g}_{1}\boldsymbol{g}_{2}\boldsymbol{g}_{3})$;
\item Table 5.1:  defining relations, classes, character tables, and matrix representations of the relevant abstract groups;
\item Table 5.7 (6.13):  the single-valued (double-valued) reps of the 230 SGs;
\item Table 6.7:  the effect of  point-group operators on spin.
\end{itemize}

It should be notice that there exist some typos in  Ref. \cite{Bradley2009Mathematical-Oxford}, which have been pointed out and corrected by  Liu et. al. \cite{GBLiu_2020}. We establish the corep tables of type II MSG   based on the corrected results.

(ii) The momentum ${\bm k}$ used in effective Hamiltonian of the degeneracies are referred to the right-hand  orthogonal set of axes $Ok_xk_yk_z$. Moreover, we define \begin{eqnarray}
\left\{ p_{x},p_{y},p_{z}\right\}  & = & \frac{1}{\sqrt{12}}\left\{ -2\sqrt{2}k_{x}-\sqrt{2}k_{y}+\sqrt{2}k_{z},\sqrt{6}k_{y}+\sqrt{6}k_{z},2k_{x}-2k_{y}+2k_{z}\right\} ,\\
\left\{ q_{x},q_{y},q_{z}\right\}  & = & \frac{1}{\sqrt{12}}\left\{ -\sqrt{2}k_{x}-\sqrt{2}k_{y}+2\sqrt{2}k_{z},\sqrt{6}k_{x}-\sqrt{6}k_{y},2k_{x}+2k_{y}+2k_{z}\right\},
\end{eqnarray}
which would appear in certain effective Hamiltonian.
The coordinates of ${\bm g}_1$, ${\bm g}_2$ and ${\bm g}_3$ with respect to $k_x$, $k_y$ and $k_z$ axes for each Bravais lattice are listed in Table 3.3 of Ref. \cite{Bradley2009Mathematical-Oxford}.

(iii) As discussed in Ref. \cite{Bradley2009Mathematical-Oxford}, the reps of  the little group  $(\text{\textbf{G}}^{\boldsymbol{k}})$ for  high-symmetry line are obtained with the aid of the  central extension of the corresponding little co-group $(\bar{\text{\textbf{G}}}^{\boldsymbol{k*}})$. The relationship between the small reps $\Gamma^{\boldsymbol{k}}_p$  of $\text{\textbf{G}}^{\boldsymbol{k}}$ and  the reps $\text{\textbf{D}}^{\boldsymbol{k}}_p$ of $\bar{\text{\textbf{G}}}^{\boldsymbol{k*}}$ is
\begin{equation}
\Gamma^{\boldsymbol{k}}_p[\{R|\boldsymbol{v}\}]=e^{-i\boldsymbol{k \cdot v}}\text{\textbf{D}}^{\boldsymbol{k}}_p[(R,0)],
\end{equation}
with $(R,\alpha)$ ($\alpha$ an integer) the element of $\bar{\text{\textbf{G}}}^{\boldsymbol{k*}}$ (see sections 3.7 and 3.8 in  Ref. \cite{Bradley2009Mathematical-Oxford}).

For the tables in Sec. \ref{S3} and Sec. \ref{S4}, the matrix representation of generating elements on high-symmetry line are given in the form of   $\Gamma^{\boldsymbol{k}}_p[\{R|\boldsymbol{v}\}]$.
As the coordinate $\boldsymbol{k}$ of high-symmetry line is  a function of $\alpha$ with  $0<\alpha<\frac{1}{2}$ [such as the coordinate of $T$ in SG 92 is $\boldsymbol{k}=(\alpha\frac{1}{2}\frac{1}{2})$], then generally the matrix representation of $\{R|\boldsymbol{v}\}$ for rep $\Gamma^{\boldsymbol{k}}_p$  also is a function of $\alpha$.
However, for the sake of simplicity, we omit the factor involving $\alpha$ in the reps tables in  Sec. \ref{S3} and Sec. \ref{S4}, or in other word, we set $\alpha=0$ for the phase $e^{-i\boldsymbol{k \cdot v}}$.

(iv) As  discussed in Sec. \ref{MSG-corep}, there exist three possible cases (a-c) for the corep derived from a rep $R_i$.  The corep is label as  $R_i$ for case (a), $\{R_i,R_i\}$ for  case (b) and  $\{R_i,R_j\}$ for case (c) in Sec. \ref{S3} and Sec. \ref{S4}.

(v) Abbreviations used in the Tables of Sec. \ref{S3} and Sec. \ref{S4}:
\begin{itemize}
\item NP/NL/NS: nodal point/line/surface;
\item P-WNL$_{AB}$: a nodal point resides on a Weyl nodal line, which occurs along high-symmetry line $AB$, or at the joint point of multiple  Weyl nodal lines, which all  connected to WNL$_{AB}$ by symmetry operators.
\item P-WNL: a nodal point resides on a Weyl nodal line, which  however does not occurs along any high-symmetry line;
\item P-WNLs: a nodal point resides at the joint point of multiple  Weyl nodal lines;
\item P-DNL$_{AB}$: a nodal point resides on a Dirac nodal line, which occurs along high-symmetry line $AB$, or at the joint point of multiple  Dirac nodal lines, which all connected to DNL$_{AB}$ by symmetry operators.
\item P-DNL: a nodal point resides on a Dirac nodal line, which however  does not occurs along any high-symmetry line;
\item P-DNLs: a nodal  point resides at the joint point of multiple  Dirac nodal lines;
\item P-NS$_{ABCD}$: a nodal point resides on a nodal surface, which  occurs  in  high-symmetry plane $ABCD$;
\item P-NSs: a nodal point resides at the intersection of  two or three nodal surfaces.
\item P-WNL/NS: a nodal point resides at the  intersection of  a nodal line and a nodal surface.
\item L-NS$_{ABCD}$: a Weyl nodal line resides on a nodal surface, which  occurs  in  high-symmetry plane $ABCD$;
\item L-NSs: a nodal line resides at the hinge  of two nodal surfaces.
\end{itemize}

\subsection{Defined matrixes}

\subsubsection{Two-dimensional matrixes}

We define
\begin{equation*}
\sigma_0=\left(
\begin{array}{cc}
 1 & 0 \\
 0 & 1 \\
\end{array}
\right),\sigma _1=\left(
\begin{array}{cc}
 0 & 1 \\
 1 & 0 \\
\end{array}
\right),\sigma _2=\left(
\begin{array}{cc}
 0 & -i \\
 i & 0 \\
\end{array}
\right),\sigma _3=\left(
\begin{array}{cc}
 1 & 0 \\
 0 & -1 \\
\end{array}
\right),
\end{equation*}
$\sigma_{\pm}=(\sigma_{1}\pm i\sigma_{2})/2$  and
\begin{eqnarray*}
 && \sigma_{4}=\frac{-\sigma_{0}+i\left(\sqrt{2}\sigma_{1}+\sigma_{2}\right)}{2},\sigma_{5}=\frac{\sigma_{1}-\sqrt{2}\sigma_{2}-\sigma_{3}}{2i},\sigma_{6}=\frac{-\sigma_{0}-i\left(\sigma_{1}-\sigma_{2}-\sigma_{3}\right)}{2},\sigma_{7}=\frac{i\left(\sigma_{1}-\sigma_{2}+\sigma_{3}\right)}{\sqrt{3}},\\
 && \sigma_{8}=\frac{\left(3+\sqrt{3}\right)\sigma_{1}-\left(\sqrt{3}-3\right)\sigma_{2}-2\sqrt{3}\sigma_{3}}{6i},\sigma_{9}=\frac{\left((-1)^{1/3}-1\right)\sigma_{0}+\left(1+(-1)^{1/3}\right)\sigma_{3}}{2},\\
 && \sigma_{10}=\frac{\left(-1-(-1)^{2/3}\right)\sigma_{0}+\left((-1)^{2/3}-1\right)\sigma_{3}}{2},\sigma_{11}=\frac{\left(\sqrt{3}+3i\right)\sigma_{0}+\left(\sqrt{3}-i\right)\sigma_{3}}{4},\\
 && \sigma_{12}=\frac{\left(\sqrt{3}-i\right)\sigma_{0}+\left(\sqrt{3}+3i\right)\sigma_{3}}{4},\sigma_{13}=\frac{2\left((-1)^{5/6}+i\right)\sigma_{0}+\left(\sqrt{3}+i\right)\sigma_{3}}{4},\\
 && \sigma_{14}=\frac{2(-1)^{5/6}\sigma_{0}+\left(\sqrt{3}+3i\right)\sigma_{3}}{4},\sigma_{15}=\frac{\left(\sqrt{3}+i\right)\sigma_{0}+2\left((-1)^{5/6}+i\right)\sigma_{3}}{4},\\
 && \sigma_{16}=\frac{-(-1)^{1/6}\sigma_{0}+\left((-1)^{5/6}+i\right)\sigma_{3}}{2},\sigma_{17}=\frac{2(-1)^{5/6}\sigma_{3}+\left(-\sqrt{3}-3i\right)\sigma_{0}}{4},\\
 && \sigma_{18}=\frac{-2(-1)^{1/6}\sigma_{3}+\left(\sqrt{3}-3i\right)\sigma_{0}}{4},\sigma_{19}=\frac{\left(\sqrt{3}-i\right)\sigma_{0}-\left(\sqrt{3}+3i\right)\sigma_{3}}{4},\\
 && \sigma_{20}=\frac{-2(-1)^{1/6}\sigma_{0}+\left(\sqrt{3}-3i\right)\sigma_{3}}{4},\sigma_{21}=-\frac{\sigma_{1}+\sigma_{2}+\sigma_{3}}{\sqrt{3}},\\
 && \sigma_{22}=\frac{\left(3-\sqrt{3}\right)\sigma_{1}-\left(3+\sqrt{3}\right)\sigma_{2}+2\sqrt{3}\sigma_{3}}{6},\sigma_{23}=\frac{\sqrt{2}\left(\sqrt{3}-3\right)\sigma_{0}-4i\sqrt{6}\sigma_{1}+2i\sqrt{3\left(2+\sqrt{3}\right)}\sigma_{3}}{12},\\
 && \sigma_{24}=\frac{\left(3-\sqrt{3}\right)\sigma_{1}+\left(3+\sqrt{3}\right)\sigma_{2}+2\sqrt{3}\sigma_{3}}{6},\\
 && \sigma_{25}=\frac{(6+6i)\sigma_{0}+(1-i)\left[\left(\sqrt{3}-3\right)\sigma_{1}-\left(3+\sqrt{3}\right)\sigma_{2}-2\sqrt{3}\sigma_{3}\right]}{12},\\
 && \sigma_{26}=\frac{(-1+i)\left[6i\sigma_{0}+\left(\sqrt{3}-3\right)\sigma_{1}-\left(3+\sqrt{3}\right)\sigma_{2}-2\sqrt{3}\sigma_{3}\right]}{12}.
\end{eqnarray*}

\clearpage
\subsubsection{Three-dimensional matrixes}

We define
\begin{eqnarray*}
 &  & A_0=\left(
\begin{array}{ccc}
 1 & 0 & 0 \\
 0 & 1 & 0 \\
 0 & 0 & 1 \\
\end{array}
\right), A_1=\left(
\begin{array}{ccc}
 0 & -i & 0 \\
 i & 0 & 0 \\
 0 & 0 & 0 \\
\end{array}
\right), A_2=\left(
\begin{array}{ccc}
 0 & 0 & -i \\
 0 & 0 & 0 \\
 i & 0 & 0 \\
\end{array}
\right), A_3=\left(
\begin{array}{ccc}
 0 & 0 & 0 \\
 0 & 0 & -i \\
 0 & i & 0 \\
\end{array}
\right), A_4=\left(
\begin{array}{ccc}
 0 & 1 & 0 \\
 1 & 0 & 0 \\
 0 & 0 & 0 \\
\end{array}
\right),\\
&&A_5=\left(
\begin{array}{ccc}
 1 & 0 & 0 \\
 0 & -1 & 0 \\
 0 & 0 & 0 \\
\end{array}
\right), A_6=\left(
\begin{array}{ccc}
 0 & 0 & 1 \\
 0 & 0 & 0 \\
 1 & 0 & 0 \\
\end{array}
\right), A_7=\left(
\begin{array}{ccc}
 0 & 0 & 0 \\
 0 & 0 & 1 \\
 0 & 1 & 0 \\
\end{array}
\right), A_8=\frac{1}{\sqrt{3}}\left(
\begin{array}{ccc}
 1 & 0 & 0 \\
 0 & 1 & 0 \\
 0 & 0 & -2 \\
\end{array}
\right),
\end{eqnarray*}
and
\begin{eqnarray*}
 &  & A_{9}=\frac{-iA_{1}+iA_{2}-iA_{3}+A_{4}+A_{6}+A_{7}}{2},A_{10}=-\frac{A_{0}}{3}-A_{5}+\frac{A_{8}}{\sqrt{3}},A_{11}=\frac{iA_{1}-iA_{2}+iA_{3}+A_{4}+A_{6}+A_{7}}{2},\\
 &  & A_{12}=-\frac{A_{0}}{3}+A_{4}+\frac{A_{8}}{\sqrt{3}},A_{13}=-\frac{A_{0}}{3}+A_{5}+\frac{A_{8}}{\sqrt{3}},A_{14}=\frac{A_{7}+i\left[A_{1}-A_{2}-A_{3}+i\left(A_{4}+A_{6}\right)\right]}{2},\\
 &  & A_{15}=A_{6}+\frac{-2A_{0}+3A_{5}-\sqrt{3}A_{8}}{6},A_{16}=-\frac{A_{0}}{3}+iA_{1}+\frac{A_{8}}{\sqrt{3}},A_{17}=A_{7}+\frac{2A_{0}+3A_{5}+\sqrt{3}A_{8}}{6},\\
 &  & A_{18}=\frac{2A_{0}+6iA_{2}-3A_{5}+\sqrt{3}A_{8}}{6},A_{19}=\frac{2A_{0}-6iA_{3}+3A_{5}+\sqrt{3}A_{8}}{6},A_{20}=\frac{A_{0}}{3}+A_{4}-\frac{A_{8}}{\sqrt{3}},\\
 &  & A_{21}=\frac{2A_{0}-6iA_{2}-3A_{5}+\sqrt{3}A_{8}}{6},A_{22}=\frac{A_{1}+A_{2}+A_{3}-i\left(A_{4}-A_{6}+A_{7}\right)}{2i},\\
 &  & A_{23}=\frac{\left(-2-2i\sqrt{3}\right)A_{0}-12(-1)^{5/6}A_{2}+\left(3+3i\sqrt{3}\right)A_{5}-\left(\sqrt{3}+3i\right)A_{8}}{12},\\
 &  & A_{24}=\frac{-2i\sqrt{3}A_{3}-3A_{5}-\sqrt{3}A_{8}}{4},A_{25}=\frac{4A_{0}-3\left(\sqrt{3}+3i\right)A_{1}+6\left[1+(-1)^{2/3}\right]A_{4}-4\sqrt{3}A_{8}}{12},\\
 &  & A_{26}=\frac{2A_{0}+(3-3i)A_{5}+\sqrt{-24+18i}A_{8}}{6},A_{27}=-\frac{A_{0}}{3}+iA_{5}+\frac{A_{8}}{\sqrt{3}},A_{28}=\frac{i\left(A_{0}+3iA_{5}-\sqrt{3}A_{8}\right)}{3},\\
 &  & A_{29}=\frac{2A_{0}+6iA_{3}+3A_{5}+\sqrt{3}A_{8}}{6},A_{30}=A_{7}+\frac{i\left(2A_{0}+3A_{5}+\sqrt{3}A_{8}\right)}{6},\\
 &  & A_{31}=A_{7}-\frac{i\left(2A_{0}+3A_{5}+\sqrt{3}A_{8}\right)}{6},A_{32}=\frac{2A_{0}+(3+3i)A_{5}+\sqrt{-24-18i}A_{8}}{6},\\
 &  & A_{33}=\frac{-2\left((-1)^{2/3}-1\right)A_{5}+\left(\sqrt{3}+3i\right)A_{8}}{4},A_{34}=\frac{4A_{0}+3\left(\sqrt{3}-3i\right)A_{3}+6A_{5}-6\left((-1)^{1/3}-1\right)A_{7}+2\sqrt{3}A_{8}}{12},\\
 &  & A_{35}=\frac{8A_{0}-6\left(\sqrt[3]{-1}-1\right)A_{5}+\left(\sqrt{3}+9i\right)A_{8}}{12},A_{36}=\frac{-2i\sqrt{3}A_{3}+3A_{5}+\sqrt{3}A_{8}}{4},\\
 &  & A_{37}=-\frac{i\left(2A_{0}+6A_{3}+3A_{5}+\sqrt{3}A_{8}\right)}{6},A_{38}=\frac{i\left(2A_{0}-6A_{3}+3A_{5}+\sqrt{3}A_{8}\right)}{6},\\
 &  & A_{39}=\frac{8A_{0}-6i\sqrt{3}A_{3}+3A_{5}+\sqrt{3}A_{8}}{12},A_{40}=\frac{-8A_{0}-6i\sqrt{3}A_{3}-3A_{5}-\sqrt{3}A_{8}}{12}.
\end{eqnarray*}

\clearpage
\subsubsection{Four-dimensional matrixes}

We define $\Gamma_{i,j}=\sigma_i\otimes\sigma_j$ with $i,j=0,1,2,3,\pm$ and
\begin{eqnarray*}
 &  & \Gamma_{1}=\frac{\Gamma_{0,0}-i\sqrt{2}\Gamma_{0,2}-i\Gamma_{3,1}+\sqrt{3}\left(\sqrt{2}\Gamma_{1,0}+i\Gamma_{1,2}-\Gamma_{2,3}\right)}{4},\Gamma_{2}=i\frac{\Gamma_{0,3}-\sqrt{2}\Gamma_{0,1}+\Gamma_{3,2}+\sqrt{3}\left(\Gamma_{1,1}+\sqrt{2}\Gamma_{1,3}-\Gamma_{2,0}\right)}{4},\\
 &  & \Gamma_{3}=\frac{\Gamma_{0,0}-i\left(\Gamma_{0,1}-\Gamma_{0,2}+\Gamma_{0,3}\right)}{2},\Gamma_{4}=-\frac{\sqrt{3}\Gamma_{0,1}+\Gamma_{0,3}-i\left(\sqrt{3}\Gamma_{1,1}+\Gamma_{1,3}\right)}{2\sqrt{2}},\\
 &  & \Gamma_{5}=\frac{\Gamma_{0,0}-i\sqrt{3}\Gamma_{0,2}+i\Gamma_{1,0}+\sqrt{3}\Gamma_{1,2}-i\Gamma_{2,0}-\sqrt{3}\Gamma_{2,2}-i\Gamma_{3,0}-\sqrt{3}\Gamma_{3,2}}{4},\\
 &  & \Gamma_{6}=-\frac{i\left(\sqrt{6}\Gamma_{0,2}+3\sqrt{2}\Gamma_{3,1}-2\sqrt{3}\Gamma_{3,3}\right)}{6},\Gamma_{7}=-\frac{i\left(\sqrt{6}\Gamma_{0,2}-3\sqrt{2}\Gamma_{3,1}-2\sqrt{3}\Gamma_{3,3}\right)}{6},\\
 &  & \Gamma_{8}=\frac{-\Gamma_{0,0}+3\Gamma_{0,3}+i\sqrt{3}\left(\Gamma_{3,0}+\Gamma_{3,3}\right)}{4},\Gamma_{9}=\frac{\Gamma_{0,2}-\Gamma_{3,1}-\Gamma_{3,3}}{i\sqrt{3}},\Gamma_{10}=i\frac{\left(\sqrt{3}-3\right)\Gamma_{0,2}-\left(3+\sqrt{3}\right)\Gamma_{3,1}+2\sqrt{3}\Gamma_{3,3}}{6},\\
 &  & \Gamma_{11}=\frac{-\Gamma_{0,0}+i\Gamma_{0,2}-i\Gamma_{3,1}+i\Gamma_{3,3}+\sqrt{3}\left(\Gamma_{0,3}+i\Gamma_{3,0}-\Gamma_{0,1}+\Gamma_{3,2}\right)}{4},\Gamma_{12}=\frac{-\Gamma_{0,0}+i\sqrt{3}\left(\Gamma_{0,3}-i\sqrt{3}\Gamma_{3,0}+\Gamma_{3,3}\right)}{4},\\
 &  & \Gamma_{13}=\frac{-\Gamma_{0,0}-i\sqrt{3}\Gamma_{0,2}-3\Gamma_{3,0}+i\sqrt{3}\Gamma_{3,2}}{4},\Gamma_{14}=\frac{\sqrt{3}\Gamma_{0,0}+3i\Gamma_{0,3}+\sqrt{3}\Gamma_{3,0}-i\Gamma_{3,3}}{4},\\
 &  & \Gamma_{15}=\frac{-\sqrt{3}\Gamma_{0,0}+3i\Gamma_{0,3}+\sqrt{3}\Gamma_{3,0}+i\Gamma_{3,3}}{4},\Gamma_{16}=\frac{\sqrt{3}\Gamma_{0,0}+i\left(3\Gamma_{0,3}+i\sqrt{3}\Gamma_{3,0}+\Gamma_{3,3}\right)}{4},\\
 &  & \Gamma_{17}=\frac{-i\Gamma_{0,0}+\sqrt{3}\Gamma_{0,3}-3i\Gamma_{3,0}-\sqrt{3}\Gamma_{3,3}}{4},\Gamma_{18}=\frac{-\Gamma_{0,0}-\Gamma_{0,3}+\Gamma_{3,0}-\Gamma_{3,3}}{2},\Gamma_{19}=\frac{\Gamma_{0,1}-i\Gamma_{0,2}+\Gamma_{3,1}+i\Gamma_{3,2}}{2},\\
 &  & \Gamma_{20}=\frac{\Gamma_{0,0}-\Gamma_{0,3}-\Gamma_{3,0}-\Gamma_{3,3}}{2},\Gamma_{21}=\frac{\Gamma_{0,2}+\Gamma_{0,3}-\Gamma_{3,2}+\Gamma_{3,3}}{2i},\Gamma_{22}=\frac{\Gamma_{0,0}+\Gamma_{0,3}+\Gamma_{3,0}-\Gamma_{3,3}}{2},\\
 &  & \Gamma_{23}=\frac{-\Gamma_{0,0}+\Gamma_{0,1}+\Gamma_{3,0}+\Gamma_{3,1}}{2},\Gamma_{24}=-\frac{i\left(\Gamma_{0,2}-\Gamma_{0,3}-\Gamma_{3,2}-\Gamma_{3,3}\right)}{2},\Gamma_{25}=\frac{-\Gamma_{0,0}+\Gamma_{0,3}-\Gamma_{3,0}-\Gamma_{3,3}}{2},\\
 &  & \Gamma_{26}=\frac{i\Gamma_{0,2}+\Gamma_{0,3}-i\Gamma_{3,2}+\Gamma_{3,3}}{2},\Gamma_{27}=\frac{\Gamma_{0,0}+i\left(\Gamma_{0,1}-\Gamma_{0,2}-\Gamma_{0,3}\right)}{2},\Gamma_{28}=\frac{i\left(\Gamma_{0,1}-\Gamma_{0,2}+\Gamma_{0,3}\right)}{\sqrt{3}},\\
 &  & \Gamma_{29}=\frac{\left(3+\sqrt{3}\right)\Gamma_{0,1}-\left(\sqrt{3}-3\right)\Gamma_{0,2}-2\sqrt{3}\Gamma_{0,3}}{6i},\Gamma_{30}=-\frac{\Gamma_{0,1}+\Gamma_{0,2}+\Gamma_{0,3}}{\sqrt{3}},\\
 &  & \Gamma_{31}=\frac{-\left(\sqrt{3}-3\right)\Gamma_{0,1}-\left(3+\sqrt{3}\right)\Gamma_{0,2}+2\sqrt{3}\Gamma_{0,3}}{6},\Gamma_{32}=-\frac{\Gamma_{0,2}+\Gamma_{3,1}+\Gamma_{3,3}}{\sqrt{3}},\\
 &  & \Gamma_{33}=\frac{-\left(3+\sqrt{3}\right)\Gamma_{0,2}-\left(\sqrt{3}-3\right)\Gamma_{3,1}+2\sqrt{3}\Gamma_{3,3}}{6},\Gamma_{34}=-\frac{i\left(\sqrt{6}\Gamma_{0,2}-3\sqrt{2}\Gamma_{3,1}+2\sqrt{3}\Gamma_{3,3}\right)}{6},\\
 &  & \Gamma_{35}=-\frac{i\left(\sqrt{6}\Gamma_{0,2}+3\sqrt{2}\Gamma_{3,1}+2\sqrt{3}\Gamma_{3,3}\right)}{6},\\
 &  & \Gamma_{36}=\frac{2i\Gamma_{0,0}+2\Gamma_{0,1}+2\Gamma_{0,2}+2\Gamma_{0,3}+\sqrt{6}\left[\Gamma_{1,0}-i\left(\Gamma_{1,1}+\Gamma_{1,2}+\Gamma_{1,3}-i\Gamma_{2,0}-\Gamma_{2,1}-\Gamma_{2,2}-\Gamma_{2,3}\right)\right]}{8},\\
 &  & \Gamma_{37}=\frac{\left(3+\sqrt{3}\right)\Gamma_{0,2}-\left(\sqrt{3}-3\right)\Gamma_{3,1}+2\sqrt{3}\Gamma_{3,3}}{6},\\
 &  & \Gamma_{38}=\frac{(1+i)\left[-i\left(\sqrt{3}-3\right)\Gamma_{0,1}+2i\sqrt{3}\Gamma_{0,3}+6\Gamma_{3,0}+i\left(3+\sqrt{3}\right)\Gamma_{3,2}\right]}{12}.
\end{eqnarray*}

\clearpage

\subsubsection{Six-dimensional matrixes}

We define $S_{i,j}=\sigma_i\otimes A_j$ with $i=0,1,2,3$ and $j=0,1,2,3,4,5,7,8$, and
\begin{eqnarray*}
 &  & S_{1}=\frac{2i\left(2S_{2,0}+\sqrt{3}S_{2,8}\right)+3i\left(S_{0,2}-S_{0,3}-S_{1,2}-S_{1,3}+S_{2,6}-S_{2,7}+S_{3,2}+S_{3,3}\right)}{12}\\
 &  & +\frac{S_{0,6}-S_{0,7}-2S_{1,5}+S_{1,6}+S_{1,7}+S_{2,2}-S_{2,3}+S_{3,6}+S_{3,7}}{4},\\
 &  & S_{2}=\frac{-S_{0,0}+\sqrt{3}S_{0,8}+2S_{3,0}+\sqrt{3}S_{3,8}}{3},S_{3}=\frac{-2S_{0,0}-3S_{0,5}-\sqrt{3}S_{0,8}+3S_{3,5}-3\sqrt{3}S_{3,8}}{6},\\
 &  & S_{4}=\frac{-S_{0,5}-S_{0,7}-S_{1,6}-S_{2,2}-S_{3,5}+S_{3,7}}{2},S_{5}=\frac{S_{0,4}-S_{0,6}+S_{1,7}+S_{2,3}+S_{3,4}+S_{3,6}}{2},\\
 &  & S_{6}=i\frac{S_{0,1}-S_{0,2}+S_{0,3}+i\left(S_{0,4}+S_{0,6}+S_{0,7}\right)}{2},S_{7}=\frac{S_{0,0}-3S_{0,5}-\sqrt{3}S_{0,8}}{3},\\
 &  & S_{8}=S_{3,6}+\frac{-2S_{3,0}+3S_{3,5}-\sqrt{3}S_{3,8}}{6},\\
 &  & S_{9}=\frac{3\left[S_{0,5}-S_{3,5}+\sqrt{3}\left(S_{0,8}-S_{3,8}\right)-2i\left(S_{0,2}+S_{3,2}\right)\right]+2i\left(2S_{2,0}-3S_{2,5}+\sqrt{3}S_{2,8}\right)}{12},\\
 &  & S_{10}=\frac{2\left(2S_{1,0}+\sqrt{3}S_{1,8}\right)+3\left(S_{0,6}+S_{0,7}-S_{1,6}-S_{1,7}-S_{2,2}+S_{2,3}+S_{3,6}-S_{3,7}\right)}{12}\\
 &  & +\frac{i\left(S_{0,2}+S_{0,3}+S_{1,2}+S_{1,3}-2S_{2,5}-S_{2,6}+S_{2,7}+S_{3,2}-S_{3,3}\right)}{4},\\
 &  & S_{11}=\frac{S_{0,0}+3S_{0,5}-\sqrt{3}S_{0,8}}{3i},S_{12}=-\frac{(-1)^{1/4}\left(2S_{3,0}+6iS_{3,2}-3S_{3,5}+\sqrt{3}S_{3,8}\right)}{6},\\
 &  & S_{13}=\frac{-iS_{0,1}+iS_{0,2}+iS_{0,3}+S_{0,4}+S_{0,6}-S_{0,7}}{2},S_{14}=\frac{i\left(-S_{3,0}+3S_{3,4}+\sqrt{3}S_{3,8}\right)}{3},\\
 &  & S_{15}=\frac{i\left(2S_{3,0}+3S_{3,5}+6S_{3,7}+\sqrt{3}S_{3,8}\right)}{6}.
\end{eqnarray*}

\subsubsection{Eight-dimensional matrixes}

We define $Q_{i,j,k}=\sigma_i\otimes \sigma_j \otimes \sigma_k$ with $i,j,k=0,1,2,3$, and
\begin{eqnarray*}
 &  & Q_{1}=\frac{-Q_{0,0,0}-iQ_{0,0,2}+\sqrt{3}Q_{0,3,1}+\sqrt{3}Q_{0,3,3}+i\left[Q_{3,0,1}+Q_{3,0,3}+\sqrt{3}\left(Q_{3,3,0}+iQ_{3,3,2}\right)\right]}{4},\\
 &  & Q_{2}=\frac{Q_{0,0,0}-i\sqrt{2}Q_{0,0,2}+\sqrt{6}Q_{0,1,0}+i\sqrt{3}Q_{0,1,2}-\sqrt{3}Q_{0,2,3}-iQ_{0,3,1}}{4}.
\end{eqnarray*}

\clearpage

\section{Encyclopedia of emergent particles in 3D  crystals without SOC effect} \label{S3}
\subsection{The single-valued corepresentations of the 230 type-II MSGs and the essential degeneracies} \label{S3B}
%
\subsubsection{Notes to Sec. \ref{S3B} }

(i) For each table in Sec. \ref{S3B}, the first two lines  present the  SG number,  the BZ type, the generating elements of the type II MSG (translations are not included here), whether centrosymmetry is contained in the group and whether SOC is considered.

(ii) Below the first two lines,  the columns from left to right (separated by the semicolons)  are the high-symmetry momentum $\bm{k}$, the location  of $\bm{k}$ [with respect to reciprocal vectors $(\boldsymbol{g}_{1},\boldsymbol{g}_{2},\boldsymbol{g}_{3})$], the generating elements of the little group at $\bm{k}$ (only point-group operators are presented and a full expression of each  generating element can be found in Ref. \cite{Bradley2009Mathematical-Oxford} and in Sec. \ref{SGs-opers}),  the deduced corep of the little group at $\bm{k}$, the dimension  of the corep, the matrix representations of the generating elements,  the species and the topological charge of the essential  degeneracy.

(iii) A correspondence between the notation of the corep used here ($R_i$) and the band-representation notations can be found in Refs. \cite{Bradley2009Mathematical-Oxford, GBLiu_2020}. Moreover, Ref. \cite{GBLiu_2020} has established a SpaceGroupIrep package  to analyze the band representation based on the notation of Ref. \cite{Bradley2009Mathematical-Oxford}.

\subsubsection{SG 1-10}

{
\global\long\def\arraystretch{1.4}%
\begin{table}[H]
\raggedright{}%
\begin{tabular}{lllllllll}
\multicolumn{8}{l}{SG 1} & \tabularnewline
\hline
\multicolumn{8}{l}{$\Gamma_{t}$; $\{E|000\},\mathcal{T}$; Non-Centrosymmetric; without
SOC } & \tabularnewline
\multicolumn{8}{l}{ } & \tabularnewline
\multirow{1}{*}{$\Gamma$;} & \multirow{1}{*}{$(000)$;} & \multirow{1}{*}{$E$,$\mathcal{T}$;} & $R_{1};$  & $1;$  & $1,1;$  &  &  & \tabularnewline
\multirow{1}{*}{$B$;} & \multirow{1}{*}{$(\frac{1}{2}00)$;} & \multirow{1}{*}{$E$,$\mathcal{T}$;} & $R_{1};$  & $1;$  & $1,1;$  &  &  & \tabularnewline
\multirow{1}{*}{$F$;} & \multirow{1}{*}{$(0\frac{1}{2}0)$;} & \multirow{1}{*}{$E$,$\mathcal{T}$;} & $R_{1};$  & $1;$  & $1,1;$  &  &  & \tabularnewline
\multirow{1}{*}{$G$;} & \multirow{1}{*}{$(00\frac{1}{2})$;} & \multirow{1}{*}{$E$,$\mathcal{T}$;} & $R_{1};$  & $1;$  & $1,1;$  &  &  & \tabularnewline
\end{tabular}
\end{table}

}

{
\global\long\def\arraystretch{1.4}%
\begin{table}[H]
\raggedright{}%
\begin{tabular}{lllllllll}
\multicolumn{8}{l}{SG 2} & \tabularnewline
\hline
\multicolumn{8}{l}{$\Gamma_{t}$; $\{I|000\},\mathcal{T}$; Centrosymmetric; without
SOC } & \tabularnewline
\multicolumn{8}{l}{ } & \tabularnewline
\multicolumn{1}{l}{$\Gamma$;} & \multicolumn{1}{l}{$(000)$;} & \multicolumn{1}{l}{$I$,$\mathcal{T}$;} & $R_{1};$  & $1;$  & $1,1;$  &  &  & \tabularnewline
 &  &  & $R_{2};$  & $1;$  & $-1,1;$  &  &  & \tabularnewline
\multicolumn{1}{l}{$B$;} & \multicolumn{1}{l}{$(\frac{1}{2}00)$;} & \multicolumn{1}{l}{$I$,$\mathcal{T}$;} & $R_{1};$  & $1;$  & $1,1;$  &  &  & \tabularnewline
 &  &  & $R_{2};$  & $1;$  & $-1,1;$  &  &  & \tabularnewline
\multicolumn{1}{l}{$F$;} & \multicolumn{1}{l}{$(0\frac{1}{2}0)$;} & \multicolumn{1}{l}{$I$,$\mathcal{T}$;} & $R_{1};$  & $1;$  & $1,1;$  &  &  & \tabularnewline
 &  &  & $R_{2};$  & $1;$  & $-1,1;$  &  &  & \tabularnewline
\multicolumn{1}{l}{$G$;} & \multicolumn{1}{l}{$(00\frac{1}{2})$;} & \multicolumn{1}{l}{$I$,$\mathcal{T}$;} & $R_{1};$  & $1;$  & $1,1;$  &  &  & \tabularnewline
 &  &  & $R_{2};$  & $1;$  & $-1,1;$  &  &  & \tabularnewline
\end{tabular}
\end{table}

}

{
\global\long\def\arraystretch{1.4}%
\begin{table}[H]
\raggedright{}%
\begin{tabular}{lllllllll}
\multicolumn{8}{l}{SG 3} & \tabularnewline
\hline
\multicolumn{8}{l}{$\Gamma_{m}$; $\{C_{2z}|000\},\mathcal{T}$; Non-Centrosymmetric;
without SOC } & \tabularnewline
\multicolumn{8}{l}{ } & \tabularnewline
\multicolumn{1}{l}{$\Gamma$;} & \multicolumn{1}{l}{$(000)$;} & \multicolumn{1}{l}{$C_{2z}$,$\mathcal{T}$;} & $R_{1};$  & $1;$  & $1,1;$  &  &  & \tabularnewline
 &  &  & $R_{2};$  & $1;$  & $-1,1;$  &  &  & \tabularnewline
\multicolumn{1}{l}{$B$;} & \multicolumn{1}{l}{$(\bar{\frac{1}{2}}00)$;} & \multicolumn{1}{l}{$C_{2z}$,$\mathcal{T}$;} & $R_{1};$  & $1;$  & $1,1;$  &  &  & \tabularnewline
 &  &  & $R_{2};$  & $1;$  & $-1,1;$  &  &  & \tabularnewline
\multicolumn{1}{l}{$Y$;} & \multicolumn{1}{l}{$(0\frac{1}{2}0)$;} & \multicolumn{1}{l}{$C_{2z}$,$\mathcal{T}$;} & $R_{1};$  & $1;$  & $1,1;$  &  &  & \tabularnewline
\end{tabular}
\end{table}

}

{
\global\long\def\arraystretch{1.4}%
\begin{table}[H]
\raggedright{}%
\begin{tabular}{lllllllll}
\multicolumn{8}{l}{SG 4} & \tabularnewline
\hline
\multicolumn{8}{l}{$\Gamma_{m}$; $\{C_{2z}|00\frac{1}{2}\},\mathcal{T}$; Non-Centrosymmetric;
without SOC } & \tabularnewline
\multicolumn{8}{l}{ } & \tabularnewline
\multicolumn{1}{l}{$\Gamma$;} & \multicolumn{1}{l}{$(000)$;} & \multicolumn{1}{l}{$C_{2z}$,$\mathcal{T}$;} & $R_{1};$  & $1;$  & $1,1;$  &  &  & \tabularnewline
 &  &  & $R_{2};$  & $1;$  & $-1,1;$  &  &  & \tabularnewline
\multicolumn{1}{l}{$B$;} & \multicolumn{1}{l}{$(\bar{\frac{1}{2}}00)$;} & \multicolumn{1}{l}{$C_{2z}$,$\mathcal{T}$;} & $R_{1};$  & $1;$  & $1,1;$  &  &  & \tabularnewline
 &  &  & $R_{2};$  & $1;$  & $-1,1;$  &  &  & \tabularnewline
\multicolumn{1}{l}{$Y$;} & \multicolumn{1}{l}{$(0\frac{1}{2}0)$;} & \multicolumn{1}{l}{$C_{2z}$,$\mathcal{T}$;} & $R_{1};$  & $1;$  & $1,1;$  &  &  & \tabularnewline
\end{tabular}
\end{table}
}

{
\global\long\def\arraystretch{1.4}%
\begin{table}[H]
\raggedright{}%
\begin{tabular}{lllllllll}
\multicolumn{8}{l}{SG 5} & \tabularnewline
\hline
\multicolumn{8}{l}{$\Gamma_{m}^{b}$; $\{C_{2z}|000\},\mathcal{T}$; Non-Centrosymmetric;
without SOC } & \tabularnewline
\multicolumn{8}{l}{ } & \tabularnewline
\multicolumn{1}{l}{$\Gamma$;} & \multicolumn{1}{l}{$(000)$;} & \multicolumn{1}{l}{$C_{2z}$,$\mathcal{T}$;} & $R_{1};$  & $1;$  & $1,1;$  &  &  & \tabularnewline
 &  &  & $R_{2};$  & $1;$  & $-1,1;$  &  &  & \tabularnewline
\multicolumn{1}{l}{$A$;} & \multicolumn{1}{l}{$(\bar{\frac{1}{2}}00)$;} & \multirow{1}{*}{$C_{2z}$,$\mathcal{T}$;} & $R_{1};$  & $1;$  & $1,1;$  &  &  & \tabularnewline
 &  &  & $R_{2};$  & $1;$  & $-1,1;$  &  &  & \tabularnewline
\multicolumn{1}{l}{$Z$;} & \multicolumn{1}{l}{$(0\bar{\frac{1}{2}}\frac{1}{2})$;} & \multicolumn{1}{l}{$C_{2z}$,$\mathcal{T}$;} & $R_{1};$  & $1;$  & $1,1;$  &  &  & \tabularnewline
 &  &  & $R_{2};$  & $1;$  & $-1,1;$  &  &  & \tabularnewline
\multicolumn{1}{l}{$M$;} & \multicolumn{1}{l}{$(\bar{\frac{1}{2}}\bar{\frac{1}{2}}\frac{1}{2})$;} & \multicolumn{1}{l}{$C_{2z}$,$\mathcal{T}$;} & $R_{1};$  & $1;$  & $1,1;$  &  &  & \tabularnewline
 &  &  & $R_{2};$  & $1;$  & $-1,1;$  &  &  & \tabularnewline
\multirow{1}{*}{$L$;} & \multirow{1}{*}{$(\bar{\frac{1}{2}}0\frac{1}{2})$;} & \multirow{1}{*}{$E$,$\mathcal{T}$;} & $R_{1};$  & $1;$  & $1,1;$  &  &  & \tabularnewline
\multirow{1}{*}{$V$;} & \multirow{1}{*}{$(00\frac{1}{2})$;} & \multirow{1}{*}{$E$,$\mathcal{T}$;} & $R_{1};$  & $1;$  & $1,1;$  &  &  & \tabularnewline
\multicolumn{1}{l}{$\Lambda$;} & \multicolumn{1}{l}{$\text{\ensuremath{\Gamma}Z}$;} & \multirow{1}{*}{$C_{2z}$;} & $R_{1};$  & $1;$  & $1;$  &  &  & \tabularnewline
 &  &  & $R_{2};$  & $1;$  & $-1;$  &  &  & \tabularnewline
\multicolumn{1}{l}{$U$;} & \multicolumn{1}{l}{$\text{AM}$;} & \multirow{1}{*}{$C_{2z}$;} & $R_{1};$  & $1;$  & $1;$  &  &  & \tabularnewline
 &  &  & $R_{2};$  & $1;$  & $-1;$  &  &  & \tabularnewline
\end{tabular}
\end{table}

}


{
\global\long\def\arraystretch{1.4}%
\begin{table}[H]
\raggedright{}%
\begin{tabular}{lllllllll}
\multicolumn{8}{l}{SG 6} & \tabularnewline
\hline
\multicolumn{8}{l}{$\Gamma_{m}$; $\{\sigma_{z}|000\},\mathcal{T}$; Non-Centrosymmetric;
without SOC } & \tabularnewline
\multicolumn{8}{l}{ } & \tabularnewline
\multicolumn{1}{l}{$\Gamma$;} & \multicolumn{1}{l}{$(000)$;} & \multicolumn{1}{l}{$\sigma_{z}$,$\mathcal{T}$;} & $R_{1};$  & $1;$  & $1,1;$  &  &  & \tabularnewline
 &  &  & $R_{2};$  & $1;$  & $-1,1;$  &  &  & \tabularnewline
\multicolumn{1}{l}{$B$;} & \multicolumn{1}{l}{$(\bar{\frac{1}{2}}00)$;} & \multicolumn{1}{l}{$\sigma_{z}$,$\mathcal{T}$;} & $R_{1};$  & $1;$  & $1,1;$  &  &  & \tabularnewline
 &  &  & $R_{2};$  & $1;$  & $-1,1;$  &  &  & \tabularnewline
\multicolumn{1}{l}{$Y$;} & \multicolumn{1}{l}{$(0\frac{1}{2}0)$;} & \multicolumn{1}{l}{$\sigma_{z}$,$\mathcal{T}$;} & $R_{1};$  & $1;$  & $1,1;$  &  &  & \tabularnewline
\end{tabular}
\end{table}

}
{
\global\long\def\arraystretch{1.4}%
\begin{table}[H]
\raggedright{}%
\begin{tabular}{lllllllll}
\multicolumn{8}{l}{SG 7} & \tabularnewline
\hline
\multicolumn{8}{l}{$\Gamma_{m}$; $\{\sigma_{z}|\frac{1}{2}00\},\mathcal{T}$; Non-Centrosymmetric;
without SOC } & \tabularnewline
\multicolumn{8}{l}{ } & \tabularnewline
\multicolumn{1}{l}{$\Gamma$;} & \multicolumn{1}{l}{$(000)$;} & \multicolumn{1}{l}{$\sigma_{z}$,$\mathcal{T}$;} & $R_{1};$  & $1;$  & $1,1;$  &  &  & \tabularnewline
 &  &  & $R_{2};$  & $1;$  & $-1,1;$  &  &  & \tabularnewline
\multicolumn{1}{l}{$B$;} & \multirow{1}{*}{$(\bar{\frac{1}{2}}00)$;} & \multirow{1}{*}{$\sigma_{z}$,$\mathcal{T}$;} & $\{R_{2},R_{4}\};$  & $2;$  & $i\sigma_{3},\sigma_{1};$  & P-WNLs; &  & \tabularnewline
\multicolumn{1}{l}{$Y$;} & \multicolumn{1}{l}{$(0\frac{1}{2}0)$;} & \multicolumn{1}{l}{$\sigma_{z}$,$\mathcal{T}$;} & $R_{1};$  & $1;$  & $1,1;$  &  &  & \tabularnewline
 &  &  & $R_{2};$  & $1;$  & $-1,1;$  &  &  & \tabularnewline
\end{tabular}
\end{table}

}
{
\global\long\def\arraystretch{1.4}%
\begin{table}[H]
\raggedright{}%
\begin{tabular}{lllllllll}
\multicolumn{8}{l}{SG 8} & \tabularnewline
\hline
\multicolumn{8}{l}{$\Gamma_{m}^{b}$; $\{\sigma_{z}|000\},\mathcal{T}$; Non-Centrosymmetric;
without SOC } & \tabularnewline
\multicolumn{8}{l}{ } & \tabularnewline
\multicolumn{1}{l}{$\Gamma$;} & \multicolumn{1}{l}{$(000)$;} & \multicolumn{1}{l}{$\sigma_{z}$,$\mathcal{T}$;} & $R_{1};$  & $1;$  & $1,1;$  &  &  & \tabularnewline
 &  &  & $R_{2};$  & $1;$  & $-1,1;$  &  &  & \tabularnewline
\multicolumn{1}{l}{$A$;} & \multicolumn{1}{l}{$(\bar{\frac{1}{2}}00)$;} & \multicolumn{1}{l}{$\sigma_{z}$,$\mathcal{T}$;} & $R_{1};$  & $1;$  & $1,1;$  &  &  & \tabularnewline
 &  &  & $R_{2};$  & $1;$  & $-1,1;$  &  &  & \tabularnewline
\multicolumn{1}{l}{$Z$;} & \multicolumn{1}{l}{$(0\bar{\frac{1}{2}}\frac{1}{2})$;} & \multicolumn{1}{l}{$\sigma_{z}$,$\mathcal{T}$;} & $R_{1};$  & $1;$  & $1,1;$  &  &  & \tabularnewline
 &  &  & $R_{2};$  & $1;$  & $-1,1;$  &  &  & \tabularnewline
\multicolumn{1}{l}{$M$;} & \multicolumn{1}{l}{$(\bar{\frac{1}{2}}\bar{\frac{1}{2}}\frac{1}{2})$;} & \multicolumn{1}{l}{$\sigma_{z}$,$\mathcal{T}$;} & $R_{1};$  & $1;$  & $1,1;$  &  &  & \tabularnewline
 &  &  & $R_{2};$  & $1;$  & $-1,1;$  &  &  & \tabularnewline
\multirow{1}{*}{$L$;} & \multirow{1}{*}{$(\bar{\frac{1}{2}}0\frac{1}{2})$;} & \multirow{1}{*}{$E$,$\mathcal{T}$;} & $R_{1};$  & $1;$  & $1,1;$  &  &  & \tabularnewline
\multirow{1}{*}{$V$;} & \multirow{1}{*}{$(00\frac{1}{2})$;} & \multirow{1}{*}{$E$,$\mathcal{T}$;} & $R_{1};$  & $1;$  & $1,1;$  &  &  & \tabularnewline
\multirow{1}{*}{$\Lambda$;} & \multirow{1}{*}{$\text{\ensuremath{\Gamma}Z}$;} & \multirow{1}{*}{$E$,$\mathcal{T}\sigma_{z}$;} & $R_{1};$  & $1;$  & $1,1;$  &  &  & \tabularnewline
\multirow{1}{*}{$U$;} & \multirow{1}{*}{$\text{AM}$;} & \multirow{1}{*}{$E$,$\mathcal{T}\sigma_{z}$;} & $R_{1};$  & $1;$  & $1,1;$  &  &  & \tabularnewline
\end{tabular}
\end{table}

}
{
\global\long\def\arraystretch{1.4}%
\begin{table}[H]
\raggedright{}%
\begin{tabular}{lllllllll}
\multicolumn{8}{l}{SG 9} & \tabularnewline
\hline
\multicolumn{8}{l}{$\Gamma_{m}^{b}$; $\{\sigma_{z}|\frac{1}{2}00\},\mathcal{T}$; Non-Centrosymmetric;
without SOC } & \tabularnewline
\multicolumn{8}{l}{ } & \tabularnewline
\multicolumn{1}{l}{$\Gamma$;} & \multicolumn{1}{l}{$(000)$;} & \multicolumn{1}{l}{$\sigma_{z}$,$\mathcal{T}$;} & $R_{1};$  & $1;$  & $1,1;$  &  &  & \tabularnewline
 &  &  & $R_{2};$  & $1;$  & $-1,1;$  &  &  & \tabularnewline
\multirow{1}{*}{$A$;} & \multirow{1}{*}{$(\bar{\frac{1}{2}}00)$;} & \multirow{1}{*}{$\sigma_{z}$,$\mathcal{T}$;} & $\{R_{2},R_{4}\};$  & $2;$  & $i\sigma_{3},\sigma_{1};$  & P-WNLs; &  & \tabularnewline
\multicolumn{1}{l}{$Z$;} & \multicolumn{1}{l}{$(0\bar{\frac{1}{2}}\frac{1}{2})$;} & \multicolumn{1}{l}{$\sigma_{z}$,$\mathcal{T}$;} & $R_{1};$  & $1;$  & $1,1;$  &  &  & \tabularnewline
 &  &  & $R_{2};$  & $1;$  & $-1,1;$  &  &  & \tabularnewline
\multirow{1}{*}{$M$;} & \multirow{1}{*}{$(\bar{\frac{1}{2}}\bar{\frac{1}{2}}\frac{1}{2})$;} & \multirow{1}{*}{$\sigma_{z}$,$\mathcal{T}$;} & $\{R_{2},R_{4}\};$  & $2;$  & $i\sigma_{3},\sigma_{1};$  & P-WNLs; &  & \tabularnewline
\multirow{1}{*}{$L$;} & \multirow{1}{*}{$(\bar{\frac{1}{2}}0\frac{1}{2})$;} & \multirow{1}{*}{$E$,$\mathcal{T}$;} & $R_{1};$  & $1;$  & $1,1;$  &  &  & \tabularnewline
\multirow{1}{*}{$V$;} & \multirow{1}{*}{$(00\frac{1}{2})$;} & \multirow{1}{*}{$E$,$\mathcal{T}$;} & $R_{1};$  & $1;$  & $1,1;$  &  &  & \tabularnewline
\multirow{1}{*}{$\Lambda$;} & \multirow{1}{*}{$\text{\ensuremath{\Gamma}Z}$;} & \multirow{1}{*}{$E$,$\mathcal{T}\sigma_{z}$;} & $R_{1};$  & $1;$  & $1,1;$  &  &  & \tabularnewline
\multirow{1}{*}{$U$;} & \multirow{1}{*}{$\text{AM}$;} & \multirow{1}{*}{$E$,$\mathcal{T}\sigma_{z}$;} & $\{R_{1},R_{1}\};$  & $2;$  & $\sigma_{0},-i\sigma_{2};$  & WNL; & $\pi$  & \tabularnewline
\end{tabular}
\end{table}

}
{
\global\long\def\arraystretch{1.4}%
\begin{table}[H]
\raggedright{}%
\begin{tabular}{lllllllll}
\multicolumn{8}{l}{SG 10} & \tabularnewline
\hline
\multicolumn{8}{l}{$\Gamma_{m}$; $\{C_{2z}|000\},\{I|000\},\mathcal{T}$; Centrosymmetric;
without SOC } & \tabularnewline
\multicolumn{8}{l}{ } & \tabularnewline
\multicolumn{1}{l}{$\Gamma$;} & \multicolumn{1}{l}{$(000)$;} & \multicolumn{1}{l}{$C_{2z}$,$I$,$\mathcal{T}$;} & $R_{1};$  & $1;$  & $1,1,1;$  &  &  & \tabularnewline
 &  &  & $R_{2};$  & $1;$  & $1,-1,1;$  &  &  & \tabularnewline
 &  &  & $R_{3};$  & $1;$  & $-1,1,1;$  &  &  & \tabularnewline
 &  &  & $R_{4};$  & $1;$  & $-1,-1,1;$  &  &  & \tabularnewline
\multicolumn{1}{l}{$B$;} & \multicolumn{1}{l}{$(\bar{\frac{1}{2}}00)$;} & \multicolumn{1}{l}{$C_{2z}$,$I$,$\mathcal{T}$;} & $R_{1};$  & $1;$  & $1,1,1;$  &  &  & \tabularnewline
\end{tabular}
\end{table}

}

\subsubsection{SG 11-20}

{
\global\long\def\arraystretch{1.4}%
\begin{table}[H]
\raggedright{}%
\begin{tabular}{lllllllll}
\multicolumn{8}{l}{SG 11} & \tabularnewline
\hline
\multicolumn{8}{l}{$\Gamma_{m}$; $\{C_{2z}|00\frac{1}{2}\},\{I|00\frac{1}{2}\},\mathcal{T}$;
Centrosymmetric; without SOC } & \tabularnewline
\multicolumn{8}{l}{ } & \tabularnewline
\multicolumn{1}{l}{$\Gamma$;} & \multicolumn{1}{l}{$(000)$;} & \multicolumn{1}{l}{$C_{2z}$,$I$,$\mathcal{T}$;} & $R_{1};$  & $1;$  & $1,1,1;$  &  &  & \tabularnewline
 &  &  & $R_{2};$  & $1;$  & $1,-1,1;$  &  &  & \tabularnewline
 &  &  & $R_{3};$  & $1;$  & $-1,1,1;$  &  &  & \tabularnewline
 &  &  & $R_{4};$  & $1;$  & $-1,-1,1;$  &  &  & \tabularnewline
\multicolumn{1}{l}{$B$;} & \multicolumn{1}{l}{$(\bar{\frac{1}{2}}00)$;} & \multicolumn{1}{l}{$C_{2z}$,$I$,$\mathcal{T}$;} & $R_{1};$  & $1;$  & $1,1,1;$  &  &  & \tabularnewline
\end{tabular}
\end{table}

}

{
\global\long\def\arraystretch{1.4}%
\begin{table}[H]
\raggedright{}%
\begin{tabular}{lllllllll}
\multicolumn{8}{l}{SG 12} & \tabularnewline
\hline
\multicolumn{8}{l}{$\Gamma_{m}^{b}$; $\{C_{2z}|000\},\{I|000\},\mathcal{T}$; Centrosymmetric;
without SOC } & \tabularnewline
\multicolumn{8}{l}{} & \tabularnewline
\multicolumn{1}{l}{$\Gamma$;} & \multicolumn{1}{l}{$(000)$;} & \multicolumn{1}{l}{$C_{2z}$,$I$,$\mathcal{T}$;} & $R_{1};$  & $1;$  & $1,1,1;$  &  &  & \tabularnewline
 &  &  & $R_{2};$  & $1;$  & $1,-1,1;$  &  &  & \tabularnewline
 &  &  & $R_{3};$  & $1;$  & $-1,1,1;$  &  &  & \tabularnewline
 &  &  & $R_{4};$  & $1;$  & $-1,-1,1;$  &  &  & \tabularnewline
\multicolumn{1}{l}{$A$;} & \multicolumn{1}{l}{$(\bar{\frac{1}{2}}00)$;} & \multicolumn{1}{l}{$C_{2z}$,$I$,$\mathcal{T}$;} & $R_{1};$  & $1;$  & $1,1,1;$  &  &  & \tabularnewline
\end{tabular}
\end{table}

}

{
\global\long\def\arraystretch{1.4}%
\begin{table}[H]
\raggedright{}%
\begin{tabular}{lllllllll}
\multicolumn{8}{l}{SG 13} & \tabularnewline
\hline
\multicolumn{8}{l}{$\Gamma_{m}$; $\{C_{2z}|000\},\{I|\frac{1}{2}00\},\mathcal{T}$;
Centrosymmetric; without SOC } & \tabularnewline
\multicolumn{8}{l}{ } & \tabularnewline
\multicolumn{1}{l}{$\Gamma$;} & \multicolumn{1}{l}{$(000)$;} & \multicolumn{1}{l}{$C_{2z}$,$I$,$\mathcal{T}$;} & $R_{1};$  & $1;$  & $1,1,1;$  &  &  & \tabularnewline
 &  &  & $R_{2};$  & $1;$  & $1,-1,1;$  &  &  & \tabularnewline
 &  &  & $R_{3};$  & $1;$  & $-1,1,1;$  &  &  & \tabularnewline
 &  &  & $R_{4};$  & $1;$  & $-1,-1,1;$  &  &  & \tabularnewline
\multirow{1}{*}{$B$;} & \multirow{1}{*}{$(\bar{\frac{1}{2}}00)$;} & \multirow{1}{*}{$\sigma_{z}$,$I$,$\mathcal{T}$;} & $R_{5};$  & $2;$  & $i\sigma_{2},\sigma_{3},-\sigma_{0};$  & P-WNLs; &  & \tabularnewline
\end{tabular}
\end{table}

}

{
\global\long\def\arraystretch{1.4}%
\begin{table}[H]
\raggedright{}%
\begin{tabular}{lllllllll}
\multicolumn{8}{l}{SG 14} & \tabularnewline
\hline
\multicolumn{8}{l}{$\Gamma_{m}$; $\{C_{2z}|00\frac{1}{2}\},\{I|\frac{1}{2}0\frac{1}{2}\},\mathcal{T}$;
Centrosymmetric; without SOC } & \tabularnewline
\multicolumn{8}{l}{ } & \tabularnewline
\multicolumn{1}{l}{$\Gamma$;} & \multicolumn{1}{l}{$(000)$;} & \multicolumn{1}{l}{$C_{2z}$,$I$,$\mathcal{T}$;} & $R_{1};$  & $1;$  & $1,1,1;$  &  &  & \tabularnewline
\multicolumn{1}{l}{} &  &  & $R_{2};$  & $1;$  & $1,-1,1;$  &  &  & \tabularnewline
 &  &  & $R_{3};$  & $1;$  & $-1,1,1;$  &  &  & \tabularnewline
 &  &  & $R_{4};$  & $1;$  & $-1,-1,1;$  &  &  & \tabularnewline
\multirow{1}{*}{$B$;} & \multirow{1}{*}{$(\bar{\frac{1}{2}}00)$;} & \multirow{1}{*}{$\sigma_{z}$,$I$,$\mathcal{T}$;} & $R_{5};$  & $2;$  & $i\sigma_{2},\sigma_{3},-\sigma_{0};$  & P-WNLs; &  & \tabularnewline
\end{tabular}
\end{table}

}

{
\global\long\def\arraystretch{1.4}%
\begin{table}[H]
\raggedright{}%
\begin{tabular}{lllllllll}
\multicolumn{8}{l}{SG 15} & \tabularnewline
\hline
\multicolumn{8}{l}{$\Gamma_{m}^{b}$; $\{C_{2z}|000\},\{I|\frac{1}{2}00\},\mathcal{T}$;
Centrosymmetric; without SOC } & \tabularnewline
\multicolumn{8}{l}{ } & \tabularnewline
\multicolumn{1}{l}{$\Gamma$;} & \multicolumn{1}{l}{$(000)$;} & \multicolumn{1}{l}{$C_{2z}$,$I$,$\mathcal{T}$;} & $R_{1};$  & $1;$  & $1,1,1;$  &  &  & \tabularnewline
 &  &  & $R_{2};$  & $1;$  & $1,-1,1;$  &  &  & \tabularnewline
 &  &  & $R_{3};$  & $1;$  & $-1,1,1;$  &  &  & \tabularnewline
 &  &  & $R_{4};$  & $1;$  & $-1,-1,1;$  &  &  & \tabularnewline
\multirow{1}{*}{$A$;} & \multicolumn{1}{l}{$(\bar{\frac{1}{2}}00)$;} & \multirow{1}{*}{$\sigma_{z}$,$I$,$\mathcal{T}$;} & $R_{5};$  & $2;$  & $i\sigma_{2},\sigma_{3},-\sigma_{0};$  & P-WNLs; &  & \tabularnewline
\end{tabular}
\end{table}

}

{
\global\long\def\arraystretch{1.4}%
\begin{table}[H]
\raggedright{}%
\begin{tabular}{lllllllll}
\multicolumn{8}{l}{SG 16} & \tabularnewline
\hline
\multicolumn{8}{l}{$\Gamma_{o}$; $\{C_{2z}|000\},\{C_{2y}|000\},\mathcal{T}$; Non-Centrosymmetric;
without SOC } & \tabularnewline
\multicolumn{8}{l}{} & \tabularnewline
\multicolumn{1}{l}{$\Gamma$;} & $(000)$;  & \multicolumn{1}{l}{$C_{2z}$,$C_{2y}$,$\mathcal{T}$;} & $R_{1};$  & $1;$  & $1,1,1;$  &  &  & \tabularnewline
 &  &  & $R_{2};$  & $1;$  & $1,-1,1;$  &  &  & \tabularnewline
 &  &  & $R_{3};$  & $1;$  & $-1,1,1;$  &  &  & \tabularnewline
 &  &  & $R_{4};$  & $1;$  & $-1,-1,1;$  &  &  & \tabularnewline
\multicolumn{1}{l}{$Y$;} & \multicolumn{1}{l}{$(\bar{\frac{1}{2}}00)$;} & \multicolumn{1}{l}{$C_{2z}$,$C_{2y}$,$\mathcal{T}$;} & $R_{1};$  & $1;$  & $1,1,1;$  &  &  & \tabularnewline
\end{tabular}
\end{table}

}
{
\global\long\def\arraystretch{1.4}%
\begin{table}[H]
\raggedright{}%
\begin{tabular}{lllllllll}
\multicolumn{1}{l}{$C$;} & \multicolumn{1}{l}{$\text{YS}$;} & \multicolumn{1}{l}{$C_{2x}$,$\mathcal{T}C_{2z}$;} & $R_{1};$  & $1;$  & $1,1;$  &  &  & \tabularnewline
 &  &  & $R_{2};$  & $1;$  & $-1,1;$  &  &  & \tabularnewline
\multicolumn{1}{l}{$E$;} & \multicolumn{1}{l}{$\text{TR}$;} & \multicolumn{1}{l}{$C_{2x}$,$\mathcal{T}C_{2z}$;} & $R_{1};$  & $1;$  & $1,1;$  &  &  & \tabularnewline
 &  &  & $R_{2};$  & $1;$  & $-1,1;$  &  &  & \tabularnewline
\multicolumn{1}{l}{$A$;} & \multicolumn{1}{l}{$\text{ZU}$;} & \multicolumn{1}{l}{$C_{2x}$,$\mathcal{T}C_{2z}$;} & $R_{1};$  & $1;$  & $1,1;$  &  &  & \tabularnewline
 &  &  & $R_{2};$  & $1;$  & $-1,1;$  &  &  & \tabularnewline
$\Lambda$;  & \multicolumn{1}{l}{$\text{\ensuremath{\Gamma}Z}$;} & \multicolumn{1}{l}{$C_{2z}$,$\mathcal{T}C_{2x}$;} & $R_{1};$  & $1;$  & $1,1;$  &  &  & \tabularnewline
 &  &  & $R_{2};$  & $1;$  & $-1,1;$  &  &  & \tabularnewline
\multicolumn{1}{l}{$H$;} & $\text{YT}$;  & \multicolumn{1}{l}{$C_{2z}$,$\mathcal{T}C_{2x}$;} & $R_{1};$  & $1;$  & $1,1;$  &  &  & \tabularnewline
 &  &  & $R_{2};$  & $1;$  & $-1,1;$  &  &  & \tabularnewline
\multicolumn{1}{l}{$Q$;} & \multicolumn{1}{l}{$\text{SR}$;} & \multicolumn{1}{l}{$C_{2z}$,$\mathcal{T}C_{2x}$;} & $R_{1};$  & $1;$  & $1,1;$  &  &  & \tabularnewline
 &  &  & $R_{2};$  & $1;$  & $-1,1;$  &  &  & \tabularnewline
\multicolumn{1}{l}{$G$;} & \multicolumn{1}{l}{$\text{XU}$;} & \multicolumn{1}{l}{$C_{2z}$,$\mathcal{T}C_{2x}$;} & $R_{1};$  & $1;$  & $1,1;$  &  &  & \tabularnewline
 &  &  & $R_{2};$  & $1;$  & $-1,1;$  &  &  & \tabularnewline
\end{tabular}
\end{table}

}

{
\global\long\def\arraystretch{1.4}%
\begin{table}[H]
\raggedright{}%
\begin{tabular}{lllllllll}
\multicolumn{8}{l}{SG 17} & \tabularnewline
\hline
\multicolumn{8}{l}{$\Gamma_{o}$; $\{C_{2z}|00\frac{1}{2}\},\{C_{2y}|000\},\mathcal{T}$;
Non-Centrosymmetric; without SOC } & \tabularnewline
\multicolumn{8}{l}{} & \tabularnewline
\multicolumn{1}{l}{$\Gamma$;} & \multicolumn{1}{l}{$(000)$;} & \multicolumn{1}{l}{$C_{2z}$,$C_{2y}$,$\mathcal{T}$;} & $R_{1};$  & $1;$  & $1,1,1;$  &  &  & \tabularnewline
 &  &  & $R_{2};$  & $1;$  & $1,-1,1;$  &  &  & \tabularnewline
 &  &  & $R_{3};$  & $1;$  & $-1,1,1;$  &  &  & \tabularnewline
 &  &  & $R_{4};$  & $1;$  & $-1,-1,1;$  &  &  & \tabularnewline
\multicolumn{1}{l}{$Y$;} & \multicolumn{1}{l}{$(\bar{\frac{1}{2}}00)$;} & \multicolumn{1}{l}{$C_{2z}$,$C_{2y}$,$\mathcal{T}$;} & $R_{1};$  & $1;$  & $1,1,1;$  &  &  & \tabularnewline
\end{tabular}
\end{table}

}

{
\global\long\def\arraystretch{1.4}%
\begin{table}[H]
\raggedright{}%
\begin{tabular}{lllllllll}
\multicolumn{8}{l}{SG 18} & \tabularnewline
\hline
\multicolumn{8}{l}{$\Gamma_{o}$; $\{C_{2z}|000\},\{C_{2y}|\frac{1}{2}\frac{1}{2}0\},\mathcal{T}$;
Non-Centrosymmetric; without SOC } & \tabularnewline
\multicolumn{8}{l}{} & \tabularnewline
\multicolumn{1}{l}{$\Gamma$;} & \multicolumn{1}{l}{$(000)$;} & \multicolumn{1}{l}{$C_{2z}$,$C_{2y}$,$\mathcal{T}$;} & $R_{1};$  & $1;$  & $1,1,1;$  &  &  & \tabularnewline
 &  &  & $R_{2};$  & $1;$  & $1,-1,1;$  &  &  & \tabularnewline
 &  &  & $R_{3};$  & $1;$  & $-1,1,1;$  &  &  & \tabularnewline
 &  &  & $R_{4};$  & $1;$  & $-1,-1,1;$  &  &  & \tabularnewline
\multicolumn{1}{l}{$Y$;} & \multirow{1}{*}{$(\bar{\frac{1}{2}}00)$;} & \multirow{1}{*}{$C_{2y}$,$C_{2z}$,$\mathcal{T}$;} & $R_{5};$  & $2;$  & $i\sigma_{2},\sigma_{3},-\sigma_{0};$  & P-NS$_{YTSR}$;  &  & \tabularnewline
\end{tabular}
\end{table}

}

{
\global\long\def\arraystretch{1.4}%
\begin{table}[H]
\raggedright{}%
\begin{tabular}{lllllllll}
\multicolumn{8}{l}{SG 19} & \tabularnewline
\hline
\multicolumn{8}{l}{$\Gamma_{o}$; $\{C_{2z}|\frac{1}{2}0\frac{1}{2}\},\{C_{2y}|\frac{1}{2}\frac{1}{2}0\},\mathcal{T}$;
Non-Centrosymmetric; without SOC } & \tabularnewline
\multicolumn{8}{l}{} & \tabularnewline
\multicolumn{1}{l}{$\Gamma$;} & \multicolumn{1}{l}{$(000)$;} & \multicolumn{1}{l}{$C_{2z}$,$C_{2y}$,$\mathcal{T}$;} & $R_{1};$  & $1;$  & $1,1,1;$  &  &  & \tabularnewline
 &  &  & $R_{2};$  & $1;$  & $1,-1,1;$  &  &  & \tabularnewline
 &  &  & $R_{3};$  & $1;$  & $-1,1,1;$  &  &  & \tabularnewline
 &  &  & $R_{4};$  & $1;$  & $-1,-1,1;$  &  &  & \tabularnewline
\multicolumn{1}{l}{$Y$;} & \multicolumn{1}{l}{$(\bar{\frac{1}{2}}00)$;} & \multirow{1}{*}{$C_{2y}$,$C_{2z}$,$\mathcal{T}$;} & $R_{5};$  & $2;$  & $i\sigma_{2},\sigma_{3},-\sigma_{0};$  & P-NS$_{YTSR}$;  &  & \tabularnewline
\end{tabular}
\end{table}

}

{
\global\long\def\arraystretch{1.4}%
\begin{table}[H]
\raggedright{}%
\begin{tabular}{lllllllll}
\multicolumn{8}{l}{SG 20} & \tabularnewline
\hline
\multicolumn{8}{l}{$\Gamma_{o}^{b}$; $\{C_{2z}|00\frac{1}{2}\},\{C_{2y}|00\frac{1}{2}\},\mathcal{T}$;
Non-Centrosymmetric; without SOC } & \tabularnewline
\multicolumn{8}{l}{} & \tabularnewline
\multicolumn{1}{l}{$\Gamma$;} & \multicolumn{1}{l}{$(000)$;} & \multicolumn{1}{l}{$C_{2z}$,$C_{2y}$,$\mathcal{T}$;} & $R_{1};$  & $1;$  & $1,1,1;$  &  &  & \tabularnewline
 &  &  & $R_{2};$  & $1;$  & $1,-1,1;$  &  &  & \tabularnewline
 &  &  & $R_{3};$  & $1;$  & $-1,1,1;$  &  &  & \tabularnewline
 &  &  & $R_{4};$  & $1;$  & $-1,-1,1;$  &  &  & \tabularnewline
\multicolumn{1}{l}{$Y$;} & \multicolumn{1}{l}{$(\frac{1}{2}\frac{1}{2}0)$;} & \multicolumn{1}{l}{$C_{2z}$,$C_{2y}$,$\mathcal{T}$;} & $R_{1};$  & $1;$  & $1,1,1;$  &  &  & \tabularnewline
\end{tabular}
\end{table}

}
\subsubsection{SG 21-30}
{
\global\long\def\arraystretch{1.4}%
\begin{table}[H]
\raggedright{}%
\begin{tabular}{lllllllll}
\multicolumn{8}{l}{SG 21} & \tabularnewline
\hline
\multicolumn{8}{l}{$\Gamma_{o}^{b}$; $\{C_{2z}|000\},\{C_{2y}|000\},\mathcal{T}$; Non-Centrosymmetric;
without SOC } & \tabularnewline
\multicolumn{8}{l}{} & \tabularnewline
\multicolumn{1}{l}{$\Gamma$;} & \multicolumn{1}{l}{$(000)$;} & \multicolumn{1}{l}{$C_{2z}$,$C_{2y}$,$\mathcal{T}$;} & $R_{1};$  & $1;$  & $1,1,1;$  &  &  & \tabularnewline
 &  &  & $R_{2};$  & $1;$  & $1,-1,1;$  &  &  & \tabularnewline
 &  &  & $R_{3};$  & $1;$  & $-1,1,1;$  &  &  & \tabularnewline
 &  &  & $R_{4};$  & $1;$  & $-1,-1,1;$  &  &  & \tabularnewline
\multicolumn{1}{l}{$Y$;} & \multicolumn{1}{l}{$(\frac{1}{2}\frac{1}{2}0)$;} & \multicolumn{1}{l}{$C_{2z}$,$C_{2y}$,$\mathcal{T}$;} & $R_{1};$  & $1;$  & $1,1,1;$  &  &  & \tabularnewline
\end{tabular}
\end{table}

}

{
\global\long\def\arraystretch{1.4}%
\begin{table}[H]
\raggedright{}%
\begin{tabular}{lllllllll}
\multicolumn{8}{l}{SG 22} & \tabularnewline
\hline
\multicolumn{8}{l}{$\Gamma_{o}^{f}$; $\{C_{2z}|000\},\{C_{2y}|000\},\mathcal{T}$; Non-Centrosymmetric;
without SOC } & \tabularnewline
\multicolumn{8}{l}{ } & \tabularnewline
\multicolumn{1}{l}{$\Gamma$;} & \multicolumn{1}{l}{$(000)$;} & \multicolumn{1}{l}{$C_{2z}$,$C_{2y}$,$\mathcal{T}$;} & $R_{1};$  & $1;$  & $1,1,1;$  &  &  & \tabularnewline
 &  &  & $R_{2};$  & $1;$  & $1,-1,1;$  &  &  & \tabularnewline
 &  &  & $R_{3};$  & $1;$  & $-1,1,1;$  &  &  & \tabularnewline
 &  &  & $R_{4};$  & $1;$  & $-1,-1,1;$  &  &  & \tabularnewline
\multicolumn{1}{l}{$Y$;} & \multicolumn{1}{l}{$(0\bar{\frac{1}{2}}\bar{\frac{1}{2}})$;} & \multicolumn{1}{l}{$C_{2z}$,$C_{2y}$,$\mathcal{T}$;} & $R_{1};$  & $1;$  & $1,1,1;$  &  &  & \tabularnewline
\end{tabular}
\end{table}

}

{
\global\long\def\arraystretch{1.4}%
\begin{table}[H]
\raggedright{}%
\begin{tabular}{lllllllll}
\multicolumn{8}{l}{SG 23} & \tabularnewline
\hline
\multicolumn{8}{l}{$\Gamma_{o}^{v}$; $\{C_{2z}|000\},\{C_{2y}|000\},\mathcal{T}$; Non-Centrosymmetric;
without SOC } & \tabularnewline
\multicolumn{8}{l}{} & \tabularnewline
\multicolumn{1}{l}{$\Gamma$;} & \multicolumn{1}{l}{$(000)$;} & \multicolumn{1}{l}{$C_{2z}$,$C_{2y}$,$\mathcal{T}$;} & $R_{1};$  & $1;$  & $1,1,1;$  &  &  & \tabularnewline
 &  &  & $R_{2};$  & $1;$  & $1,-1,1;$  &  &  & \tabularnewline
 &  &  & $R_{3};$  & $1;$  & $-1,1,1;$  &  &  & \tabularnewline
 &  &  & $R_{4};$  & $1;$  & $-1,-1,1;$  &  &  & \tabularnewline
\multicolumn{1}{l}{$X$;} & \multicolumn{1}{l}{$(\frac{1}{2}\bar{\frac{1}{2}}\frac{1}{2})$;} & \multicolumn{1}{l}{$C_{2z}$,$C_{2y}$,$\mathcal{T}$;} & $R_{1};$  & $1;$  & $1,1,1;$  &  &  & \tabularnewline
\end{tabular}
\end{table}

}

{
\global\long\def\arraystretch{1.4}%
\begin{table}[H]
\raggedright{}%
\begin{tabular}{lllllllll}
\multicolumn{8}{l}{SG 24} & \tabularnewline
\hline
\multicolumn{8}{l}{$\Gamma_{o}^{v}$; $\{C_{2z}|\frac{1}{2}0\frac{1}{2}\},\{C_{2y}|\frac{1}{2}\frac{1}{2}0\},\mathcal{T}$;
Non-Centrosymmetric; without SOC } & \tabularnewline
\multicolumn{8}{l}{ } & \tabularnewline
\multicolumn{1}{l}{$\Gamma$;} & \multicolumn{1}{l}{$(000)$;} & \multicolumn{1}{l}{$C_{2z}$,$C_{2y}$,$\mathcal{T}$;} & $R_{1};$  & $1;$  & $1,1,1;$  &  &  & \tabularnewline
 &  &  & $R_{2};$  & $1;$  & $1,-1,1;$  &  &  & \tabularnewline
 &  &  & $R_{3};$  & $1;$  & $-1,1,1;$  &  &  & \tabularnewline
 &  &  & $R_{4};$  & $1;$  & $-1,-1,1;$  &  &  & \tabularnewline
\multicolumn{1}{l}{$X$;} & \multicolumn{1}{l}{$(\frac{1}{2}\bar{\frac{1}{2}}\frac{1}{2})$;} & \multicolumn{1}{l}{$C_{2z}$,$C_{2y}$,$\mathcal{T}$;} & $R_{1};$  & $1;$  & $1,1,1;$  &  &  & \tabularnewline
\end{tabular}
\end{table}

}

{
\global\long\def\arraystretch{1.4}%
\begin{table}[H]
\raggedright{}%
\begin{tabular}{lllllllll}
\multicolumn{8}{l}{SG 25} & \tabularnewline
\hline
\multicolumn{8}{l}{$\Gamma_{o}$; $\{C_{2z}|000\},\{\sigma_{y}|000\},\mathcal{T}$; Non-Centrosymmetric;
without SOC } & \tabularnewline
\multicolumn{8}{l}{ } & \tabularnewline
\multicolumn{1}{l}{$\Gamma$;} & \multicolumn{1}{l}{$(000)$;} & \multicolumn{1}{l}{$C_{2z}$,$\sigma_{y}$,$\mathcal{T}$;} & $R_{1};$  & $1;$  & $1,1,1;$  &  &  & \tabularnewline
 &  &  & $R_{2};$  & $1;$  & $1,-1,1;$  &  &  & \tabularnewline
 &  &  & $R_{3};$  & $1;$  & $-1,1,1;$  &  &  & \tabularnewline
 &  &  & $R_{4};$  & $1;$  & $-1,-1,1;$  &  &  & \tabularnewline
\multicolumn{1}{l}{$Y$;} & \multicolumn{1}{l}{$(\bar{\frac{1}{2}}00)$;} & \multicolumn{1}{l}{$C_{2z}$,$\sigma_{y}$,$\mathcal{T}$;} & $R_{1};$  & $1;$  & $1,1,1;$  &  &  & \tabularnewline
\end{tabular}
\end{table}

}
{
\global\long\def\arraystretch{1.4}%
\begin{table}[H]
\raggedright{}%
\begin{tabular}{lllllllll}
\multicolumn{1}{l}{$A$;} & \multicolumn{1}{l}{$\text{ZU}$;} & \multicolumn{1}{l}{$\sigma_{y}$,$\mathcal{T}\sigma_{x}$;} & $R_{1};$  & $1;$  & $1,1;$  &  &  & \tabularnewline
 &  &  & $R_{2};$  & $1;$  & $-1,1;$  &  &  & \tabularnewline
\multicolumn{1}{l}{$\Lambda$;} & \multicolumn{1}{l}{$\text{\ensuremath{\Gamma}Z}$;} & \multicolumn{1}{l}{$C_{2z}$,$\sigma_{y}$;} & $R_{1};$  & $1;$  & $1,1;$  &  &  & \tabularnewline
 &  &  & $R_{2};$  & $1;$  & $1,-1;$  &  &  & \tabularnewline
 &  &  & $R_{3};$  & $1;$  & $-1,1;$  &  &  & \tabularnewline
 &  &  & $R_{4};$  & $1;$  & $-1,-1;$  &  &  & \tabularnewline
\multicolumn{1}{l}{$H$;} & \multicolumn{1}{l}{$\text{YT}$;} & \multicolumn{1}{l}{$C_{2z}$,$\sigma_{y}$;} & $R_{1};$  & $1;$  & $1,1;$  &  &  & \tabularnewline
 &  &  & $R_{2};$  & $1;$  & $1,-1;$  &  &  & \tabularnewline
 &  &  & $R_{3};$  & $1;$  & $-1,1;$  &  &  & \tabularnewline
 &  &  & $R_{4};$  & $1;$  & $-1,-1;$  &  &  & \tabularnewline
\multicolumn{1}{l}{$Q$;} & \multicolumn{1}{l}{$\text{SR}$;} & \multicolumn{1}{l}{$C_{2z}$,$\sigma_{y}$;} & $R_{1};$  & $1;$  & $1,1;$  &  &  & \tabularnewline
 &  &  & $R_{2};$  & $1;$  & $1,-1;$  &  &  & \tabularnewline
 &  &  & $R_{3};$  & $1;$  & $-1,1;$  &  &  & \tabularnewline
 &  &  & $R_{4};$  & $1;$  & $-1,-1;$  &  &  & \tabularnewline
\multicolumn{1}{l}{$G$;} & \multicolumn{1}{l}{$\text{XU}$;} & \multicolumn{1}{l}{$C_{2z}$,$\sigma_{y}$;} & $R_{1};$  & $1;$  & $1,1;$  &  &  & \tabularnewline
 &  &  & $R_{2};$  & $1;$  & $1,-1;$  &  &  & \tabularnewline
 &  &  & $R_{3};$  & $1;$  & $-1,1;$  &  &  & \tabularnewline
 &  &  & $R_{4};$  & $1;$  & $-1,-1;$  &  &  & \tabularnewline
\end{tabular}
\end{table}

}

{
\global\long\def\arraystretch{1.4}%
\begin{table}[H]
\raggedright{}%
\begin{tabular}{lllllllll}
\multicolumn{8}{l}{SG 26} & \tabularnewline
\hline
\multicolumn{8}{l}{$\Gamma_{o}$; $\{C_{2z}|00\frac{1}{2}\},\{\sigma_{y}|00\frac{1}{2}\},\mathcal{T}$;
Non-Centrosymmetric; without SOC } & \tabularnewline
\multicolumn{8}{l}{} & \tabularnewline
\multicolumn{1}{l}{$\Gamma$;} & \multicolumn{1}{l}{$(000)$;} & \multicolumn{1}{l}{$C_{2z}$,$\sigma_{y}$,$\mathcal{T}$;} & $R_{1};$  & $1;$  & $1,1,1;$  &  &  & \tabularnewline
 &  &  & $R_{2};$  & $1;$  & $1,-1,1;$  &  &  & \tabularnewline
 &  &  & $R_{3};$  & $1;$  & $-1,1,1;$  &  &  & \tabularnewline
 &  &  & $R_{4};$  & $1;$  & $-1,-1,1;$  &  &  & \tabularnewline
\multicolumn{1}{l}{$Y$;} & \multicolumn{1}{l}{$(\bar{\frac{1}{2}}00)$;} & \multicolumn{1}{l}{$C_{2z}$,$\sigma_{y}$,$\mathcal{T}$;} & $R_{1};$  & $1;$  & $1,1,1;$  &  &  & \tabularnewline
\end{tabular}
\end{table}

}
{
\global\long\def\arraystretch{1.4}%
\begin{table}[H]
\raggedright{}%
\begin{tabular}{lllllllll}
\multicolumn{1}{l}{$Q$;} & \multicolumn{1}{l}{$\text{SR}$;} & \multicolumn{1}{l}{$C_{2z}$,$\sigma_{y}$;} & $R_{1};$  & $1;$  & $1,1;$  &  &  & \tabularnewline
 &  &  & $R_{2};$  & $1;$  & $1,-1;$  &  &  & \tabularnewline
 &  &  & $R_{3};$  & $1;$  & $-1,1;$  &  &  & \tabularnewline
 &  &  & $R_{4};$  & $1;$  & $-1,-1;$  &  &  & \tabularnewline
\multicolumn{1}{l}{$G$;} & \multicolumn{1}{l}{$\text{XU}$;} & \multicolumn{1}{l}{$C_{2z}$,$\sigma_{y}$;} & $R_{1};$  & $1;$  & $1,1;$  &  &  & \tabularnewline
 &  &  & $R_{2};$  & $1;$  & $1,-1;$  &  &  & \tabularnewline
 &  &  & $R_{3};$  & $1;$  & $-1,1;$  &  &  & \tabularnewline
 &  &  & $R_{4};$  & $1;$  & $-1,-1;$  &  &  & \tabularnewline
\end{tabular}
\end{table}

}

{
\global\long\def\arraystretch{1.4}%
\begin{table}[H]
\raggedright{}%
\begin{tabular}{lllllllll}
\multicolumn{8}{l}{SG 27} & \tabularnewline
\hline
\multicolumn{8}{l}{$\Gamma_{o}$; $\{C_{2z}|000\},\{\sigma_{y}|00\frac{1}{2}\},\mathcal{T}$;
Non-Centrosymmetric; without SOC } & \tabularnewline
\multicolumn{8}{l}{ } & \tabularnewline
\multicolumn{1}{l}{$\Gamma$;} & \multicolumn{1}{l}{$(000)$;} & \multicolumn{1}{l}{$C_{2z}$,$\sigma_{y}$,$\mathcal{T}$;} & $R_{1};$  & $1;$  & $1,1,1;$  &  &  & \tabularnewline
 &  &  & $R_{2};$  & $1;$  & $1,-1,1;$  &  &  & \tabularnewline
 &  &  & $R_{3};$  & $1;$  & $-1,1,1;$  &  &  & \tabularnewline
 &  &  & $R_{4};$  & $1;$  & $-1,-1,1;$  &  &  & \tabularnewline
\multicolumn{1}{l}{$Y$;} & \multicolumn{1}{l}{$(\bar{\frac{1}{2}}00)$;} & \multicolumn{1}{l}{$C_{2z}$,$\sigma_{y}$,$\mathcal{T}$;} & $R_{1};$  & $1;$  & $1,1,1;$  &  &  & \tabularnewline
\end{tabular}
\end{table}

}
{
\global\long\def\arraystretch{1.4}%
\begin{table}[H]
\raggedright{}%
\begin{tabular}{lllllllll}
\multicolumn{1}{l}{$Q$;} & \multicolumn{1}{l}{$\text{SR}$;} & \multicolumn{1}{l}{$C_{2z}$,$\sigma_{y}$;} & $R_{1};$  & $1;$  & $1,1;$  &  &  & \tabularnewline
 &  &  & $R_{2};$  & $1;$  & $1,-1;$  &  &  & \tabularnewline
 &  &  & $R_{3};$  & $1;$  & $-1,1;$  &  &  & \tabularnewline
 &  &  & $R_{4};$  & $1;$  & $-1,-1;$  &  &  & \tabularnewline
\multicolumn{1}{l}{$G$;} & \multicolumn{1}{l}{$\text{XU}$;} & \multicolumn{1}{l}{$C_{2z}$,$\sigma_{y}$;} & $R_{1};$  & $1;$  & $1,1;$  &  &  & \tabularnewline
 &  &  & $R_{2};$  & $1;$  & $1,-1;$  &  &  & \tabularnewline
 &  &  & $R_{3};$  & $1;$  & $-1,1;$  &  &  & \tabularnewline
 &  &  & $R_{4};$  & $1;$  & $-1,-1;$  &  &  & \tabularnewline
\end{tabular}
\end{table}

}

{
\global\long\def\arraystretch{1.4}%
\begin{table}[H]
\raggedright{}%
\begin{tabular}{lllllllll}
\multicolumn{8}{l}{SG 28} & \tabularnewline
\hline
\multicolumn{8}{l}{$\Gamma_{o}$; $\{C_{2z}|\frac{1}{2}00\},\{\sigma_{y}|000\},\mathcal{T}$;
Non-Centrosymmetric; without SOC } & \tabularnewline
\multicolumn{8}{l}{} & \tabularnewline
\multicolumn{1}{l}{$\Gamma$;} & \multicolumn{1}{l}{$(000)$;} & \multicolumn{1}{l}{$C_{2z}$,$\sigma_{y}$,$\mathcal{T}$;} & $R_{1};$  & $1;$  & $1,1,1;$  &  &  & \tabularnewline
 &  &  & $R_{2};$  & $1;$  & $1,-1,1;$  &  &  & \tabularnewline
 &  &  & $R_{3};$  & $1;$  & $-1,1,1;$  &  &  & \tabularnewline
 &  &  & $R_{4};$  & $1;$  & $-1,-1,1;$  &  &  & \tabularnewline
\multicolumn{1}{l}{$Y$;} & \multirow{1}{*}{$(\bar{\frac{1}{2}}00)$;} & \multirow{1}{*}{$\sigma_{x}$,$\sigma_{y}$,$\mathcal{T}$;} & $R_{5};$  & $2;$  & $i\sigma_{2},\sigma_{3},-\sigma_{0};$  & P-WNLs; &  & \tabularnewline
\end{tabular}
\end{table}

}

{
\global\long\def\arraystretch{1.4}%
\begin{table}[H]
\raggedright{}%
\begin{tabular}{lllllllll}
\multicolumn{8}{l}{SG 29} & \tabularnewline
\hline
\multicolumn{8}{l}{$\Gamma_{o}$; $\{C_{2z}|\frac{1}{2}0\frac{1}{2}\},\{\sigma_{y}|00\frac{1}{2}\},\mathcal{T}$;
Non-Centrosymmetric; without SOC } & \tabularnewline
\multicolumn{8}{l}{} & \tabularnewline
\multicolumn{1}{l}{$\Gamma$;} & \multicolumn{1}{l}{$(000)$;} & \multicolumn{1}{l}{$C_{2z}$,$\sigma_{y}$,$\mathcal{T}$;} & $R_{1};$  & $1;$  & $1,1,1;$  &  &  & \tabularnewline
 &  &  & $R_{2};$  & $1;$  & $1,-1,1;$  &  &  & \tabularnewline
 &  &  & $R_{3};$  & $1;$  & $-1,1,1;$  &  &  & \tabularnewline
 &  &  & $R_{4};$  & $1;$  & $-1,-1,1;$  &  &  & \tabularnewline
\multicolumn{1}{l}{$Y$;} & \multirow{1}{*}{$(\bar{\frac{1}{2}}00)$;} & \multirow{1}{*}{$\sigma_{x}$,$C_{2z}$,$\mathcal{T}$;} & $R_{5};$  & $2;$  & $i\sigma_{2},\sigma_{3},-\sigma_{0};$  & P-WNLs; &  & \tabularnewline
\end{tabular}
\end{table}

}

{
\global\long\def\arraystretch{1.4}%
\begin{table}[H]
\raggedright{}%
\begin{tabular}{lllllllll}
\multicolumn{8}{l}{SG 30} & \tabularnewline
\hline
\multicolumn{8}{l}{$\Gamma_{o}$; $\{C_{2z}|\frac{1}{2}00\},\{\sigma_{y}|00\frac{1}{2}\},\mathcal{T}$;
Non-Centrosymmetric; without SOC } & \tabularnewline
\multicolumn{8}{l}{} & \tabularnewline
\multicolumn{1}{l}{$\Gamma$;} & \multicolumn{1}{l}{$(000)$;} & \multicolumn{1}{l}{$C_{2z}$,$\sigma_{y}$,$\mathcal{T}$;} & $R_{1};$  & $1;$  & $1,1,1;$  &  &  & \tabularnewline
 &  &  & $R_{2};$  & $1;$  & $1,-1,1;$  &  &  & \tabularnewline
 &  &  & $R_{3};$  & $1;$  & $-1,1,1;$  &  &  & \tabularnewline
 &  &  & $R_{4};$  & $1;$  & $-1,-1,1;$  &  &  & \tabularnewline
\multicolumn{1}{l}{$Y$;} & \multirow{1}{*}{$(\bar{\frac{1}{2}}00)$;} & \multirow{1}{*}{$\sigma_{x}$,$\sigma_{y}$,$\mathcal{T}$;} & $R_{5};$  & $2;$  & $i\sigma_{2},\sigma_{3},-\sigma_{0};$  & P-WNLs; &  & \tabularnewline
\end{tabular}
\end{table}

}
\subsubsection{SG 31-40}

{
\global\long\def\arraystretch{1.4}%
\begin{table}[H]
\raggedright{}%
\begin{tabular}{lllllllll}
\multicolumn{8}{l}{SG 31} & \tabularnewline
\hline
\multicolumn{8}{l}{$\Gamma_{o}$; $\{C_{2z}|\frac{1}{2}0\frac{1}{2}\},\{\sigma_{y}|000\},\mathcal{T}$;
Non-Centrosymmetric; without SOC } & \tabularnewline
\multicolumn{8}{l}{ } & \tabularnewline
\multirow{1}{*}{$\Gamma$;} & \multirow{1}{*}{$(000)$;} & \multirow{1}{*}{$C_{2z}$,$\sigma_{y}$,$\mathcal{T}$;} & $R_{1};$  & $1;$  & $1,1,1;$  &  &  & \tabularnewline
 &  &  & $R_{2};$  & $1;$  & $1,-1,1;$  &  &  & \tabularnewline
 &  &  & $R_{3};$  & $1;$  & $-1,1,1;$  &  &  & \tabularnewline
 &  &  & $R_{4};$  & $1;$  & $-1,-1,1;$  &  &  & \tabularnewline
\multirow{1}{*}{$Y$;} & \multirow{1}{*}{$(\bar{\frac{1}{2}}00)$;} & \multirow{1}{*}{$\sigma_{x}$,$\sigma_{y}$,$\mathcal{T}$;} & $R_{5};$  & $2;$  & $i\sigma_{2},\sigma_{3},-\sigma_{0};$  & P-WNLs; &  & \tabularnewline
\end{tabular}
\end{table}

}

{
\global\long\def\arraystretch{1.4}%
\begin{table}[H]
\raggedright{}%
\begin{tabular}{lllllllll}
\multicolumn{8}{l}{SG 32} & \tabularnewline
\hline
\multicolumn{8}{l}{$\Gamma_{o}$; $\{C_{2z}|\frac{1}{2}\frac{1}{2}0\},\{\sigma_{y}|0\frac{1}{2}0\},\mathcal{T}$;
Non-Centrosymmetric; without SOC } & \tabularnewline
\multicolumn{8}{l}{ } & \tabularnewline
\multirow{1}{*}{$\Gamma$;} & \multirow{1}{*}{$(000)$;} & \multirow{1}{*}{$C_{2z}$,$\sigma_{y}$,$\mathcal{T}$;} & $R_{1};$  & $1;$  & $1,1,1;$  &  &  & \tabularnewline
 &  &  & $R_{2};$  & $1;$  & $1,-1,1;$  &  &  & \tabularnewline
 &  &  & $R_{3};$  & $1;$  & $-1,1,1;$  &  &  & \tabularnewline
 &  &  & $R_{4};$  & $1;$  & $-1,-1,1;$  &  &  & \tabularnewline
\multirow{1}{*}{$Y$;} & \multirow{1}{*}{$(\bar{\frac{1}{2}}00)$;} & \multirow{1}{*}{$\sigma_{x}$,$C_{2z}$,$\mathcal{T}$;} & $R_{5};$  & $2;$  & $i\sigma_{2},\sigma_{3},-\sigma_{0};$  & P-WNLs; &  & \tabularnewline
\end{tabular}
\end{table}

}

{
\global\long\def\arraystretch{1.4}%
\begin{table}[H]
\raggedright{}%
\begin{tabular}{lllllllll}
\multicolumn{8}{l}{SG 33} & \tabularnewline
\hline
\multicolumn{8}{l}{$\Gamma_{o}$; $\{C_{2z}|\frac{1}{2}\frac{1}{2}\frac{1}{2}\},\{\sigma_{y}|0\frac{1}{2}\frac{1}{2}\},\mathcal{T}$;
Non-Centrosymmetric; without SOC } & \tabularnewline
\multicolumn{8}{l}{ } & \tabularnewline
\multirow{1}{*}{$\Gamma$;} & \multirow{1}{*}{$(000)$;} & \multirow{1}{*}{$C_{2z}$,$\sigma_{y}$,$\mathcal{T}$;} & $R_{1};$  & $1;$  & $1,1,1;$  &  &  & \tabularnewline
 &  &  & $R_{2};$  & $1;$  & $1,-1,1;$  &  &  & \tabularnewline
 &  &  & $R_{3};$  & $1;$  & $-1,1,1;$  &  &  & \tabularnewline
 &  &  & $R_{4};$  & $1;$  & $-1,-1,1;$  &  &  & \tabularnewline
\multirow{1}{*}{$Y$;} & \multirow{1}{*}{$(\bar{\frac{1}{2}}00)$;} & \multirow{1}{*}{$\sigma_{x}$,$C_{2z}$,$\mathcal{T}$;} & $R_{5};$  & $2;$  & $i\sigma_{2},\sigma_{3},-\sigma_{0};$  & P-WNLs; &  & \tabularnewline
\end{tabular}
\end{table}

}

{
\global\long\def\arraystretch{1.4}%
\begin{table}[H]
\raggedright{}%
\begin{tabular}{lllllllll}
\multicolumn{8}{l}{SG 34} & \tabularnewline
\hline
\multicolumn{8}{l}{$\Gamma_{o}$; $\{C_{2z}|\frac{1}{2}\frac{1}{2}0\},\{\sigma_{y}|0\frac{1}{2}\frac{1}{2}\},\mathcal{T}$;
Non-Centrosymmetric; without SOC } & \tabularnewline
\multicolumn{8}{l}{ } & \tabularnewline
\multirow{1}{*}{$\Gamma$;} & \multirow{1}{*}{$(000)$;} & \multirow{1}{*}{$C_{2z}$,$\sigma_{y}$,$\mathcal{T}$;} & $R_{1};$  & $1;$  & $1,1,1;$  &  &  & \tabularnewline
 &  &  & $R_{2};$  & $1;$  & $1,-1,1;$  &  &  & \tabularnewline
 &  &  & $R_{3};$  & $1;$  & $-1,1,1;$  &  &  & \tabularnewline
 &  &  & $R_{4};$  & $1;$  & $-1,-1,1;$  &  &  & \tabularnewline
\multirow{1}{*}{$Y$;} & \multirow{1}{*}{$(\bar{\frac{1}{2}}00)$;} & \multirow{1}{*}{$\sigma_{x}$,$C_{2z}$,$\mathcal{T}$;} & $R_{5};$  & $2;$  & $i\sigma_{2},\sigma_{3},-\sigma_{0};$  & P-WNLs; &  & \tabularnewline
\end{tabular}
\end{table}

}

{
\global\long\def\arraystretch{1.4}%
\begin{table}[H]
\raggedright{}%
\begin{tabular}{lllllllll}
\multicolumn{8}{l}{SG 35} & \tabularnewline
\hline
\multicolumn{8}{l}{$\Gamma_{o}^{b}$; $\{C_{2z}|000\},\{\sigma_{y}|000\},\mathcal{T}$;
Non-Centrosymmetric; without SOC } & \tabularnewline
\multicolumn{8}{l}{ } & \tabularnewline
\multirow{1}{*}{$\Gamma$;} & \multirow{1}{*}{$(000)$;} & \multirow{1}{*}{$C_{2z}$,$\sigma_{y}$,$\mathcal{T}$;} & $R_{1};$  & $1;$  & $1,1,1;$  &  &  & \tabularnewline
 &  &  & $R_{2};$  & $1;$  & $1,-1,1;$  &  &  & \tabularnewline
 &  &  & $R_{3};$  & $1;$  & $-1,1,1;$  &  &  & \tabularnewline
 &  &  & $R_{4};$  & $1;$  & $-1,-1,1;$  &  &  & \tabularnewline
\multirow{1}{*}{$Y$;} & \multirow{1}{*}{$(\frac{1}{2}\frac{1}{2}0)$;} & \multirow{1}{*}{$C_{2z}$,$\sigma_{y}$,$\mathcal{T}$;} & $R_{1};$  & $1;$  & $1,1,1;$  &  &  & \tabularnewline
\end{tabular}
\end{table}

}

{
\global\long\def\arraystretch{1.4}%
\begin{table}[H]
\raggedright{}%
\begin{tabular}{lllllllll}
\multicolumn{8}{l}{SG 36} & \tabularnewline
\hline
\multicolumn{8}{l}{$\Gamma_{o}^{b}$; $\{C_{2z}|00\frac{1}{2}\},\{\sigma_{y}|000\},\mathcal{T}$;
Non-Centrosymmetric; without SOC } & \tabularnewline
\multicolumn{8}{l}{ } & \tabularnewline
\multirow{1}{*}{$\Gamma$;} & \multirow{1}{*}{$(000)$;} & \multirow{1}{*}{$C_{2z}$,$\sigma_{y}$,$\mathcal{T}$;} & $R_{1};$  & $1;$  & $1,1,1;$  &  &  & \tabularnewline
 &  &  & $R_{2};$  & $1;$  & $1,-1,1;$  &  &  & \tabularnewline
 &  &  & $R_{3};$  & $1;$  & $-1,1,1;$  &  &  & \tabularnewline
 &  &  & $R_{4};$  & $1;$  & $-1,-1,1;$  &  &  & \tabularnewline
\multirow{1}{*}{$Y$;} & \multirow{1}{*}{$(\frac{1}{2}\frac{1}{2}0)$;} & \multirow{1}{*}{$C_{2z}$,$\sigma_{y}$,$\mathcal{T}$;} & $R_{1};$  & $1;$  & $1,1,1;$  &  &  & \tabularnewline
\end{tabular}
\end{table}

}

{
\global\long\def\arraystretch{1.4}%
\begin{table}[H]
\raggedright{}%
\begin{tabular}{lllllllll}
\multicolumn{8}{l}{SG 37} & \tabularnewline
\hline
\multicolumn{8}{l}{$\Gamma_{o}^{b}$; $\{C_{2z}|000\},\{\sigma_{y}|00\frac{1}{2}\},\mathcal{T}$;
Non-Centrosymmetric; without SOC } & \tabularnewline
\multicolumn{8}{l}{ } & \tabularnewline
\multirow{1}{*}{$\Gamma$;} & \multirow{1}{*}{$(000)$;} & \multirow{1}{*}{$C_{2z}$,$\sigma_{y}$,$\mathcal{T}$;} & $R_{1};$  & $1;$  & $1,1,1;$  &  &  & \tabularnewline
 &  &  & $R_{2};$  & $1;$  & $1,-1,1;$  &  &  & \tabularnewline
 &  &  & $R_{3};$  & $1;$  & $-1,1,1;$  &  &  & \tabularnewline
 &  &  & $R_{4};$  & $1;$  & $-1,-1,1;$  &  &  & \tabularnewline
\multirow{1}{*}{$Y$;} & \multirow{1}{*}{$(\frac{1}{2}\frac{1}{2}0)$;} & \multirow{1}{*}{$C_{2z}$,$\sigma_{y}$,$\mathcal{T}$;} & $R_{1};$  & $1;$  & $1,1,1;$  &  &  & \tabularnewline
\end{tabular}
\end{table}

}

{
\global\long\def\arraystretch{1.4}%
\begin{table}[H]
\raggedright{}%
\begin{tabular}{lllllllll}
\multicolumn{8}{l}{SG 38} & \tabularnewline
\hline
\multicolumn{8}{l}{$\Gamma_{o}^{b}$; $\{C_{2y}|000\},\{\sigma_{x}|000\},\mathcal{T}$;
Non-Centrosymmetric; without SOC } & \tabularnewline
\multicolumn{8}{l}{ } & \tabularnewline
\multirow{1}{*}{$\Gamma$;} & \multirow{1}{*}{$(000)$;} & \multirow{1}{*}{$C_{2y}$,$\sigma_{x}$,$\mathcal{T}$;} & $R_{1};$  & $1;$  & $1,1,1;$  &  &  & \tabularnewline
 &  &  & $R_{2};$  & $1;$  & $1,-1,1;$  &  &  & \tabularnewline
 &  &  & $R_{3};$  & $1;$  & $-1,1,1;$  &  &  & \tabularnewline
 &  &  & $R_{4};$  & $1;$  & $-1,-1,1;$  &  &  & \tabularnewline
\multirow{1}{*}{$Y$;} & \multirow{1}{*}{$(\frac{1}{2}\frac{1}{2}0)$;} & \multirow{1}{*}{$C_{2y}$,$\sigma_{x}$,$\mathcal{T}$;} & $R_{1};$  & $1;$  & $1,1,1;$  &  &  & \tabularnewline
\end{tabular}
\end{table}

}
{
\global\long\def\arraystretch{1.4}%
\begin{table}[H]
\raggedright{}%
\begin{tabular}{lllllllll}
\multirow{1}{*}{$E$;} & \multirow{1}{*}{$\text{TE}$;} & \multirow{1}{*}{$\sigma_{z}$,$\mathcal{T}C_{2y}$;} & $R_{1};$  & $1;$  & $1,1;$  &  &  & \tabularnewline
 &  &  & $R_{2};$  & $1;$  & $-1,1;$  &  &  & \tabularnewline
\multirow{1}{*}{$C$;} & \multirow{1}{*}{$\text{YC}$;} & \multirow{1}{*}{$\sigma_{z}$,$\mathcal{T}C_{2y}$;} & $R_{1};$  & $1;$  & $1,1;$  &  &  & \tabularnewline
 &  &  & $R_{2};$  & $1;$  & $-1,1;$  &  &  & \tabularnewline
\end{tabular}
\end{table}

}

{
\global\long\def\arraystretch{1.4}%
\begin{table}[H]
\raggedright{}%
\begin{tabular}{lllllllll}
\multicolumn{8}{l}{SG 39} & \tabularnewline
\hline
\multicolumn{8}{l}{$\Gamma_{o}^{b}$; $\{C_{2y}|000\},\{\sigma_{x}|\frac{1}{2}\frac{1}{2}0\},\mathcal{T}$;
Non-Centrosymmetric; without SOC } & \tabularnewline
\multicolumn{8}{l}{ } & \tabularnewline
\multirow{1}{*}{$\Gamma$;} & \multirow{1}{*}{$(000)$;} & \multirow{1}{*}{$C_{2y}$,$\sigma_{x}$,$\mathcal{T}$;} & $R_{1};$  & $1;$  & $1,1,1;$  &  &  & \tabularnewline
 &  &  & $R_{2};$  & $1;$  & $1,-1,1;$  &  &  & \tabularnewline
 &  &  & $R_{3};$  & $1;$  & $-1,1,1;$  &  &  & \tabularnewline
 &  &  & $R_{4};$  & $1;$  & $-1,-1,1;$  &  &  & \tabularnewline
\multirow{1}{*}{$Y$;} & \multirow{1}{*}{$(\frac{1}{2}\frac{1}{2}0)$;} & \multirow{1}{*}{$C_{2y}$,$\sigma_{x}$,$\mathcal{T}$;} & $R_{1};$  & $1;$  & $1,1,1;$  &  &  & \tabularnewline
\end{tabular}
\end{table}

}

{
\global\long\def\arraystretch{1.4}%
\begin{table}[H]
\raggedright{}%
\begin{tabular}{lllllllll}
\multicolumn{8}{l}{SG 40} & \tabularnewline
\hline
\multicolumn{8}{l}{$\Gamma_{o}^{b}$; $\{C_{2y}|000\},\{\sigma_{x}|00\frac{1}{2}\},\mathcal{T}$;
Non-Centrosymmetric; without SOC } & \tabularnewline
\multicolumn{8}{l}{ } & \tabularnewline
\multirow{1}{*}{$\Gamma$;} & \multirow{1}{*}{$(000)$;} & \multirow{1}{*}{$C_{2y}$,$\sigma_{x}$,$\mathcal{T}$;} & $R_{1};$  & $1;$  & $1,1,1;$  &  &  & \tabularnewline
 &  &  & $R_{2};$  & $1;$  & $1,-1,1;$  &  &  & \tabularnewline
 &  &  & $R_{3};$  & $1;$  & $-1,1,1;$  &  &  & \tabularnewline
 &  &  & $R_{4};$  & $1;$  & $-1,-1,1;$  &  &  & \tabularnewline
\multirow{1}{*}{$Y$;} & \multirow{1}{*}{$(\frac{1}{2}\frac{1}{2}0)$;} & \multirow{1}{*}{$C_{2y}$,$\sigma_{x}$,$\mathcal{T}$;} & $R_{1};$  & $1;$  & $1,1,1;$  &  &  & \tabularnewline
\end{tabular}
\end{table}

}
\include{basicinfo/SG41}
\include{basicinfo/SG51}
\include{basicinfo/SG61}
\include{basicinfo/SG71}
\include{basicinfo/SG81}
\include{basicinfo/SG91}
\include{basicinfo/SG101}
\include{basicinfo/SG111}
\include{basicinfo/SG121}
\include{basicinfo/SG131}
\include{basicinfo/SG141}
\include{basicinfo/SG151}
\include{basicinfo/SG161}
\include{basicinfo/SG171}
\include{basicinfo/SG181}
\include{basicinfo/SG191}
\include{basicinfo/SG201}
\include{basicinfo/SG211}
\include{basicinfo/SG221}
%
\subsection{The accidental  degeneracies on high-symmetry line} \label{S3C}
\subsubsection{Notes to Sec. \ref{S3C}}

(i) For each table in Sec. \ref{S3C}, the first  line  presents the  SG number.

(ii) Below the first  line,  the columns from left to right (separated by the semicolons)  are the high-symmetry momentum $\bm{k}$, the location  of $\bm{k}$, the generating elements of the little group at $\bm{k}$ (only point-group operators are presented and a full expression of each  generating element can be found in Sec. \ref{SGs-opers}),  the  two distinct coreps (separated by the comma) of the  bands forming the accidental degeneracy,  the degeneracy  of the accidental degeneracy, the matrix representations of the generating elements,  the species and the topological charge of the  accidental degeneracy.

(iii) We  do not list the type II MSGs that do not exhibit symmetry-protected  accidental degeneracies on high-symmetry line.

\subsubsection{SG 1-10}
\include{basicinfo/Acc_SG1}
\include{basicinfo/Acc_SG11}
\include{basicinfo/Acc_SG21}
\include{basicinfo/Acc_SG31}
\include{basicinfo/Acc_SG41}
\include{basicinfo/Acc_SG51}
\include{basicinfo/Acc_SG61}
\include{basicinfo/Acc_SG71}
\include{basicinfo/Acc_SG81}
\include{basicinfo/Acc_SG91}
\include{basicinfo/Acc_SG101}
\include{basicinfo/Acc_SG111}
\include{basicinfo/Acc_SG121}
\include{basicinfo/Acc_SG131}
\include{basicinfo/Acc_SG141}
\include{basicinfo/Acc_SG151}
\include{basicinfo/Acc_SG161}
\include{basicinfo/Acc_SG171}
\include{basicinfo/Acc_SG181}
\include{basicinfo/Acc_SG191}
\include{basicinfo/Acc_SG201}
\include{basicinfo/Acc_SG211}
\include{basicinfo/Acc_SG221}
%
\subsection{Effective Hamiltonian of   both essential and accidental  degeneracies} \label{S3D}

\subsubsection{Notes to Sec. \ref{S3D}}

(i) The top and bottom part of the tables in  Sec. \ref{S3D} lists the essential and accidental degeneracy, respectively.

(ii) For each table in Sec. \ref{S3D}, the first two lines  present the  SG number,  the BZ type, the generating elements of the type II MSG (translations are not included here), whether centrosymmetry is contained in the group, and whether SOC is considered.

(iii) Below the first two lines,  the columns from left to right (separated by the semicolons)  are the high-symmetry momentum $\bm{k}$, the   corep and the effective Hamiltonian of the symmetry-protected  degeneracies.

(iv) In effective Hamiltonian, we use Roman letters (such as $c_i$ and $c_{i,j}$) and Greek letter (such as $\alpha_i$) to denote
the real and complex parameters, respectively.

(v) We do not list the type II MSGs that do not exhibit symmetry-protected  degeneracies at  high-symmetry point and high-symmetry line.

\subsubsection{SG 1-10}
\include{basicinfo/mergeSG1}
\include{basicinfo/mergeSG11}
\include{basicinfo/mergeSG21}
\include{basicinfo/mergeSG31}
\include{basicinfo/mergeSG41}
\include{basicinfo/mergeSG51}
\include{basicinfo/mergeSG61}
\include{basicinfo/mergeSG71}
\include{basicinfo/mergeSG81}
\include{basicinfo/mergeSG91}
\include{basicinfo/mergeSG101}
\include{basicinfo/mergeSG111}
\include{basicinfo/mergeSG121}
\include{basicinfo/mergeSG131}
\include{basicinfo/mergeSG141}
\include{basicinfo/mergeSG151}
\include{basicinfo/mergeSG161}
\include{basicinfo/mergeSG171}
\include{basicinfo/mergeSG181}
\include{basicinfo/mergeSG191}
\include{basicinfo/mergeSG201}
\include{basicinfo/mergeSG211}
\include{basicinfo/mergeSG221}
%
%
%
%
\section{Encyclopedia of emergent particles in 3D  crystals with SOC effect} \label{S4}

\subsection{The double-valued corepresentations of the 230 type-II MSGs and the essential degeneracies} \label{S4B}

\subsubsection{Notes to Sec. \ref{S4B}}

(i) For each table in Sec. \ref{S4B}, the first two lines  present the  SG number,  the BZ type, the generating elements of the type II MSG (translations are not included here), whether centrosymmetry is contained in the group and whether SOC is considered.

(ii) Below the first two lines,  the columns from left to right (separated by the semicolons)  are the high-symmetry momentum $\bm{k}$, the location  of $\bm{k}$ [with respect to reciprocal vectors $(\boldsymbol{g}_{1},\boldsymbol{g}_{2},\boldsymbol{g}_{3})$], the generating elements of the little group at $\bm{k}$ (only point-group operators are presented and a full expression of each  generating element can be found in Ref. \cite{Bradley2009Mathematical-Oxford} and in Sec. \ref{SGs-opers}),  the deduced corep of the little group at $\bm{k}$, the dimension  of the corep, the matrix representations of the generating elements,  the species and the topological charge of the essential  degeneracy.

(iii) A correspondence between the notation of the corep used here ($R_i$) and the band-representation notations can be found in Refs. \cite{Bradley2009Mathematical-Oxford, GBLiu_2020}. Moreover, Ref. \cite{GBLiu_2020} has established a SpaceGroupIrep package to analyze the band representation based on the notation of Ref. \cite{Bradley2009Mathematical-Oxford}.

\subsubsection{SG 1-10}
\include{dSG/DSG1}
\include{dSG/DSG11}
\include{dSG/DSG21}
\include{dSG/DSG31}
\include{dSG/DSG41}
\include{dSG/DSG51}
\include{dSG/DSG61}
\include{dSG/DSG71}
\include{dSG/DSG81}
\include{dSG/DSG91}
\include{dSG/DSG101}
\include{dSG/DSG111}
\include{dSG/DSG121}
\include{dSG/DSG131}
\include{dSG/DSG141}
\include{dSG/DSG151}
\include{dSG/DSG161}
\include{dSG/DSG171}
\include{dSG/DSG181}
\include{dSG/DSG191}
\include{dSG/DSG201}
\include{dSG/DSG211}
\include{dSG/DSG221}
%
\subsection{The accidental  degeneracies on high-symmetry line} \label{S4C}
\subsubsection{Notes to Sec. \ref{S4C}}

(i) For each table in Sec. \ref{S4C}, the first  line  presents the  SG number.

(ii) Below the first  line,  the columns from left to right (separated by the semicolons)  are the high-symmetry momentum $\bm{k}$, the location  of $\bm{k}$, the generating elements of the little group at $\bm{k}$ (only point-group operators are presented and a full expression of each  generating element can be found in Sec. \ref{SGs-opers}),  the  two distinct coreps (separated by the comma) of the  bands forming the accidental degeneracy,  the degeneracy  of the accidental degeneracy, the matrix representations of the generating elements,  the species and the topological charge of the  accidental degeneracy.

(iii) We  do not list the type II MSGs that do not exhibit symmetry-protected  accidental degeneracies on high-symmetry line.

\subsubsection{SG 1-10}
{
\global\long\def\arraystretch{1.4}%

\begin{table}[H]
\raggedright{}%
\begin{tabular}{llllllllll}
\multicolumn{8}{l}{SG 3} & \tabularnewline
\hline
\multicolumn{8}{l}{Accidental degeneracies on high symmetry line} & \tabularnewline
\multicolumn{8}{l}{ } & \tabularnewline
\multirow{1}{*}{$\Lambda$;} & \multirow{1}{*}{$\text{\ensuremath{\Gamma}Z}$;} & \multirow{1}{*}{$C_{2z}$;} & $\left\{ R_{2}\right\} ,\left\{ R_{4}\right\} ;$  & $2;$  & $i\sigma_{3};$  & C-1 WP; & 1 & \tabularnewline
\multirow{1}{*}{$V$;} & \multirow{1}{*}{$\text{BD}$;} & \multirow{1}{*}{$C_{2z}$;} & $\left\{ R_{2}\right\} ,\left\{ R_{4}\right\} ;$  & $2;$  & $i\sigma_{3};$  & C-1 WP; & 1 & \tabularnewline
\multirow{1}{*}{$W$;} & \multirow{1}{*}{$\text{YC}$;} & \multirow{1}{*}{$C_{2z}$;} & $\left\{ R_{2}\right\} ,\left\{ R_{4}\right\} ;$  & $2;$  & $i\sigma_{3};$  & C-1 WP; & 1 & \tabularnewline
\multirow{1}{*}{$U$;} & \multirow{1}{*}{$\text{AE}$;} & \multirow{1}{*}{$C_{2z}$;} & $\left\{ R_{2}\right\} ,\left\{ R_{4}\right\} ;$  & $2;$  & $i\sigma_{3};$  & C-1 WP; & 1 & \tabularnewline
\end{tabular}
\end{table}

}{
\global\long\def\arraystretch{1.4}%

\begin{table}[H]
\raggedright{}%
\begin{tabular}{llllllllll}
\multicolumn{8}{l}{SG 4} & \tabularnewline
\hline
\multicolumn{8}{l}{Accidental degeneracies on high symmetry line} & \tabularnewline
\multicolumn{8}{l}{ } & \tabularnewline
\multirow{1}{*}{$\Lambda$;} & \multirow{1}{*}{$\text{\ensuremath{\Gamma}Z}$;} & \multirow{1}{*}{$C_{2z}$;} & $\left\{ R_{2}\right\} ,\left\{ R_{4}\right\} ;$  & $2;$  & $i\sigma_{3};$  & C-1 WP; & 1 & \tabularnewline
\multirow{1}{*}{$V$;} & \multirow{1}{*}{$\text{BD}$;} & \multirow{1}{*}{$C_{2z}$;} & $\left\{ R_{2}\right\} ,\left\{ R_{4}\right\} ;$  & $2;$  & $i\sigma_{3};$  & C-1 WP; & 1 & \tabularnewline
\multirow{1}{*}{$W$;} & \multirow{1}{*}{$\text{YC}$;} & \multirow{1}{*}{$C_{2z}$;} & $\left\{ R_{2}\right\} ,\left\{ R_{4}\right\} ;$  & $2;$  & $i\sigma_{3};$  & C-1 WP; & 1 & \tabularnewline
\multirow{1}{*}{$U$;} & \multirow{1}{*}{$\text{AE}$;} & \multirow{1}{*}{$C_{2z}$;} & $\left\{ R_{2}\right\} ,\left\{ R_{4}\right\} ;$  & $2;$  & $i\sigma_{3};$  & C-1 WP; & 1 & \tabularnewline
\end{tabular}
\end{table}

}{
\global\long\def\arraystretch{1.4}%

\begin{table}[H]
\raggedright{}%
\begin{tabular}{llllllllll}
\multicolumn{8}{l}{SG 5} & \tabularnewline
\hline
\multicolumn{8}{l}{Accidental degeneracies on high symmetry line} & \tabularnewline
\multicolumn{8}{l}{ } & \tabularnewline
\multirow{1}{*}{$\Lambda$;} & \multirow{1}{*}{$\text{\ensuremath{\Gamma}Z}$;} & \multirow{1}{*}{$C_{2z}$;} & $\left\{ R_{2}\right\} ,\left\{ R_{4}\right\} ;$  & $2;$  & $i\sigma_{3};$  & C-1 WP; & 1 & \tabularnewline
\multirow{1}{*}{$U$;} & \multirow{1}{*}{$\text{AM}$;} & \multirow{1}{*}{$C_{2z}$;} & $\left\{ R_{2}\right\} ,\left\{ R_{4}\right\} ;$  & $2;$  & $i\sigma_{3};$  & C-1 WP; & 1 & \tabularnewline
\end{tabular}
\end{table}

}
\include{dSG/Acc_DSG11}
\include{dSG/Acc_DSG21}
\include{dSG/Acc_DSG31}
\include{dSG/Acc_DSG41}
\include{dSG/Acc_DSG51}
\include{dSG/Acc_DSG61}
\include{dSG/Acc_DSG71}
\include{dSG/Acc_DSG81}
\include{dSG/Acc_DSG91}
\include{dSG/Acc_DSG101}
\include{dSG/Acc_DSG111}
\include{dSG/Acc_DSG121}
\include{dSG/Acc_DSG131}
\include{dSG/Acc_DSG141}
\include{dSG/Acc_DSG151}
\include{dSG/Acc_DSG161}
\include{dSG/Acc_DSG171}
\include{dSG/Acc_DSG181}
\include{dSG/Acc_DSG191}
\include{dSG/Acc_DSG201}
\include{dSG/Acc_DSG211}
\include{dSG/Acc_DSG221}
%
%
\subsection{Effective Hamiltonian of  both essential and accidental  degeneracies} \label{S4D}

\subsubsection{Notes to Sec. \ref{S4D}}

(i) The top and bottom part of the tables in  Sec. \ref{S4D} lists the essential and accidental degeneracy, respectively.

(ii) For each table in Sec. \ref{S4D}, the first two lines  present the  SG number,  the BZ type, the generating elements of the type II MSG (translations are not included here), whether centrosymmetry is contained in the group, and whether SOC is considered.

(iii) Below the first two lines,  the columns from left to right (separated by the semicolons)  are the high-symmetry momentum $\bm{k}$, the   corep and the effective Hamiltonian of the symmetry-protected  degeneracies.

(iv) In effective Hamiltonian, we use Roman letters (such as $c_i$ and $c_{i,j}$) and Greek letter (such as $\alpha_i$) to denote
the real and complex parameters, respectively.

(v) We do not list the type II MSGs that do not exhibit symmetry-protected  degeneracies at  high-symmetry point and high-symmetry line.

\subsubsection{SG 1-10}
\include{dSG/mergeDSG1}
\include{dSG/mergeDSG11}
\include{dSG/mergeDSG21}
\include{dSG/mergeDSG31}
\include{dSG/mergeDSG41}
\include{dSG/mergeDSG51}
\include{dSG/mergeDSG61}
\include{dSG/mergeDSG71}
\include{dSG/mergeDSG81}
\include{dSG/mergeDSG91}
\include{dSG/mergeDSG101}
\include{dSG/mergeDSG111}
\include{dSG/mergeDSG121}
\include{dSG/mergeDSG131}
\include{dSG/mergeDSG141}
\include{dSG/mergeDSG151}
\include{dSG/mergeDSG161}
\include{dSG/mergeDSG171}
\include{dSG/mergeDSG181}
\include{dSG/mergeDSG191}
\include{dSG/mergeDSG201}
\include{dSG/mergeDSG211}
\include{dSG/mergeDSG221}

\bibliography{SM_fermion_ref}